\documentclass[twocolumn]{aastex63}

\usepackage{gensymb}
\usepackage{natbib}
\usepackage{multirow}

\usepackage{longtable}
\LTcapwidth=\textwidth
\usepackage[utf8]{inputenc}
\usepackage[maxfloats=1000]{morefloats}
\shorttitle{XZ and the CSC}
\shortauthors{Sicilian et al.}

\turnoffedit

\begin{document}

\title{X-Ray Redshifts of Obscured \textit{Chandra} Source Catalog AGN} 
\email{d.sicilian@miami.edu}

\author{Dominic Sicilian}
\affiliation{University of Miami, Department of Physics, Coral Gables, FL}

\author{Francesca Civano}
\affiliation{Center for Astrophysics $|$ Harvard \& Smithsonian, 60 Garden Street, Cambridge, MA}

\author{Nico Cappelluti}
\affiliation{University of Miami, Department of Physics, Coral Gables, FL}

\author{Johannes Buchner}
\affiliation{Max Planck Institute for Extraterrestrial Physics, Giessenbachstrasse, D-85741 Garching, Germany}

\author{Alessandro Peca}
\affiliation{University of Miami, Department of Physics, Coral Gables, FL}

\begin{abstract}

We have computed obscured AGN redshifts using the XZ method, adopting a broad treatment in which we employed a wide-ranging data set and worked primarily at the XZ counts sensitivity threshold, culminating with a redshift catalog containing 121 sources that lack documented redshifts. We considered 363 obscured AGN from the \textit{Chandra} Source Catalog Release 2.0, 59 of which were selected using multiwavelength criteria while 304 were X-ray selected. One-third of the data set had cross-matched spectroscopic or photometric redshifts. These sources, dominated by low-$z$ and low-$N_H$ AGN, were supplemented by 1000 simulations to form a data set for testing the XZ method. We used a multi-layer perceptron neural network to examine and predict cases in which XZ fails to reproduce the known redshift, yielding a classifier that can identify and discard poor redshift estimates. This classifier demonstrated a statistically significant $\sim$3$\sigma$ improvement over the existing XZ redshift information gain filter. We applied the machine learning model to sources with no documented redshifts, resulting in the 121-source new redshift catalog, all of which were X-ray selected. Our neural network's performance suggests that nearly 90\% of these redshift estimates are consistent with hypothetical spectroscopic or photometric measurements, strengthening the notion that redshifts can be reliably estimated using only X-rays, which is valuable to current and future missions such as Athena. We have also identified a possible Compton-thick candidate that warrants further investigation.

\end{abstract}


\section{Introduction}\label{intro}

Accretion of galactic material onto central supermassive black holes at an extraordinarily high rate results in the most powerful non-explosive class of objects in the Universe, known as Active Galactic Nuclei (AGN; \citealt{hickox 2018}). The population of several million documented AGN across redshifts up to $\sim$7.6 (\citealt{wang 2021}) provides a deep window into the history and evolution of galaxies. 

According to the Unified AGN Model (\citealt{antonucci 1993}; \citealt{urry 1995}; \citealt{netzer 2015}), a toroidal structure of optically thick material looms at the edge of the AGN accretion disk. The diverse geometrical orientation of galaxies relative to the line of sight results in the obscuration of accretion disk emission by this torus for many such orientations, and hence a large portion of AGN are expected to be obscured. This hypothesis is strongly supported by Cosmic X-ray Background (CXB) measurements (see, e.g., \citealt{moretti 2009}, \citealt{cappelluti 2017}), in which the CXB spectral peak at $E\sim30$ keV has been attributed to an obscured AGN population with spectra dominated by hard X-ray emission due to absorption at lower energies (\citealt{gilli 2007}; \citealt{treister 2009}; \citealt{ballantyne 2011}; \citealt{akylas 2012}; \citealt{ueda 2014}; \citealt{aird 2015a}; \citealt{aird 2015b}; \citealt{hickox 2019}; \citealt{ananna 2020}). A complete account of accretion and its role in galaxy evolution is therefore reliant on identifying and studying obscured AGN.

In particular, measuring redshift ($z$) is valuable when piecing together this cosmological history via various means such as studying AGN accretion history and further developing AGN population synthesis models via the luminosity function and space density measurements (see, e.g., \citealt{boyle 1998}, \citealt{cowie 2003}, \citealt{ueda 2003}, \citealt{gilli 2007}, \citealt{hasinger 2008}, \citealt{treister 2009}, \citealt{ueda 2014}, \citealt{aird 2015a}, \citealt{buchner 2015}, \citealt{ananna 2019}, \citealt{ananna 2020}, \citealt{kirkpatrick 2020}, \citealt{ananna 2020b}). However, computing redshifts of obscured AGN is well known for being a highly difficult process (see \citealt{simmonds 2018}, \citealt{peca 2021}, and others referenced therein), due largely to the host galaxy contributing more to the overall emission than in the case of an unobscured AGN. Spectroscopic redshifts using features in the UV to near-IR bands are ideal and were central to the discovery of obscured AGN (\citealt{khaki 1974}; \citealt{weedman 1977}; \citealt{hickox 2018}), but are generally not practical due to the considerable required observing time and sensitivity limitations of current instruments. 

Photometric redshifts, which consider spectral energy distributions (SEDs), are often used instead. These are reliable for galaxies without AGN, particularly when the SEDs have distinct features such as Lyman and Balmer breaks, emission lines, etc. \citep{simmonds 2018}. An AGN, however, can significantly contribute to the total emission of its host galaxy, making it difficult to disentangle the contributions of the two and therefore attempts to estimate the photometric redshift will often yield multiple degenerate solutions \citep{salvato 2009}.

To avoid the issues encountered by traditional spectroscopic and photometric redshifts for obscured AGN, \cite{simmonds 2018} turned to X-ray spectra, devising a method known as XZ to compute the redshift using only X-ray spectral features---namely the photoelectric cutoff and 7.1 keV Fe K$\alpha$ absorption edges and, when possible, the 6.4 keV Fe K$\alpha$ emission line. The XZ method is similar in philosophy to the approaches taken in various other works (see, e.g., \citealt{maccacaro 2004}, \citealt{braito 2005}, \citealt{civano 2005}, \citealt{iwasawa 2012}, \citealt{vignali 2015}, \citealt{iwasawa 2020}, \citealt{peca 2021}), but XZ is distinguished by its powerful statistical foundation, streamlined modeling procedure (\citealt{buchner 2014}, \citealt{simmonds 2018}), and extensive publicly-available software, which have brought the method success in estimating obscured AGN redshifts.

In addition to its practicality, deriving redshift estimates from X-ray spectra introduces another advantage to analyzing X-ray selected obscured AGN, namely that it does not initially require matching a given AGN with multiwavelength counterparts. Such matching is known to be difficult due to the potentially large uncertainties in X-ray position (see, e.g., \citealt{hsu 2014}, \citealt{salvato 2018}), and hence future X-ray telescopes with large PSFs could take advantage of this method. Moreover, future CSC releases and surveys performed by recently-launched or future X-ray missions such as eROSITA (\citealt{erosita 2010}, \citealt{erosita 2012}, \citealt{erosita 2021}), Athena (\citealt{nandra 2013}, \citealt{athena 2018}, \citealt{athena 2020}, \citealt{athena 2021}), AXIS (\citealt{axis 2019}), and Lynx (\citealt{lynx 2018a}, \citealt{lynx 2018}, \citealt{lynx 2019}) will initially lack counterparts. Therefore, it is vital to have well-tested and effective tools to characterize and analyze the physical nature of observed objects in vast, X-ray-only data sets. In particular, for X-ray AGN with no multiwavelength counterparts, an exclusively X-ray approach can yield reliable redshift estimates without the need for other surveys.

In this work, we first test the XZ method on obscured AGN with documented redshifts, as well as on a simulated data set, to explore the limits and further constrain the capabilities of XZ, ultimately formulating a machine learning-based procedure to maximize its effectiveness on broad X-ray data sets. We then apply XZ to a large set of obscured AGN without documented redshifts in the Simbad or SDSS DR12 databases to produce a catalog of redshift estimates. Finally, we examine the best-fit models and parameters of interest to identify any peculiar or interesting sources, such as Compton-thick AGN, that warrant further study.

Our implementation of machine learning comes in the midst of a growing trend in astrophysics, data science, and other fields in which researchers are using such algorithms to detect complex patterns in data sets with many-dimensional parameter spaces (see \citealt{vp 2012}, \citealt{ferguson 2018}, \citealt{baron 2019}, \citealt{schmidt 2019}, etc.). We detail the algorithm and the realization of its capabilities on analyzing XZ results, as well as on applying them to our final redshift catalog in Section \ref{machine learning}. With the power of this machine learning approach, we produced a refined set of redshift estimates for 121 obscured AGN, none of which have cross-matched documented redshifts and all of which were selected using only X-rays.

For this analysis, we considered archival data from \textit{Chandra} due to its superb angular resolution relative to other observatories. In particular, its 0.5 arcsec resolution allows it to resolve as much as $\sim$90\% of the AGN believed to compose the CXB (\citealt{hickox 2006}, \citealt{hickox 2007}, \citealt{cappelluti 2017}), providing the deepest look into AGN among all current X-ray observatories. We also employed the \textit{Chandra} Source Catalog Release 2.0 (CSC; \citealt{evans 2010}; \citealt{evans 2019}), which has been shown in several recent works to be an excellent resource for large-scale data analysis (e.g., \citealt{kovlakas 2020}, \citealt{sicilian 2020}, \citealt{yang 2021}) and is increasingly well-equipped for data science-oriented applications. We adopt a $\Lambda$CDM cosmology throughout, with $H_0 = 67.7$ km s$^{-1}$ Mpc$^{-1}$, $\Omega_M = 0.307$, and $\Omega_{\Lambda} = 0.693$, which are approximately the results of Planck 2015 \citep{planck15}.

\section{Methodology}

\cite{simmonds 2018}'s XZ is powered by the Bayesian X-ray Analysis (BXA; \citealt{buchner 2014}) software, which adopts an advanced Bayesian approach to modeling X-ray spectra\footnote{Find BXA's AGN modeling tool \texttt{XAGNFitter} here: \url{https://github.com/JohannesBuchner/BXA/blob/master/examples/sherpa/xagnfitter.py}}. In particular, BXA utilizes the Nested Sampling algorithm \citep{skilling 2004}, an MCMC-like process that yields posterior parameter probability distributions from priors while exploring the parameter space more efficiently than MCMC (see \citealt{buchner 2014b} for a demonstration of Nested Sampling's capabilities). Additionally, XZ incorporates an advanced background model that operates on principal component analysis (PCA), first published in \cite{simmonds 2018} and now available in BXA. PCA is a machine learning data analysis technique, and BXA's application of the procedure to background modeling has been established as a highly effective approach in various recent works (see, e.g., \citealt{brusa 2021}, \citealt{liu 2021}, \citealt{arcodia 2021}, \citealt{malyali 2021}, \citealt{wolf 2021}, \citealt{zhu 2021}).

\subsection{The Original XZ Method}

\cite{simmonds 2018} considered 321 sources from the \textit{Chandra} Deep Field South (CDF-S) in which all sources were accompanied by robust optical redshift measurements, unlike our data set which is dominated by X-ray selected AGN with no cross-matched counterparts. The XZ model is BXA's \texttt{torus + scattering} detailed in \cite{buchner 2014}, with \texttt{torus} referring to the \cite{brightman 2011} \texttt{TORUS} model \edit1{(which includes an absorbed power law and accounts for torus geometry, expressed here as the opening angle $\theta_{\mathrm{op}}$ and torus viewing angle $\theta_{\mathrm{view}}$; see \citealt{brightman 2011} and \citealt{buchner 2014} for additional details)} and a \texttt{scattering} power law component corresponding to an apparent leakage of AGN emission \citep{ueda 2007} in a process separate from both Compton scattering and reflection (see \citealt{krolik 1987}, \citealt{turner 1997}, \citealt{guainazzi 2007}, \citealt{buchner 2014} for further explanation; \edit1{throughout this work, as in \citealt{simmonds 2018}, we express the scattering power law normalization as a fraction $f_{\mathrm{scat}}$ of the AGN power law normalization}). \edit1{XZ spectral fitting in \cite{simmonds 2018}, as well as in this work, was done using Sherpa \citep{sherpa}.}

Galactic absorption was applied to the model, solar abundances were assumed, and uniform priors were used for both redshift and source $N_H$ (measured in units of cm$^{-2}$ hereafter, even when not explicitly specified). The posterior redshift distribution for each spectrum was screened for high Information Gain (IG), which quantifies its divergence from the uniform prior, leaving 74 sources for final evaluation. The IG was computed using the Kullback–Leibler definition (shown below, measured in bits; \citealt{kullback}) and the threshold adopted for high IG was 2 bits.
\[
    \mathrm{IG} = \int{\mathrm{Posterior}(z) \;\;   \mathrm{log}_2\left(\frac{\mathrm{Posterior}(z)}{\mathrm{Prior}(z)}\right)\; \mathrm{d}z} \;\;\; \mathrm{bits}
\]
Among these sources, both spectroscopic and photometric redshift values were reliably reproduced and constrained by XZ. The difference between the XZ estimate and known redshift value was generally very small, with $\sim$60\% of known redshift values being within 1$\sigma$ of XZ and $\sim$75\% of XZ best-fit values falling within 0.15(1 + z) of the known value. Another important evaluation of XZ was the uncertainty of the redshift estimate, $\sigma(z)$, which was $\sim$0.2 on average. As noted by \cite{simmonds 2018}, while this is not ideal, it is a sufficient result for XZ's purpose, which is to offer a good redshift estimate when other methods are unavailable or impractical. Thus, the work was able to conclude that fits with high redshift IG can produce good redshift estimates that are satisfactorily well-constrained.

\subsection{XZ Method Updates}

Our implementation of XZ remains largely faithful to the original work, with our model differing only in the use of the \cite{wilm 2000} measured ISM abundances and \cite{vern} cross sections in an \texttt{APEC} component imported from \texttt{XSPEC} \citep{arnaud 1996}. However, our approach is broader and more aggressive due to our goals of testing its limits and providing redshifts for a large collection of AGN with no reported literature value. In particular, we explored the low-counts boundary of the simulated sensitivity threshold ($\sim$150), generally considering lower-counts spectra than \cite{simmonds 2018}, and we also were not limited to sources with multiwavelength counterparts. Hence, the majority of the data set consists of low-statistics, X-ray only spectra. We also searched widely for obscured AGN by querying the entire CSC, whereas XZ's ideal sources come from deep surveys such as those from CDF-S in \cite{simmonds 2018} due to an increased obscured fraction for higher redshift and lower luminosity (see \citealt{ueda 2003}, \citealt{hasinger 2005}), making our analysis a substantially broadened and more aggressive application of XZ.

As a result, we must further explore the method's accuracy and limitations. Although BXA's nested sampling procedure employs the Cash statistic (CSTAT; \citealt{cash 1979}) which is valid for low counts without rebinning, prior to performing any analysis we rebinned all low-counts spectra such that each bin contained a minimum of 5 counts. \edit1{This was done primarily to estimate goodness of fit using the widely familiar and conventional reduced $\chi^2$ since CSTAT is distributed approximately as $\Delta\chi^2$ for counts per bin $\gg$1 \citep{lanzuisi 2013}. As shown in that work, 5 counts per bin is sufficient for this approximation.} Before producing the final redshift catalog, we calibrated our procedure on both a subset of our sources with known redshifts and a large set of simulations to further demonstrate XZ's reliability and to examine its failure rate. Here, we adopt largely the same evaluation metrics used in the original work, but apply new filtering criteria in addition to the IG screening. The final methodology, described in the ensuing sections, should serve as a guide for obtaining reliable redshift estimates for broad X-ray data sets that lack multiwavelength counterparts, which will be found particularly useful for the aforementioned new and future X-ray observatories.

\begin{figure*}[t!]
\centering
\includegraphics[width=8.5cm]{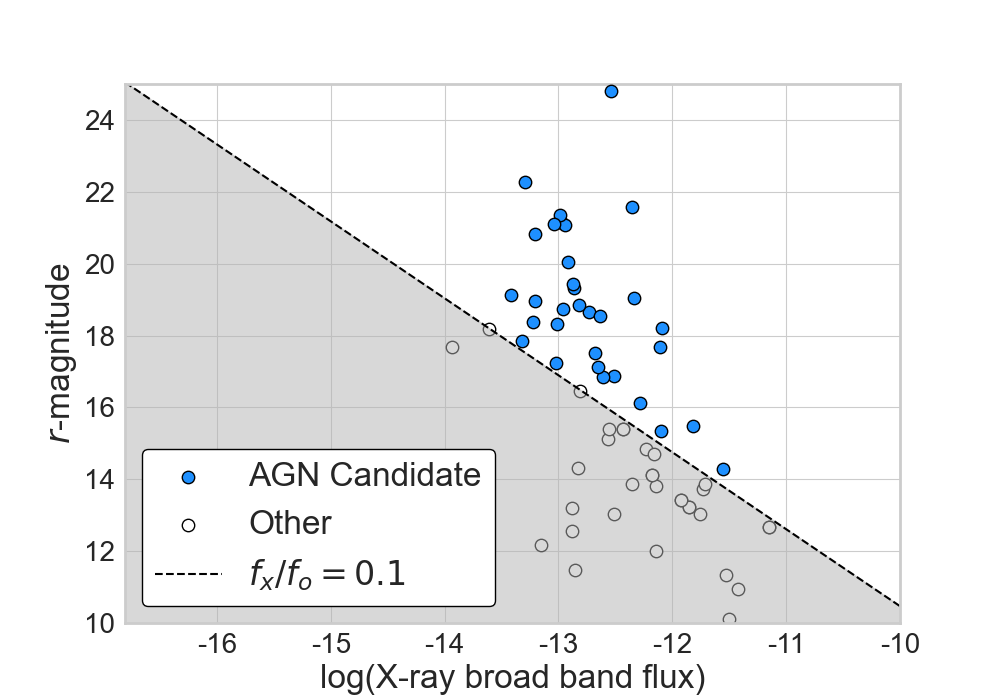}
\includegraphics[width=8.5cm]{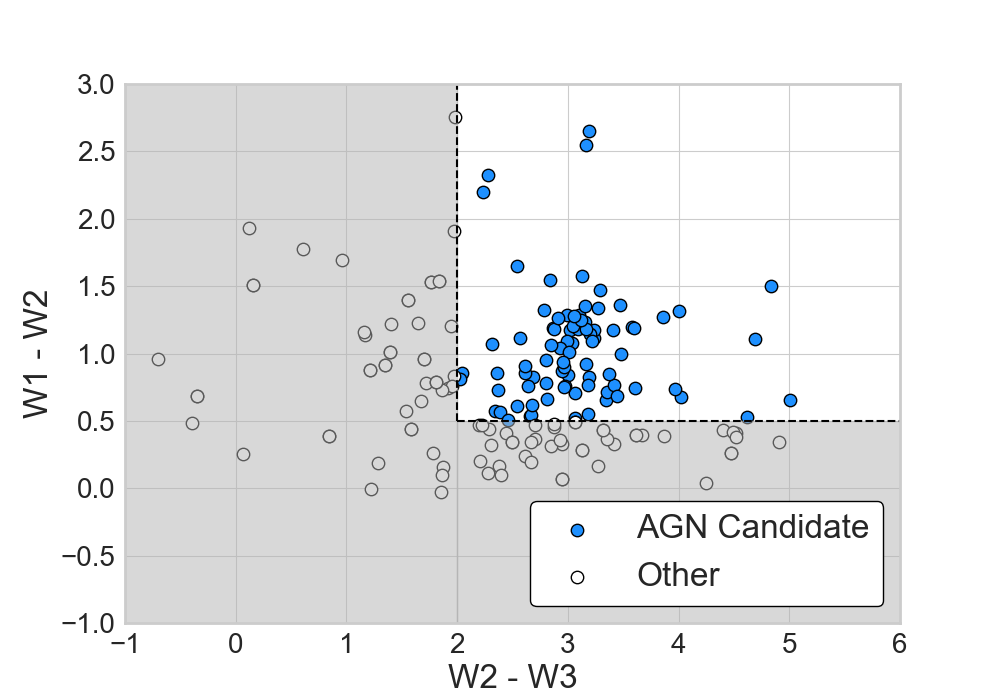}
\caption{Multiwavelength AGN selection. Gray shading denotes non-AGN regions. \textbf{Left:} SDSS $r$-band magnitude is compared to the \textit{Chandra} broad X-ray flux, which indicates the optical-to-X-ray flux ratio selection results. \textbf{Right:} WISE color analysis. In both panels, the selection boundary is marked with dotted lines and passing sources are plotted with blue.}
\label{fig:MW_selection}
\end{figure*}

\section{Data Selection}\label{data selection}

To assemble our data set, we queried the CSC for virtually all obscured AGN detected by the Advanced CCD Imaging Spectrometer (ACIS; \citealt{garmire 2003}) with sufficient counts, employing three simple selection criteria: \textit{Chandra} hard band (2.0 keV--7.0 keV) counts $> 150$ to approximately match the XZ sensitivity threshold, log$N_H > 22$ (as estimated by the CSC's simple absorbed power law model, \edit1{which assumes $z=0$}) to filter unobscured AGN, and galactic latitude $| b |\geq10$ deg to remove all sources in the Galactic Disc (using the same X-ray Galactic Disc boundary as \citealt{sicilian 2020}). This criteria may exclude some obscured AGN, since AGN fitted with parameters $\Gamma < 1.4$ and log$N_H \sim 21$ may in fact be obscured sources \citep{lanzuisi 2018}. However, as the investigation of these potential obscured candidates is outside the scope of this work, we restrict our search only to the CSC-identified obscured sources with log$N_H > 22$. This search yielded $>$700 sources with a combined $>$1600 individual spectra. Note that there are more spectra than sources due to the detection of some sources in more than one observation, hence giving multiple associated spectra.

Spectra were obtained directly from the CSC, which provides fully-reduced data products, including source spectra as well as the corresponding response and background files. For sources detected in multiple observations, all available spectral data for each observation was stacked via the \textit{Chandra} Integrated Analysis of Observations (CIAO; \citealt{antonella ciao}) software tool \texttt{combine\_spectra} to maximize statistics in both the source and background data. The issue of AGN variability among these sources, particularly the effects that stacking variable AGN spectra may have on the XZ results, is addressed in Section \ref{variability}.

This initial data set was cross-matched with several databases in search of multiwavelength counterparts, namely Gaia Data Release 2 (DR2; \citealt{gaia DR2}); the Simbad astronomical database \citep{simbad 2000}, SDSS Data Release 12 (DR12; \citealt{eisenstein 2011}; \citealt{york 2000}; \citealt{alam 2015}; \citealt{blanton 2017}), and WISE (\citealt{wright 2010}; \citealt{cutri 2012}) via the CDS X-Match Service\footnote{\url{http://cdsxmatch.u-strasbg.fr/}} (\citealt{boch 2012}; \citealt{cds 2020}). Matched counterparts offered documented redshift values for many of the sources, as well as multiwavelength AGN selection data. We define two data subsets: all sources with documented redshifts in either Simbad or SDSS DR12 compose the ``Known-z" set, and all others represent the ``No-z" set. Below we detail our AGN selection process.

\subsection{AGN Selection}

When possible, the AGN selection pipeline employed multiwavelength criteria, but in most cases such data were not available or were inconclusive, and hence for those sources we relied on X-ray only selection. Before applying our selection methods, we first made use of parallax data available in Gaia DR2. For sources with counterparts in Gaia DR2, we removed those with parallaxes $\geq$ 5$\sigma$, since all parallax measurements for 556,849 Gaia quasars are less than 5$\sigma$ \citep{gaia parallax}.

\begin{figure*}[t!]
\includegraphics[width=18.cm]{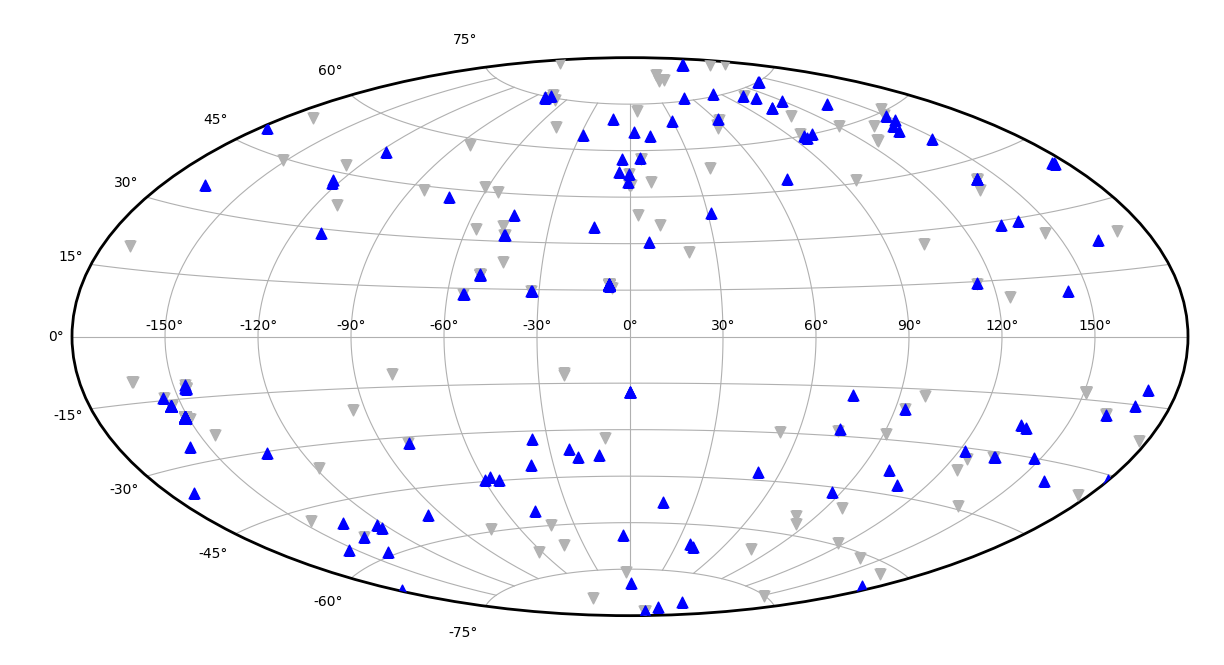}
\caption{Position in galactic coordinates of the 363 obscured AGN selected in our data set. Upward-oriented blue triangles represent the 192 AGN in the final, fully-filtered data set. Downward-oriented gray triangles represent AGN removed from the initial set of 363 during the XZ process.}
\label{fig:map}
\end{figure*}

\subsubsection{Multiwavelength}

For each source with an SDSS-matched $r$-magnitude and WISE-matched spectral data, we first considered the X-ray to optical flux ratio. As thoroughly discussed in various works (see \citealt{civano 2012}, \citealt{marchesi 2016}, \citealt{hickox 2018}, etc.), the ratio of X-ray flux ($f_x$) to optical flux ($f_o$) can be used to distinguish AGN from stars and galaxies, as the powerful emission from AGN accretion results in much higher relative X-ray flux than typical X-ray sources. Here, in accordance with various works including \cite{civano 2012}, we adopt $f_x/f_o = 0.1$ as the minimum AGN ratio. The SDSS $r$-magnitude is plotted against \textit{Chandra} broad X-ray band (0.5--7 keV) flux in Figure \ref{fig:MW_selection}, which shows the results of the flux ratio test.

We then considered WISE spectral data---in particular, flux in the W1, W2, and W3 bands (see \citealt{wright 2010})---which we then used to perform WISE color analysis. As detailed in various works (see, e.g., \citealt{mateos 2014}, \citealt{nikutta 2014}, etc.), AGN can be selected by comparing the colors W1-W2 and W2-W3. Here, we adopt the color thresholds W1-W2 $>$ 0.5 and W2-W3 $>$ 2.0 to form the boundaries of the AGN selection subspace. The results of WISE selection are also shown in Figure \ref{fig:MW_selection}.

 We required sources to meet both of these selection criteria in order to be designated as multiwavelength-selected AGN, and since those criteria each adopted conservative thresholds, we expect little to no contamination in the resulting sample. This yielded 59 multiwavelength selected sources that entered the XZ analysis pipeline, 28 of which remained in the final data set.

\subsubsection{X-ray Hardness Ratio}

A known X-ray signature of obscured AGN is a relatively high ratio of hard- to soft- X-ray flux, due to the higher rate of absorption for lower-energy X-rays \citep{mainieri 2002}. For \textit{Chandra}, these soft- and hard- bands are defined as 0.5--2.0 keV and 2.0--7.0 keV, respectively. The corresponding Hardness Ratio (HR) to compare these bands is defined as
\begin{equation}\label{HR_definition}
\mathrm{HR} = \frac{H-S}{H+S}
\end{equation}
where $H$ and $S$ represent net source counts in the Hard- and Soft- X-ray bands, respectively. Here, for each source, we used the HR provided by the CSC, computed via a Bayesian statistics approach on available \textit{Chandra} data for a given source to produce a best-fit HR value and confidence interval. The methodology is briefly described on the CSC website\footnote{\url{https://cxc.harvard.edu/csc/columns/spectral_properties.html}} and a full explanation can be found in a publicly available CSC memo\footnote{\url{https://cxc.harvard.edu/csc/memos/files/Nowak_csc2_variability_and_color.pdf}}.

As shown in various works (see, e.g., \citealt{ceca 1999}, \citealt{wang 2004}, \citealt{tajer 2007}, \citealt{brightman 2012}, \citealt{marchesi 2016b}, \citealt{peca 2021}, etc.), HR is a useful and reliable method for identifying obscured AGN. One caveat is the time-dependent nature of the \textit{Chandra} ACIS HR, which is due to the degradation of its soft-band spectral response, which has accelerated in recent years (for detailed discussions, see \citealt{plucinsky 2002} and \citealt{peca 2021}). This work considers only data present in the CSC and hence no observations taken later than 2014 are included, thus mitigating the soft-band's efficiency decline.

We conservatively adopt the threshold HR $> -0.1$ from \cite{peca 2021}, in which the data set consists of observations taken in 2017 \citep{nanni 2018} and the HR selection threshold is derived via thorough simulations of obscured AGN. Our adoption of this robust criterion, defined using 2017 \textit{Chandra} data, ensures the integrity of our selection technique, as \cite{peca 2021}'s 2017 data suffers more from the HR degradation than our pre-2015 data set.

We applied our HR criteria to the CSC's best-fit HR value for all sources that did not have sufficient data for the multiwavelength selection pipeline, as well as several sources with inconclusive multiwavelength selection results. The resulting X-ray selected AGN, particularly those with low counts, are the most noteworthy sources in the data set, as their few counts in this single band are expected to provide sufficient data for XZ to estimate their redshifts. This process yielded 304 X-ray selected AGN, giving a total of 363 obscured AGN in our initial data set. The positions of all selected AGN are shown using galactic coordinates in Figure \ref{fig:map}. The final, fully-filtered data set contains 164 X-ray selected AGN. In the following section, we test the limits of the XZ method on our initial 363 AGN.

\section{Testing XZ}

Below, we evaluate our adaptation of XZ with its low counts and more general data set. We first consider the Known-z data subset, fitting the sources and filtering the results for high IG ($\geq$2 bits), allowing a practical test on observed obscured AGN candidates. We repeat this process on a large set of simulations, then evaluate error and determine a procedure for filtering poor results out of the No-z data subset. 

\subsection{Known-z Data}\label{known-z}

The IG-filtered data set contains 302 AGN. Its Known-z subset consists of 105 sources, 85 of which are documented in the Simbad database, while 20 others have spectroscopic or photometric redshifts in SDSS DR12 (``specz" and ``photoz" hereafter, respectively) but are not yet included in Simbad.

\begin{figure}[h!]
\includegraphics[width=10.cm]{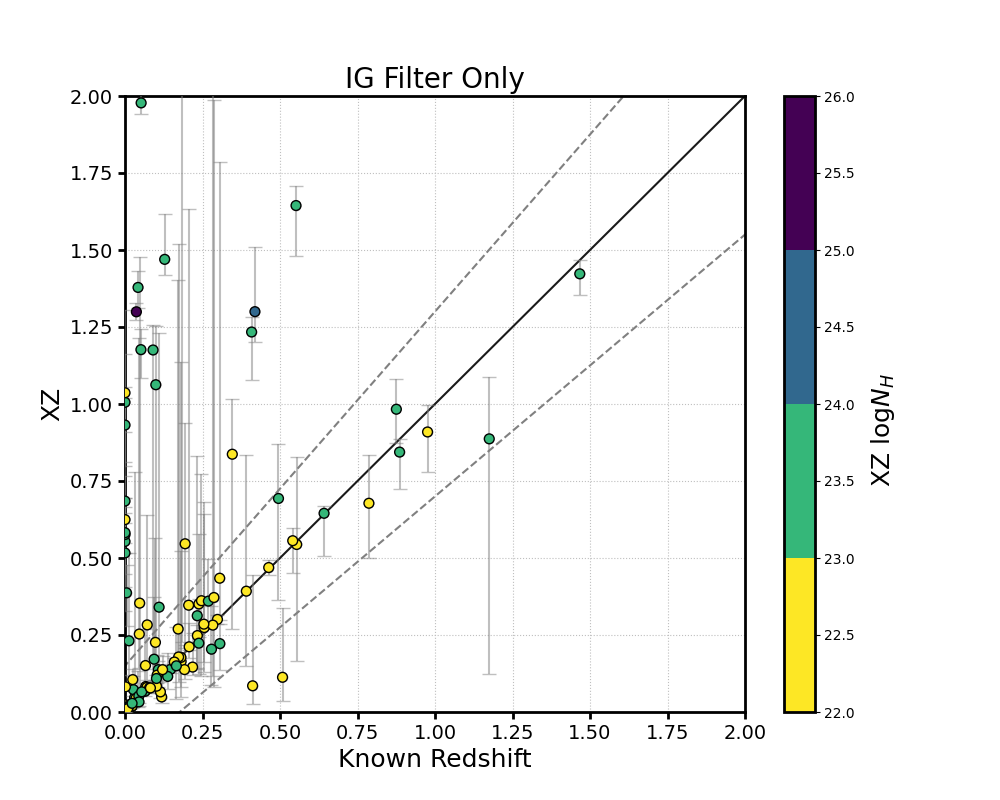}
\caption{Best-fit XZ redshifts plotted against the documented multiwavelength values for Known-z sources with IG $\geq$ 2 bits. Obscuration is indicated using the color map. A solid line shows XZ $=$ Known-z, while dotted lines bound the $\pm$0.15(1 + $z$) uncertainty region.}
\label{fig:IG_only_delz_scatter}
\end{figure}

\begin{table}
\centering
\begin{tabular}{@{\extracolsep{\fill}}c || c | c | c}

 \hline
 Tier & XZ & Requirement & Known $z$ \\
\hline 
\hline

Basic &  1$\sigma$ range & overlaps with & $\pm$0.15(1 + $z$) range \\ 

\hline

Mild & XZ best-fit & included in & $\pm$0.15(1 + $z$) range \\

\hline

Top & 1$\sigma$ range & includes & $z$ best-fit \\

 \hline

\end{tabular}
\caption{Defining the three tiers of success. For each tier, the relevant XZ value is given, as well as the relevant known redshift value, and the Requirement column describes how they must compare to meet the criteria for that tier.} 

\label{tab:XZ_metrics}
\end{table}

Figure \ref{fig:IG_only_delz_scatter} shows a redshift comparison plot similar to those found in \cite{simmonds 2018}, plotting the XZ best-fit redshift against the documented value. This figure best summarizes our measures of success rates, described in the three different success tiers defined in Table \ref{tab:XZ_metrics}. $\sim$79\% of sources achieved basic success, with an overlap between XZ's 1$\sigma$ uncertainty and the known redshift's $\pm0.15(1+z)$ uncertainty region. $\sim$68\% saw mild success, where the XZ best-fit value lies within the $\pm0.15(1+z)$ uncertainty range of the known redshift. Finally, $\sim$49\% achieved top-tier success, where XZ's 1$\sigma$ uncertainty range includes the known redshift value. These values are lower than the results seen in the original work (with mild and top-tier success rates of $\sim$75\% and $\sim$60\%, respectively), which was expected due to our broader data set and less conservative counts threshold, accordingly suggesting that further screening may be needed.

We have also defined a criteria for ``catastrophic failure," in which the 2$\sigma$ confidence interval for XZ has no overlap with the 2$\sigma$ confidence interval for the known redshift, computed as $\pm0.3(1+z)$. For these IG-filtered results, the catastrophic failure rate is $\sim$11\%.

\begin{figure*}

\includegraphics[width=9cm]{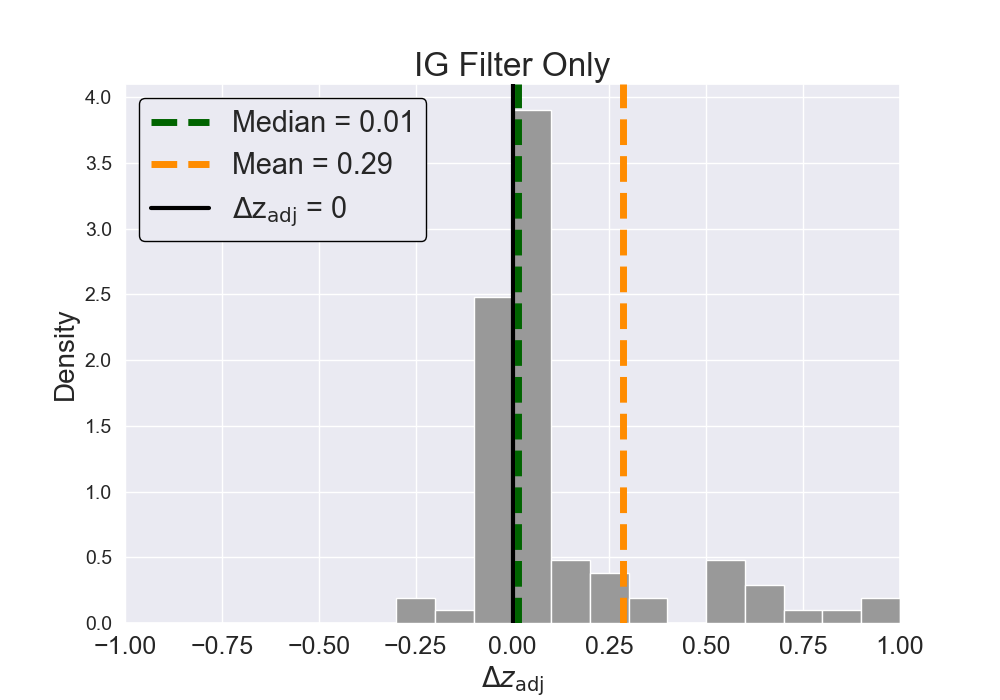}
\includegraphics[width=9.cm]{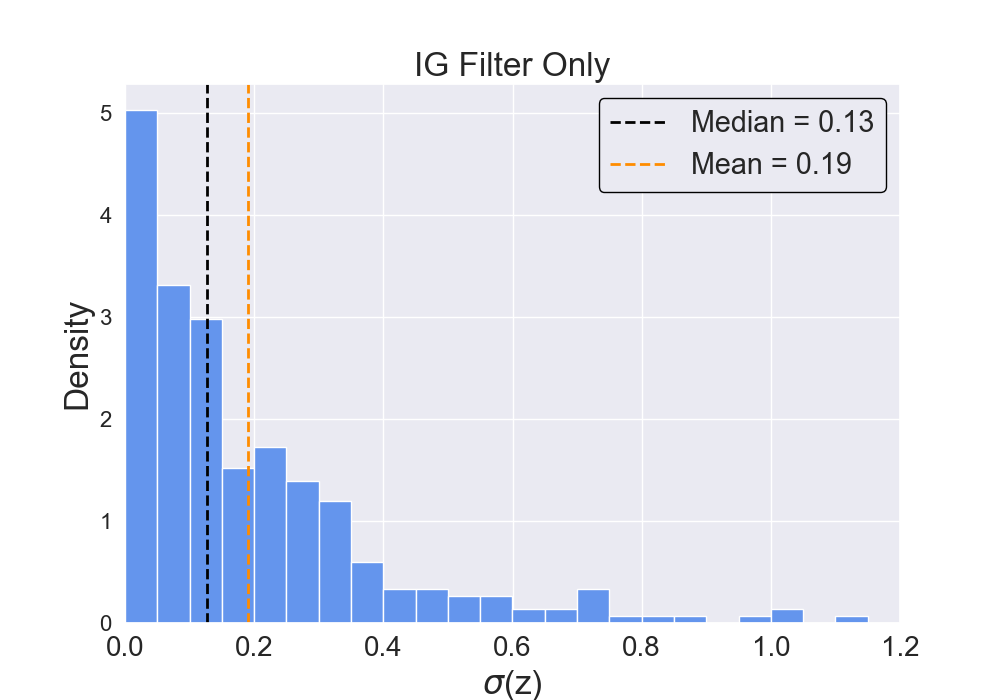}

\caption{\textbf{Left:} The distribution of $\Delta z_{\mathrm{adj}}$ for all Known-z sources with IG $\geq$ 2 bits. A black solid line shows the ideal $\Delta z_{\mathrm{adj}} = 0$, while dotted green and orange lines represent the median and mean observed values, respectively. \textbf{Right:} The distribution of XZ redshift uncertainty for all sources in both Known-z and No-z data sets with IG $\geq$ 2 bits. Dotted black and orange lines represent the median and mean values, respectively.}
\label{fig:Delta_z_histogram}
\end{figure*}

Another measure of success, a commonly adopted metric that is also seen in \cite{simmonds 2018}, is the redshift dispersion ($\Delta z_{\mathrm{adj}}$) given by
\begin{equation}
    \Delta z_{\mathrm{adj}} = \frac{\Delta z}{1 + z_{\mathrm{known}}}
\end{equation}
where $\Delta z$ is computed as
\begin{equation}
\Delta z = \mathrm{XZ} - z_{\mathrm{known}}
\end{equation}
This metric directly compares XZ to the known redshift, offering the most direct, straightforward comparison of the best-fit values. The upper panel of Figure \ref{fig:Delta_z_histogram} shows the distribution of $\Delta z_{\mathrm{adj}}$ for the high IG sources. We see $\Delta z_{\mathrm{adj}}$ peak fairly strongly around zero, with a median of $\sim$0.0, but due to the various failures we see considerable spread in the distribution and a fairly large mean of $\sim$0.3, which again seems to warrant an additional screening process.

The final evaluation metric assesses how well the model constrains the redshift by considering the posterior distribution's uncertainty, $\sigma(z)$. Here, we define $\sigma(z)$ as an average symmetric 1$\sigma$ uncertainty, i.e. one-half the $\pm$1$\sigma$ error range. As this metric refers only to the model's redshift parameter, it can be computed for all sources in both the Known-z and No-z subsets. Figure \ref{fig:Delta_z_histogram} shows the distribution of $\sigma(z)$. Here, we see a median and mean of $\sim$0.13 and $\sim$0.19, respectively, and a fairly strong tendency towards zero. These results are consistent with the $\sim$0.2 found in \cite{simmonds 2018}, which is a good result but is likely a trivial consequence of adopting the same IG threshold, as a high IG is correlated with low uncertainty, with IG $>$ 2 bits roughly corresponding to $\sigma(z) < 0.4$.

In spite of the apparent insight offered by these results, it is clear from Figure \ref{fig:IG_only_delz_scatter} that the Known-z subset is heavily dominated by low-redshift, minimally obscured sources.
This is likely the explanation for why our success rates are lower than those found in \cite{simmonds 2018}, as that work considered higher redshifts and obscuration levels by employing CDF-S. CDF-S and other deep fields are abundant with faint AGN, which are generally high-$z$ or highly-obscured. However, here we considered a counts-selected data set from the general archive which is fundamentally dominated by lower redshifts. Therefore, while we continued to take Known-z into consideration, particularly during the adoption of new filtering techniques, we turned to simulations for a more robust analysis of XZ's capabilities. 

\begin{table*}
\centering
 \begin{tabular}{ r | l | l  }
 
 \hline
 Parameter & Allowed Values & Prior \\
  Name  & \textbf{(Simulations)} & \textbf{(BXA Fitting)} \\ 
  \hline
  
  $z$ & [0.01, 0.3) in steps of 0.01 & Uniform from 0.001 to 5 \\
  log$N_H$ & [22, 26) in steps of 0.01 & Uniform from 20 to 26 \\

  \hline
  
  $\Gamma$ & 1.9 & Gaussian ($\mu = 1.95$, $\sigma = 0.15$) \\
  $f_{\mathrm{scat}}$ & $10^{-4}$ & Uniform from $10^{-7}$ to $10^{-1}$ \\
  $\theta_{\mathrm{op}}$ & 45\degree & - \\
  $\theta_{\mathrm{view}}$ & 80\degree & - \\

  \hline

 \end{tabular}
 
 \caption{The allowed parameter values in the simulated spectra, as described in the text. Free parameters are given using interval notation in the top panel, while fixed parameters are given in the bottom panel. \edit2{Torus geometry priors are absent since they are also fixed in our BXA fit procedure.}}
 \label{tab:sim_pars}
\end{table*}

\subsection{Simulations}\label{simulation section}

\begin{figure}[t!]

\includegraphics[width=9cm]{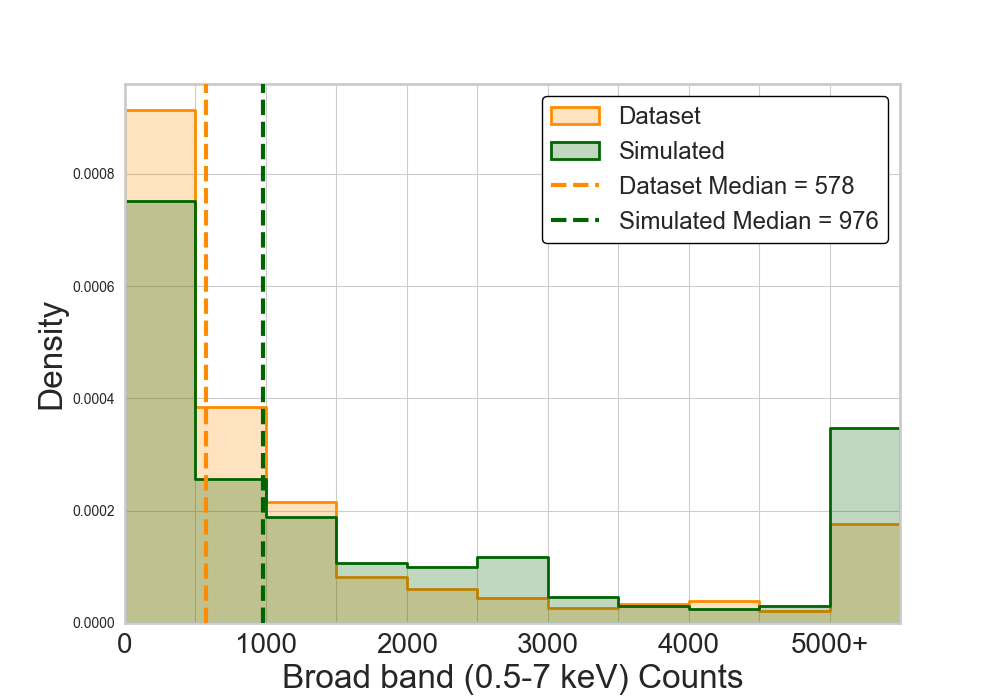}

\caption{The distributions of counts in the data set (orange) and simulations (green), showing the similarity between the two.}
\label{fig:both_counts_distributions}
\end{figure}

For a more comprehensive test of XZ, we simulated 1000 obscured AGN. We adopt the convention from both \cite{simmonds 2018} and \cite{peca 2021} in which $\Gamma$ is fixed at 1.9. The torus geometry parameters largely resemble those used in \cite{simmonds 2018}, \edit2{with the opening angle $\theta_{\mathrm{op}}$ fixed at 45\degree and the torus viewing angle $\theta_{\mathrm{view}}$ fixed at 80\degree (approximately edge-on). Note that, as in \cite{buchner 2014} and \cite{simmonds 2018}, those geometry parameters are fixed in the BXA fitting procedure as well, as neither this work or those works sought to further constrain the geometry of each obscured AGN.} \edit1{As in \cite{simmonds 2018}, we also froze the scattering fraction ($f_{\mathrm{scat}}$), but instead of using $f_{\mathrm{scat}} = 10^{-2}$ from \cite{simmonds 2018}, we chose $f_{\mathrm{scat}} = 10^{-4}$ to best reflect our data set, namely the mean ($10^{-3.7}$) and median ($10^{-4.2}$) of best-fit values for all spectra.} Each simulation was assigned a random set of \edit1{obscuration, redshift, and counts values}, constructed to \edit1{cover similar ranges of obscuration and redshift values as our data set, and to resemble its counts distribution. Specifically, we} used uniformly random values for log$N_H$ between 22 and 26 and also uniformly random values for redshift between 0.01 and 3, both in steps of 0.01. \edit2{The allowed values for simulation model parameters of interest can be found organized in Table \ref{tab:sim_pars}, which also shows the priors used for the BXA fitting method}. Counts were assigned randomly, but with a non-uniform probability designed to closely emulate the counts distribution in the data set (Figure \ref{fig:both_counts_distributions}). Background spectra were simulated by sampling a background file randomly selected from sources in the data set with multiple observations, as all such background files were stacked from the constituent observations. This ensured sufficient statistics to produce a viable and realistic background spectrum. For each simulation, we used the response files from the source whose background file was randomly selected for that simulation's background spectrum.

Using the same metrics described in Section \ref{known-z}, we can statistically quantify the capabilities of XZ. In this case, the only substantial limitation of our evaluation is systematic bias inherent and virtually unavoidable in these simulations. However, due to the large number of simulated sources, their wide parameter ranges, and the physically descriptive model on which the simulations were based, we consider the spectra to be quality test subjects for XZ. More importantly, we consider them to be better estimators of XZ's competence than the sparse results offered by smaller, less diverse data sets such as Known-z.

\begin{figure*}
\centering
\includegraphics[width=8.5cm]{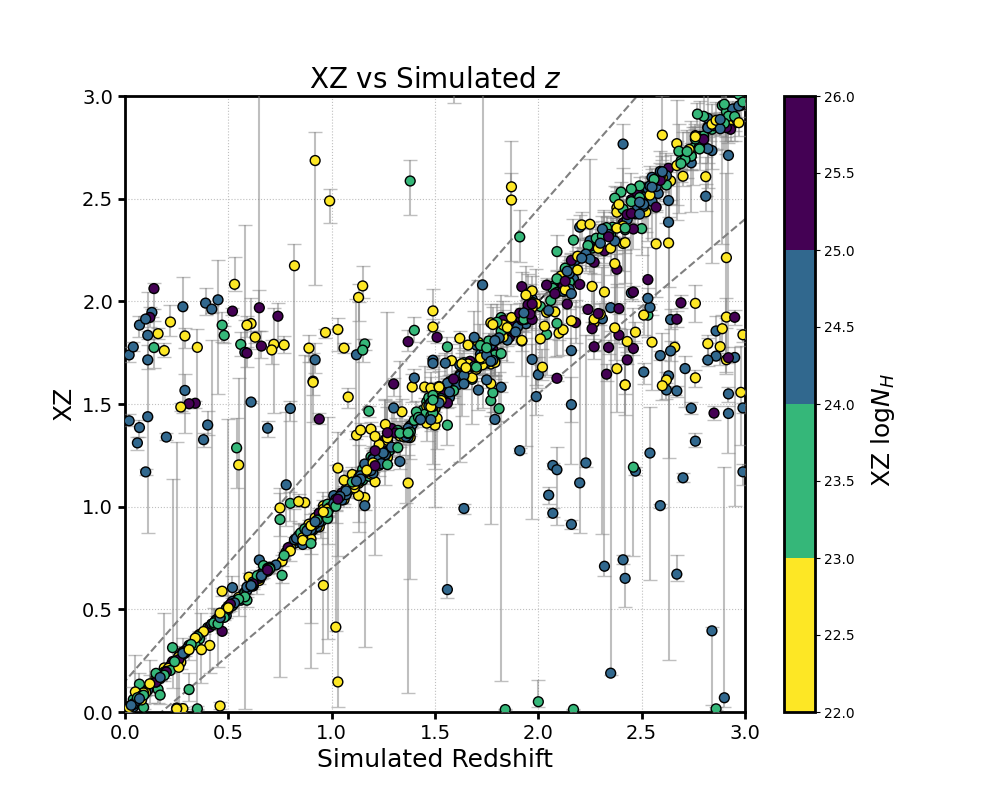}
\includegraphics[width=8.5cm]{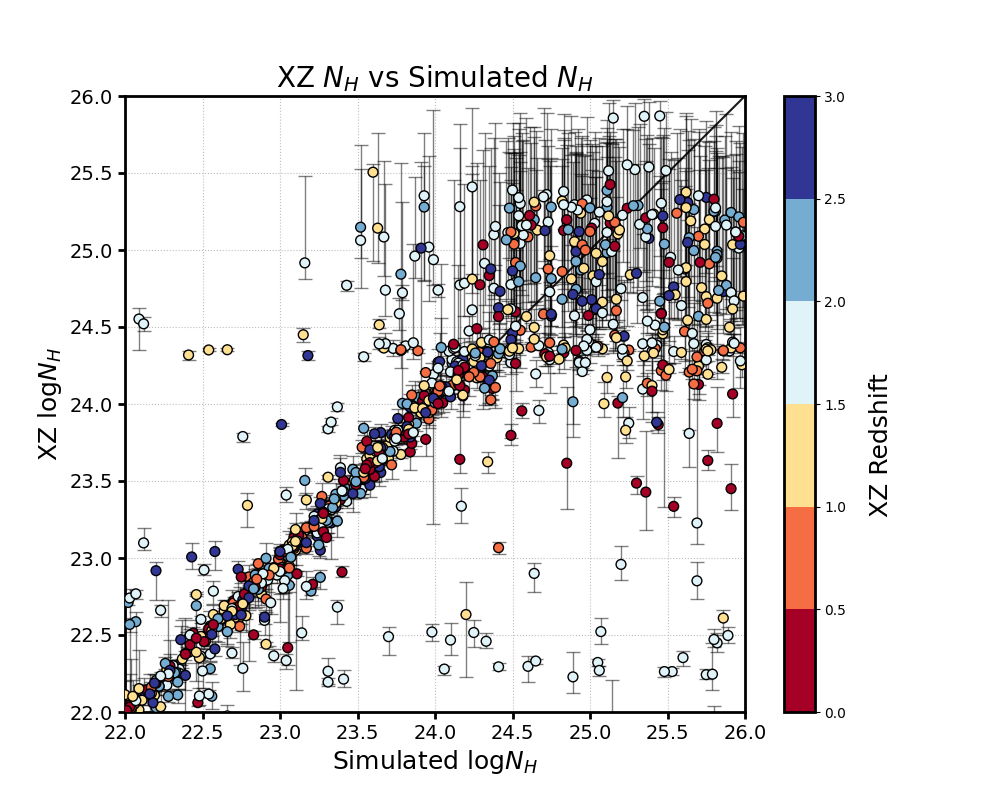}
\caption{For the simulated data set, the XZ redshift and obscuration measurements are compared to the simulated values. \textbf{Left:} XZ best-fit redshift plotted against input redshift, with the same format as Figure \ref{fig:IG_only_delz_scatter}. \textbf{Right:} XZ best-fit log$N_H$ vs. the input value. A black solid line represents XZ $N_H$ $=$ input $N_H$, while the color map indicates XZ redshift.}
\label{fig:sim_results}
\end{figure*}

\begin{figure*}

\includegraphics[width=9.cm]{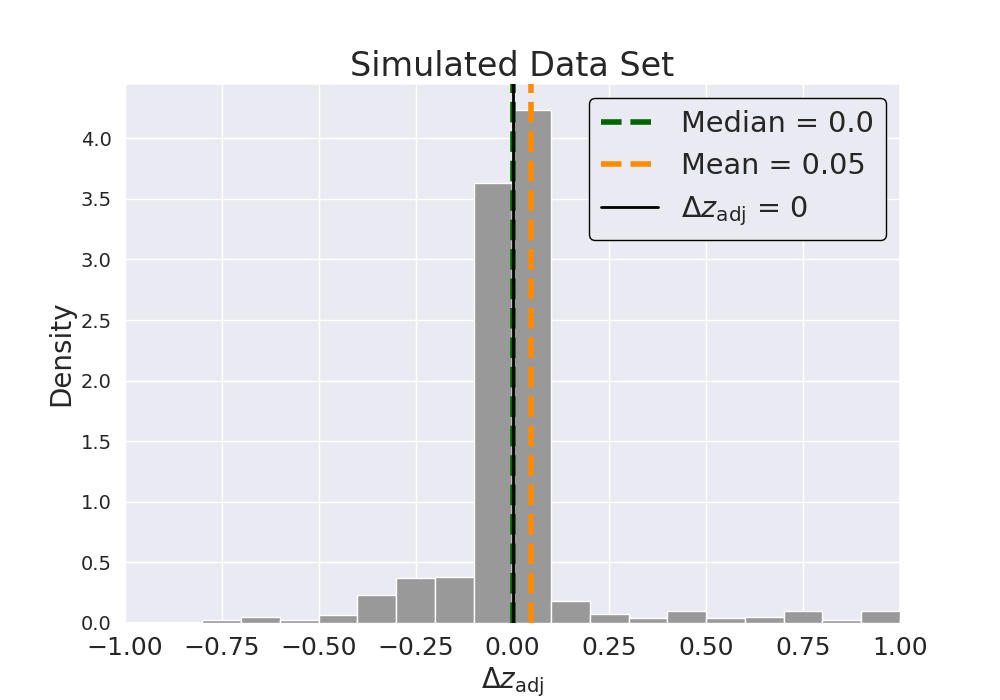}
\includegraphics[width=9.cm]{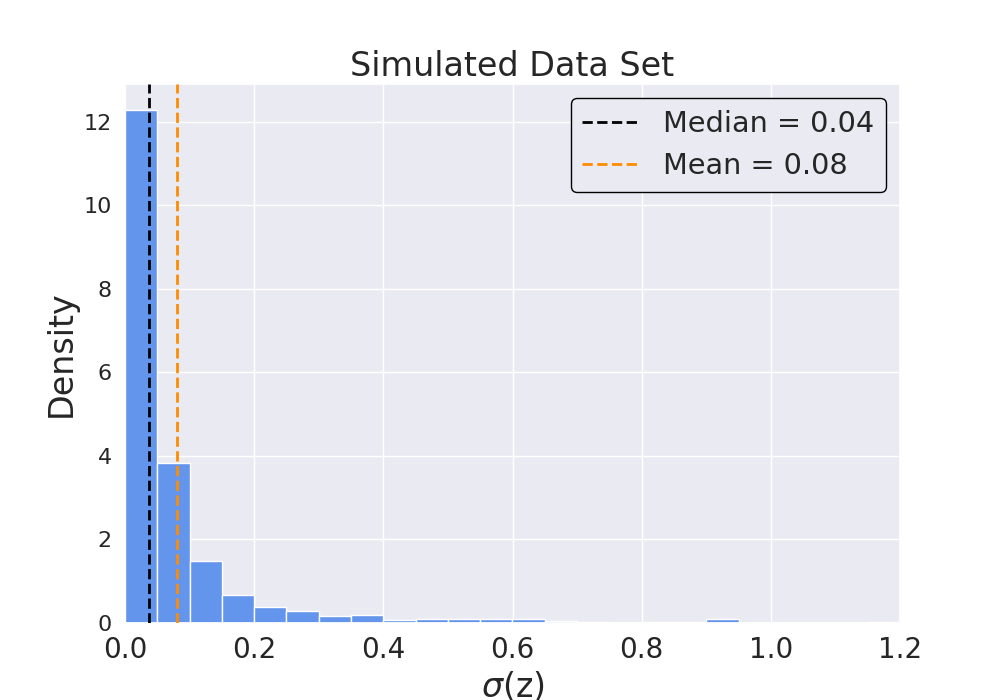}

\caption{Another visualization of XZ's accuracy on the simulated data set using the same format as Figure \ref{fig:Delta_z_histogram}. \textbf{Left:} The $\Delta z_{\mathrm{adj}}$ distribution. \ref{fig:Delta_z_histogram}. \textbf{Right:} The distribution of XZ redshift uncertainty.}
\label{fig:sim_delz_hist}
\end{figure*}

We ran the XZ procedure on the simulated data set, applying the IG $\geq$ 2 filter to the results, leaving us with 942 simulated spectra. The panels in Figure \ref{fig:sim_results} show the XZ redshift (left) and $N_H$ (right) plotted against the simulated input values. Using the same definitions of success given in \ref{known-z}, we see basic success in $\sim$85\% of the sources, mild success in 82\%, and top-tier success in $\sim$54\%. The $\Delta z_{\mathrm{adj}}$ and $\sigma(z)$ distributions are shown in Figure \ref{fig:sim_delz_hist}. \edit1{XZ's performance on the simulated data, then, is noticeably better than its performance on the Known-z sources, so it could be argued that this is evidence of a major systematic issue. However, it is critical to emphasize that XZ's accuracy on the simulated data set is fairly consistent with \cite{simmonds 2018} (with the simulations' mild success rate exceeding that of \citealt{simmonds 2018} by $\sim$7\%, but its top-tier success rate falling $\sim$6\% short of that found by \citealt{simmonds 2018}), and also that the simulated data set has a broad range of redshifts and obscurations like the deep surveys considered by \cite{simmonds 2018}. As discussed, Known-z is dominated by low-$z$ and low-$N_H$ sources, so XZ's performance on that data set was expected to be less favorable than on the simulations.}

\edit1{In spite of the promising success of XZ on the simulations}, we take a greater interest in the failures, as analyzing the nature of these failures and understanding how to predict them can be vital to interpreting results when XZ is run on sources with no known redshifts. Detailed below is our approach to this analysis, the goal of which was to establish a method for evaluating errors on XZ fits \textit{a posteriori} when \textit{a priori} information, such as known redshift, is not available.

\section{XZ Failure Analysis}\label{xz failure analysis}

To examine the reliability of XZ and the factors contributing to its success or failure, we established a single pass/fail metric to classify each fit. Namely, if the XZ result on a given source meets either the mild or top-tier success criterion (using the conventional inclusive ``or") defined in the previous sections, we classify the fit as positive (``pass", $+$). The basic success criterion is not considered sufficient due to the \edit1{large and irregular disagreement it allows between the best-fit values of XZ and known $z$}. To maximize the size of the training and testing sets, and to mitigate systematic bias from using purely simulated data, we combined the IG-filtered simulated data set with IG-filtered Known-z, producing a new master set of 1047 sources. We found the overall pass rate to be $\sim$81\%, hence making the failure rate $\sim$19\%.

\subsection{Parameter-Dependent Failure Rates}

\begin{figure*}[t!]
\centering

\includegraphics[width=6.5cm]{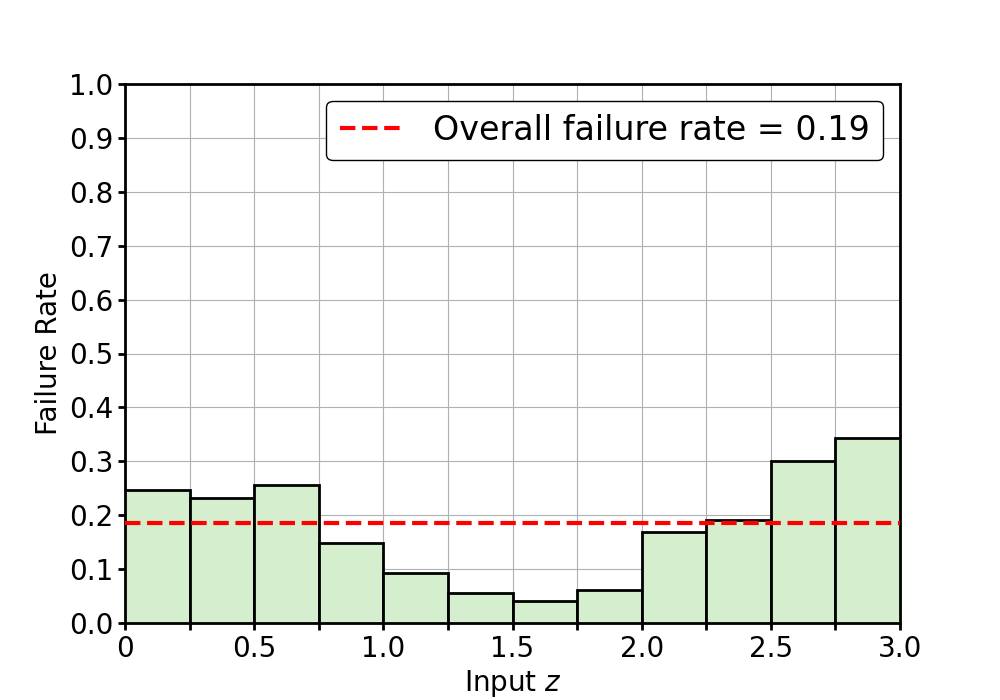}
\includegraphics[width=6.5cm]{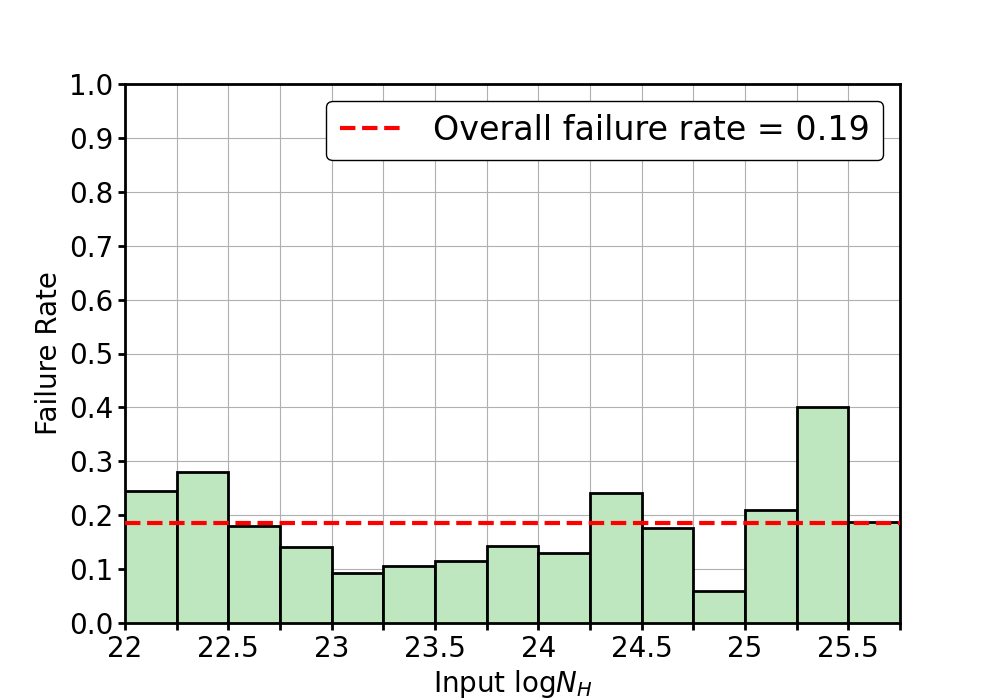}

\includegraphics[width=6.5cm]{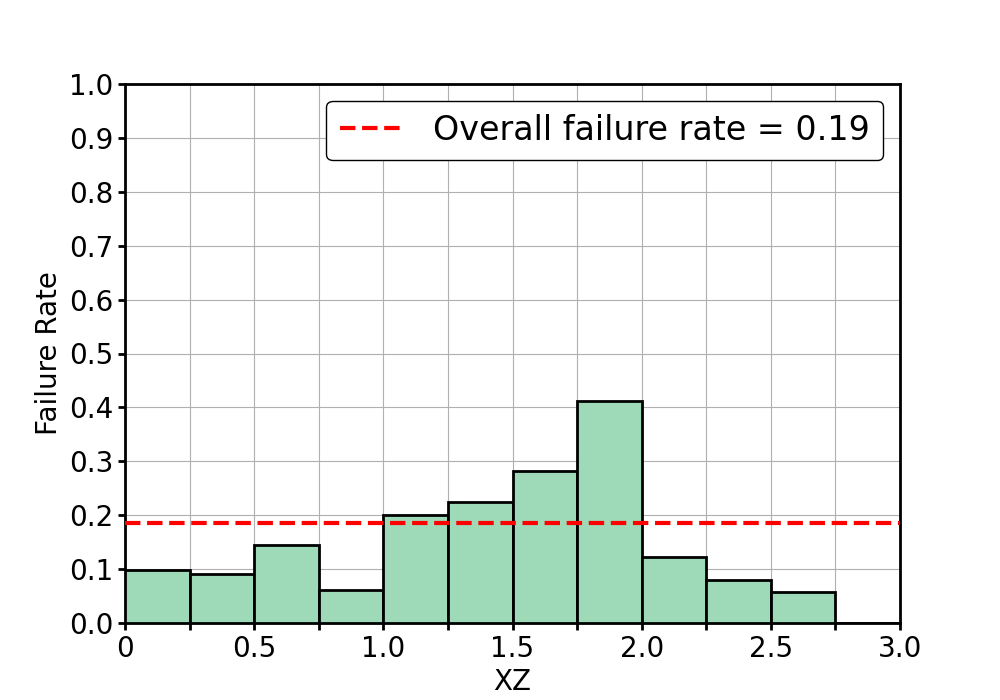}
\includegraphics[width=6.5cm]{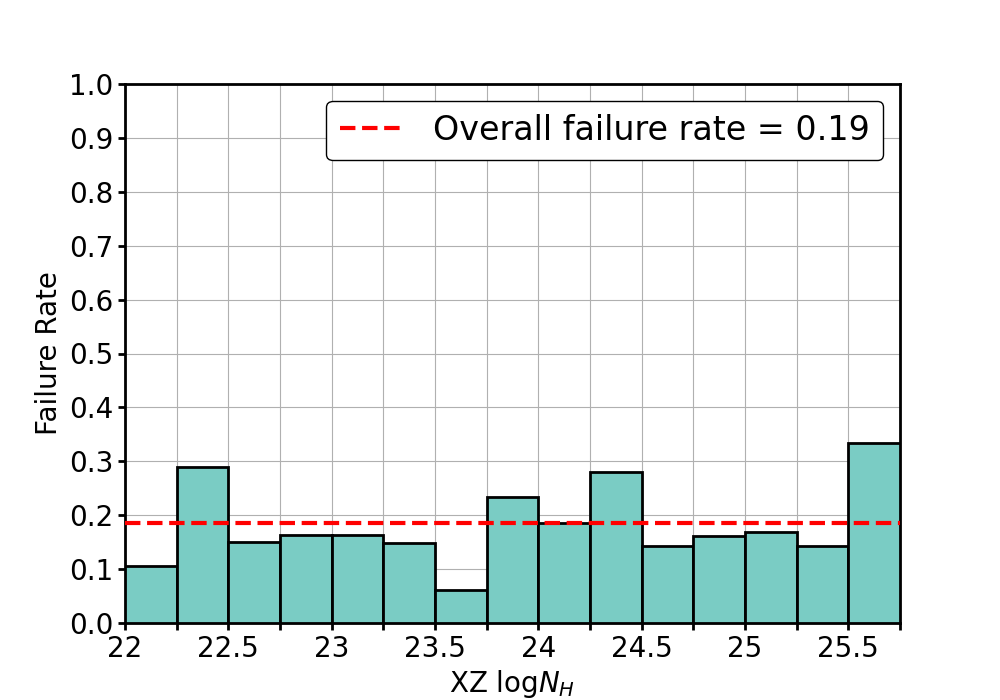}

\includegraphics[width=6.5cm]{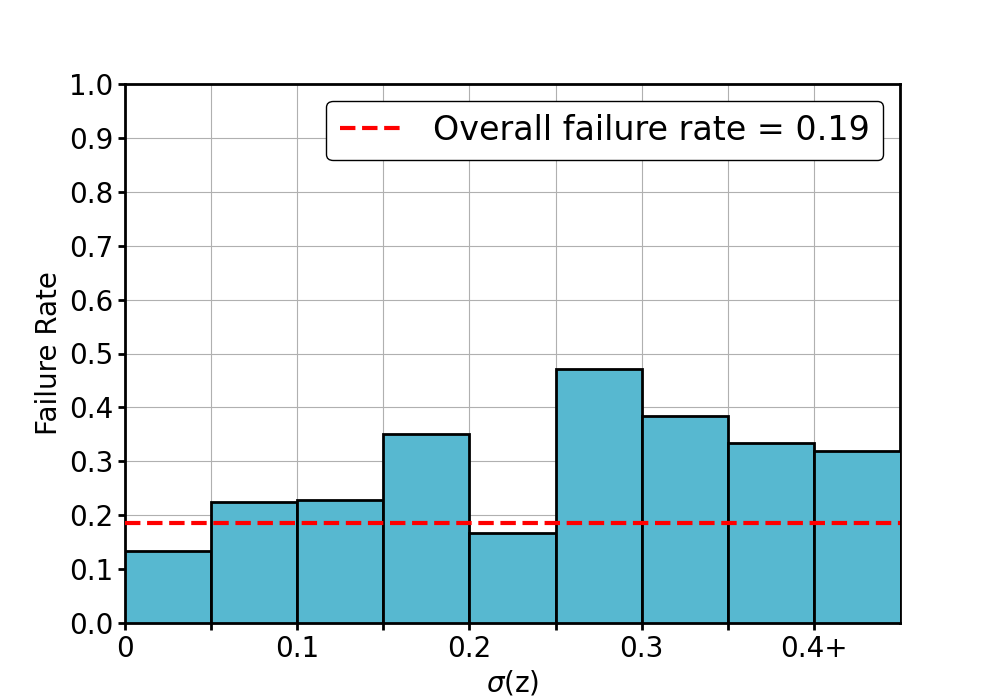}
\includegraphics[width=6.5cm]{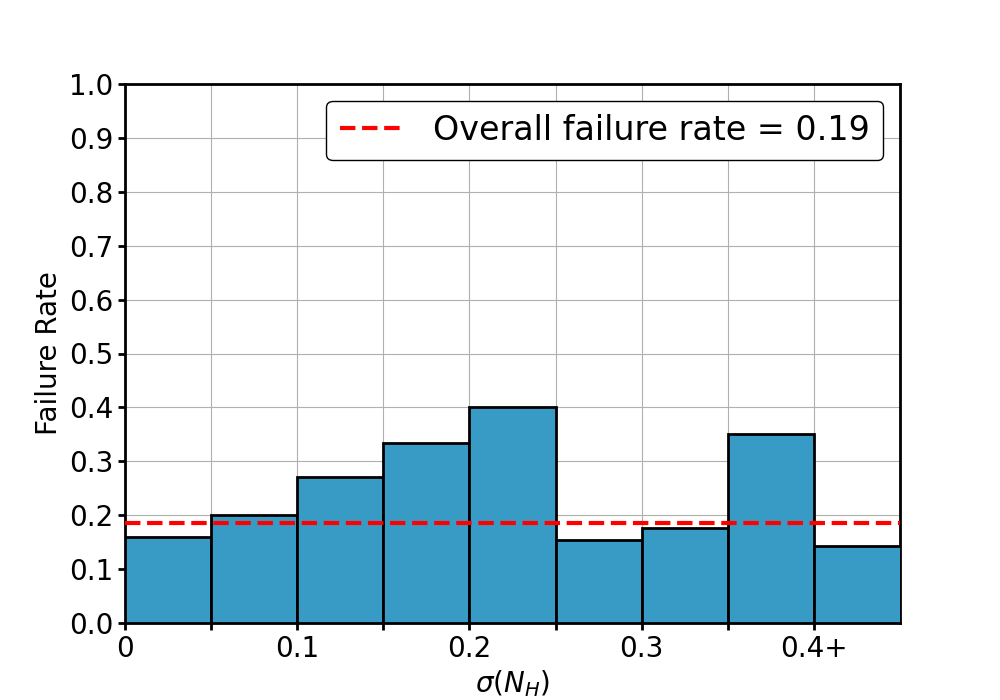}

\includegraphics[width=6.5cm]{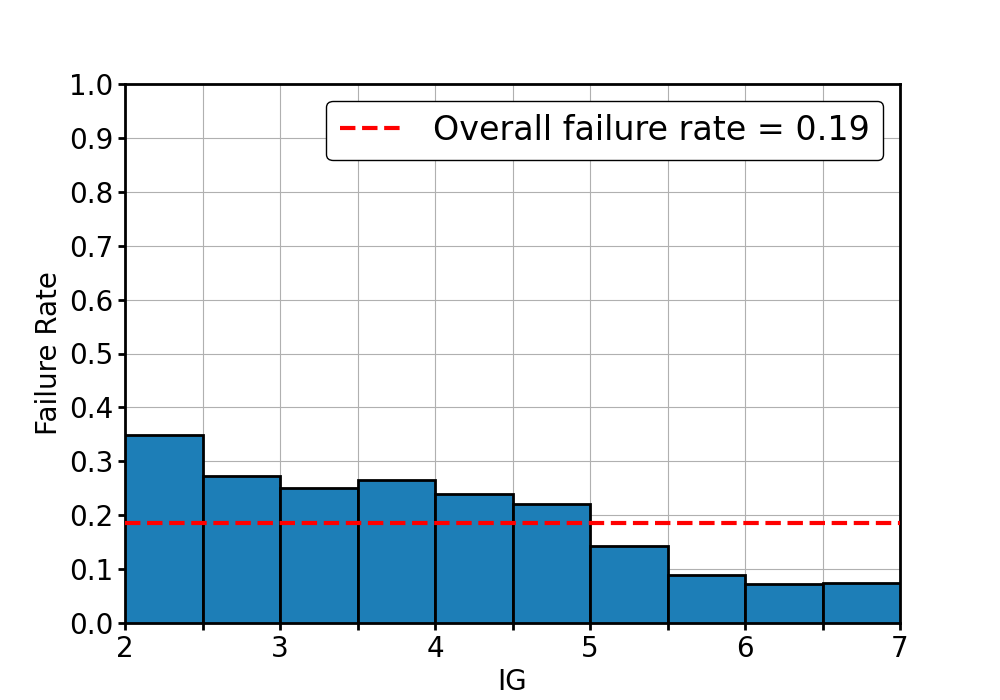}
\includegraphics[width=6.5cm]{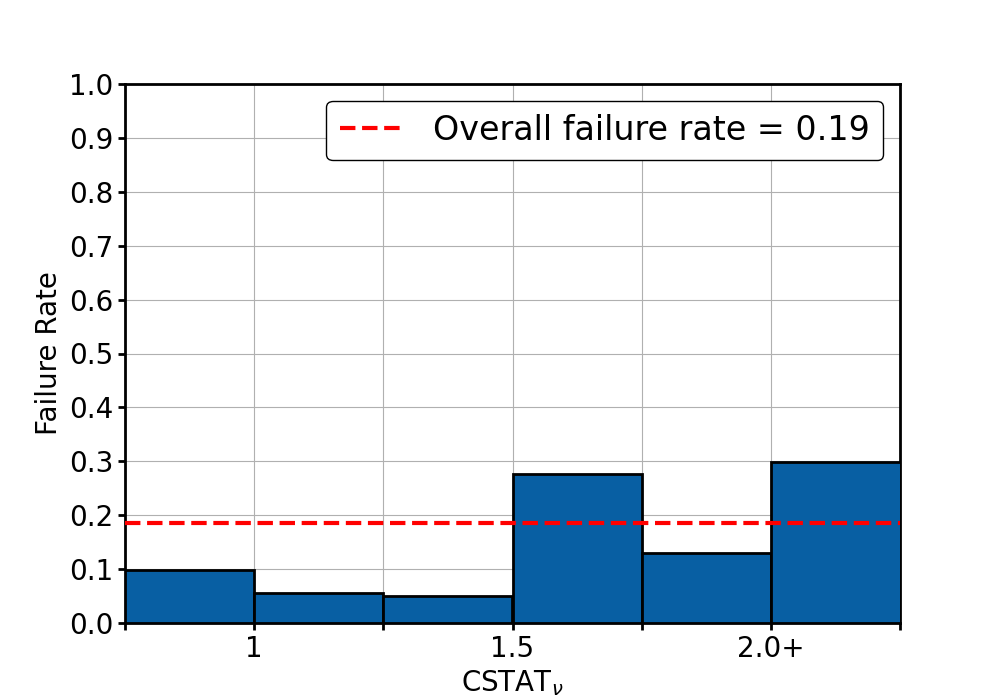}

\includegraphics[width=6.5cm]{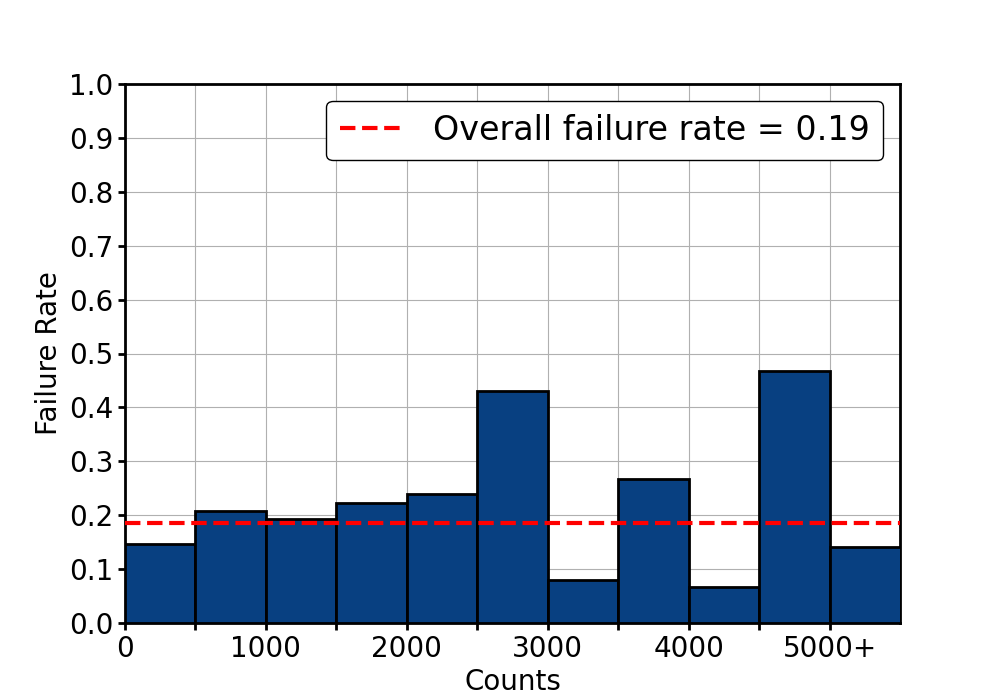}

\caption{Failure rate as a function of each spectral characteristic considered. The name of each characteristic is shown on the horizontal axis of its corresponding plot.}
\label{fig:err_hist_plots}
\end{figure*}

We first performed a basic analysis of the error rates in the combined simulated and Known-z data set as functions of various spectral characteristics, namely the known $z$ and $N_H$ values, IG, counts, reduced fit statistic, the XZ best-fit redshift and $N_H$ values, as well as $\sigma(z)$ and $\sigma(N_H)$. This set of source and fit characteristics was determined by assembling all attributes that we qualitatively expect to substantially impact the results, either physically or statistically, excluding only the power law photon index $\Gamma$ (and its associated uncertainty) due to its controlled value in the simulations. Otherwise, all characteristics potentially meaningful for the redshift results were included. Failure rates for all assessed characteristics are shown in Figure \ref{fig:err_hist_plots}. 

\subsubsection{Known Redshift and Obscuration}\label{known-z_faults}

The failure rates associated with the known $z$ and $N_H$ values are largely unsurprising. Redshifts from $\sim$0.75 to $\sim$2.5 all registered below-average failure rates, which is expected from the sensitivity limits determined in \cite{simmonds 2018}. The model is not sensitive to near-zero redshifts, and high redshift sources are typically faint with model-critical features redshifted out of \textit{Chandra}'s effective energy band.

Similarly, log$N_H$ yielded mostly below-average failure rates between 22 and 25, again consistent with the \cite{simmonds 2018} findings. Spectra with both low and extremely high $N_H$ are difficult to distinguish from an unabsorbed power law in the \textit{Chandra} energy band, depriving the model of the prominent photoelectric cutoff absorption edge vital to its redshift measurement. These results, while consistent with predictions, are not able to help make new predictions about the XZ method, particularly in cases without known values from other observations in which our goal is to deduce \textit{a posteriori} conclusions about XZ model results. Therefore, we aim to explore only features knowable or obtained via the fitting process alone.

\subsubsection{Fit Characteristics}\label{fit error rate}

The most prominent trend in XZ fit characteristics is the decrease of failure rate with an increase of redshift information gain, but otherwise only sparse relationships are evident. While we could simply apply a much stricter IG filter, increasing the threshold to $\sim$5 bits, this would leave us with just $\sim$17\% of the initial data set (and only $\sim$20\% of the IG $\geq 2$ bits data set), potentially removing hundreds of successful fits in the process. 

Another possible filtering method given the results would be to simply remove any source such that its value for a given attribute falls into one of that attribute's bins with an above-average failure rate. While this may be effective for our data set (though it would still likely suffer from removing too large a portion of the total fits), the goal in this work is to produce a more general methodology that can be applied to future large-scale X-ray redshift measurements. Therefore, our filtering procedure must be transferable to the analysis of any X-ray data set, and hence cannot be tailored to the specific errors found here.

The clearest and most plausible explanation for the absence of obvious effective filters based on these error rates is the dependence of the fit's success on the entire aggregate of these characteristics. Therefore, rather than attempt to sort out the co-dependencies of the fit attributes directly, we turned instead to machine learning to identify successful fits.

\subsection{Machine Learning Analysis}\label{machine learning}

We employed a multi-layer perceptron (MLP) neural network, part of the \texttt{Scikit-learn} Python software package \citep{pedregosa 2012}, due to the MLP's ability to solve difficult non-linear problems, as well as its effectiveness at classifying categorical variables \citep{ciaburro 2013} such as the pass/fail criteria we apply to XZ. The power of a neural network arises from its forward-propagation of input through a non-linear activation function contained in each \edit1{constituent} neuron. Neurons are connected such that a neuron in a given layer will receive input from every neuron in the previous layer, and send its output to every neuron in the ensuing layer. Together, the neurons \edit1{produce} a final model that can approximate any arbitrarily complex function \citep{hornik}. Therefore, in principle, a neural network is capable of solving any classification problem, a result known as the Universal Approximation Theorem. 

We constructed our neural network with five layers of neurons (thus meeting the general criteria for ``deep learning"), including the input, three hidden layers (containing 1200, 800, and 400 neurons, respectively), and the output, which yields the classifier. For our activation function, we used the highly successful rectified linear unit function (ReLU), the most widely used activation function in deep learning (\citealt{relu 1}, \citealt{relu 2}). For more detailed discussions of the \texttt{Scikit-learn} software and MLP neural networks, see \cite{mlp 1}, \cite{pedregosa 2012}, and \cite{mlp 2}.

To \edit1{obtain} a training set and a testing set, we performed a random split of the combined Known-z and simulated sources. We chose to allot $\sim$50\% of the data to the training set, which should provide a balanced sample as indicated by the results of \cite{training split 1} and \cite{training split 2}. We assigned a binary class to each source, the simple pass/fail as defined earlier in this section. Therefore, the MLP should produce a model that will classify a given source's XZ redshift estimate as either pass ($+$) or fail ($-$). The algorithm was trained using the seven \textit{a posteriori} source characteristics given in Section \ref{fit error rate} as the features, namely XZ redshift, XZ log$N_H$, $\sigma(z)$, $\sigma(N_H)$, CSTAT$_\nu$, IG, and counts. All were rescaled such that each feature had a mean of zero and unit variance. The resulting model was applied to the testing set to evaluate its accuracy. The results of these tests are shown in Table \ref{tab:machine_learning}. 

\subsubsection{Performance on Total Testing Set}\label{ML performance}

The most important performance metric, for our purposes of identifying good XZ redshift values, is the probability that a fit classified as successful (test$+$) is indeed a good estimate of the redshift (is$+$). This is a common metric (e.g., \citealt{covid1}, \citealt{covid2}) \edit1{known in machine learning as ``precision,"} and can be computed using Bayes' Theorem \citep{bayes} in the following manner:
\begin{equation}\label{bayes}
    \mathrm{Precision} \equiv P(\mathrm{is}+|\;\mathrm{test}+) = P(\mathrm{test}+|\; \mathrm{is}+) \frac{P(\mathrm{is}+)}{P(\mathrm{test}+)}
\end{equation}
Since precision gives the probability that a positively-classified source is truly a good redshift estimate, it is therefore the most crucial accuracy measure of our machine learning filter, \edit1{as it represents the percentage of the testing set that would correctly estimate the redshift if we filtered out all spectra classified as failures ($-$). I.e., it would be the filtered testing set's pass rate, and therefore offers an estimate for the performance we could expect on a general data set when filtered accordingly using the fitted classifier}. As shown in Table \ref{tab:machine_learning}, we find the precision to be $\sim$89\% in the testing set.

To evaluate this finding, we \edit1{can compare the precision to the testing set's original pass rate, which was $\sim$83\%, by performing a simple two-proportion z-test. Our classifier yields a statistically significant $\sim$3$\sigma$ improvement over the initial pass rate, therefore offering strong evidence and motivation for applying the MLP classifier as a new filter. Note that the pass rate of $\sim$81\% (and corresponding fail rate of $\sim$19\%) described in Section \ref{xz failure analysis} and reflected by Figure \ref{fig:err_hist_plots} refers to the initial pass rate of the entire combined set of simulations and Known-z sources, and therefore differs slightly from the initial $\sim$83\% pass rate of the testing set due to a slight statistical fluctuation in the train/test splitting process.}

\edit1{More rigorously, this approach is equivalent to considering the existing IG $\geq$ 2 filter as the null hypothesis classifier}, which would result in every source in the testing set being accepted as a good redshift estimate. In this case, we \edit1{would} compute the corresponding \edit1{precision} and compare it statistically to that of our MLP classifier. 

Furthermore, it can be shown using Equation \ref{bayes} that our choice of the null hypothesis is mathematically arbitrary because any random classifier (e.g., giving each source a statistically independent, random 50\% chance of passing) would yield the same \edit1{precision} as allowing all sources to pass (which, in this context, is equivalent to giving each source a statistically independent, ``random" 100\% chance of passing). \edit1{This is why, regardless of the approach, the null hypothesis precision is exactly equal to the pass rate of the testing set, and therefore allows for the intuitive comparison between the initial pass rate and the model precision.}



\edit1{As shown in Table \ref{tab:machine_learning}, we also considered other performance measurements to further evaluate the classifier, particularly its ability to correctly identify good redshift estimates ($+$ cases).} These performance findings on the total testing set were also favorable. The model had an \edit1{accuracy score} of $\sim$79\%. \edit1{The accuracy score of a classifier is computed by dividing the number of correct classifications by the total number of sources in the testing set, which indicates the classifier's overall ability to classify both $+$ and $-$ cases, and thus $\sim$79\% is a fairly positive result. In addition, we found the recall to be $\sim$85\%. Recall (also known as the ``true positive rate") is the probability that a good redshift estimate (is$+$) will be classified as such (test$+$). Note that this differs from precision; in particular, recall is the $P(\mathrm{test}+|\; \mathrm{is}+)$ term in Equation \ref{bayes}. Therefore, although recall is less crucial to interpreting the filtered data set than precision (which, as discussed, gives the percentage of good redshift estimates in the filtered data set), it still offers a similar insight into the classifier's performance by showing its ability to correctly identify good redshift estimates by giving the percentage of those good estimates in the data set that are classified as such by the model. An equivalent interpretation is to consider that $1 - \mathrm{recall}$ is the proportion of good redshifts that will be erroneously filtered out. Both interpretations make it clear that the recall should be close to 1, so our $\sim$85\% for the total testing set shows good model performance.} Moreover, the model also exhibited a favorable receiver operator characteristic curve (ROC curve; Figure \ref{fig:AUC_ROC}), which examines the relationship between a classifier's rates of true positives and false positives. The area under the ROC curve (AUC ROC) is $\sim$0.78, indicating the classifier's tendency towards true positives and away from false positives.

\subsubsection{Performance on Subsets of Testing Set}

Due to the small number of Known-z sources (60), as well as its mostly low-redshift sources, the classifier's Known-z performance was generally less favorable than that of the total testing set. Most notably, the crucial \edit1{precision statistic} is $\sim$83\%. While this demonstrates a noticeable improvement over the null value of $\sim$75\%, the sparse size of the Known-z testing subset renders this a statistically marginal $\sim$1$\sigma$ improvement. While this is not as statistically significant as the total testing set's \edit1{precision}, for further insight the Known-z results can be compared to simulations with comparably low redshifts, also found in Table \ref{tab:machine_learning}.

\begin{figure}
\includegraphics[width=10.cm]{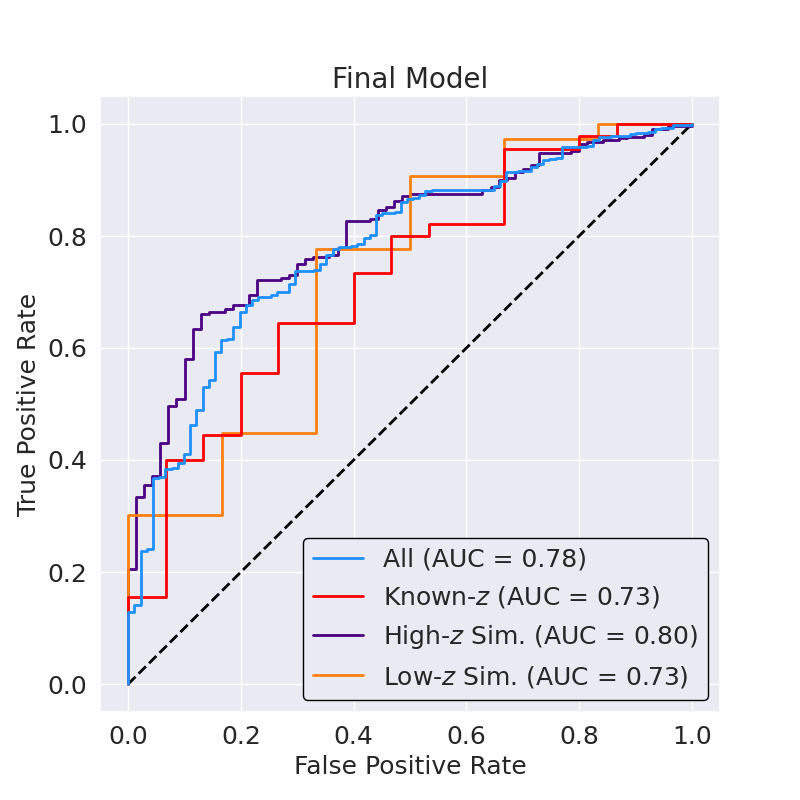}
\caption{The final classifier's ROC curves. Evaluated on the total testing set, Known-$z$ sources, low-$z$ simulations, and high-$z$ simulations. For comparison, the black dotted line represents the baseline case of AUC = 0.5, which indicates a classifier with equal tendencies towards true and false positives.}
\label{fig:AUC_ROC}
\end{figure}

We defined low-$z$ simulations such that the XZ redshift $\leq 0.593$, which is the 75$^{\mathrm{th}}$ percentile of Known-z redshifts to offer a comparison between Known-z and a set of similar simulated spectra. Like Known-z, the set of low-$z$ simulations is small, with just 80 sources. The significance of \edit1{the precision} on these low-$z$ simulations is $\sim$1$\sigma$, similar to the performance on Known-z. The high-$z$ simulations, on the other hand (which include all simulations with XZ $> 0.593$), display $\sim$3$\sigma$ significance, resembling the classifier's performance on the overall testing set. These similarities are further illustrated by Figure \ref{fig:AUC_ROC}'s ROC curves, where the total testing set's curve strongly resembles that of the high-$z$ simulations, while Known-z's curve closely resembles that of the low-$z$ simulations.

Together, this evidence suggests that the classifier's reduced performance on Known-z is largely to be expected from XZ's generally lower compatibility with low-redshift sources, as well as the small number of Known-z sources. In addition, a machine learning testing set (or meaningful testing subset) would ideally employ far more than the 60 sources Known-z contains. However, despite Known-z's small size and low redshifts, the classifier was still able to offer statistically marginal but non-negligible improvement which, in conjunction with the $\sim$3$\sigma$ improvement on the total testing set, shows that our model can serve as an effective filter on XZ results.

\begin{table*}[t!]
\centering
 \begin{tabular}{@{\extracolsep{\fill}}c || c c c c | c c c }
 
 \hline
 Testing & $N_\mathrm{src}$ & Accuracy & Recall & AUC & Model & Null & \textbf{Model} \\
 
 Subset & & Score & & ROC & Precision  & Precision & \textbf{Significance} \\
 
  \hline
  \hline
  
  
  
  Total Testing Set & 524 & 79.4\% & 85.2\% & 0.78 & 89.3\% & 82.6\% & \textbf{2.91$\sigma$} \\
  
  
  Known-z in Testing & 60 & 70.0\% & 75.6\% & 0.73 & 82.9\% & 75.0\% & \textbf{0.95$\sigma$} \\

\hline


 High-$z$ Simulations & 382 & 78.5\% & 83.7\% & 0.80 & 89.4\% & 81.7\% & \textbf{2.78$\sigma$} \\


 Low-$z$ Simulations & 82 & 89.0\% & 93.4\% & 0.73 & 94.7\% & 92.7\% & \textbf{0.51$\sigma$} \\

  \hline

 \end{tabular}
 \caption{The results of the machine learning analysis. The performance is assessed on the entire testing set, as well as on three of its subsets. $N_\mathrm{src}$ is the number of sources in each data set.}
 \label{tab:machine_learning}
\end{table*}

\subsubsection{Feature Importance Evaluation}
 
A fundamental aspect of any machine learning algorithm is the selection of features, to optimize the model by using only those that are informative and excluding features irrelevant to classification (see, e.g., \citealt{attributes}). We conducted two separate analyses of our features, first by computing the amount of classification-relevant information contained in each, then by performing an ablation analysis.

For the information content approach, we used the method presented in \cite{kubat} and described below. In that work, a feature's information content ($I_f$) is defined as:
\begin{equation}\label{info_content}
    I_f = H_\mathrm{T} - H_f
\end{equation}
where $H_T$ is the total statistical entropy of the training set and $H_f$ is the entropy of the feature. Statistical entropy ($H$) is defined in \cite{kubat} as: 
\begin{equation}
    H = -\sum_{j}P_j\;\mathrm{log}_2 P_j = -P_+\;\mathrm{log}_2 P_+  - P_-\; \mathrm{log}_2P_- 
\end{equation}
where $P_+$ and $P_-$ are the pass and fail rates of the considered data set. After computing the total entropy of the training set, $H_\mathrm{T}$, we then considered each feature. We binned that feature's values in the training set such that the number of sources in each bin ($N_i$) was at least 5. The entropy was computed for each bin ($H_i$) and a sum over all bins yielded the total statistical entropy of the feature, $H_f$, given by:
\begin{equation}
    H_f = \sum_{i} \frac{N_i}{N_\mathrm{T}} H_i
\end{equation}
where $N_T$ is the number of sources in the training set. We then obtained the relevant information contained in each feature using Equation \ref{info_content}. As shown in Table \ref{tab:attribute_info}, the information contents of our initial features are approximately equal, with all such features containing $\sim$0.2--0.3 bits. This suggests that all features in our initial set are sufficiently information-dense to warrant inclusion in the model.

\begin{figure*}
\includegraphics[width=20.cm]{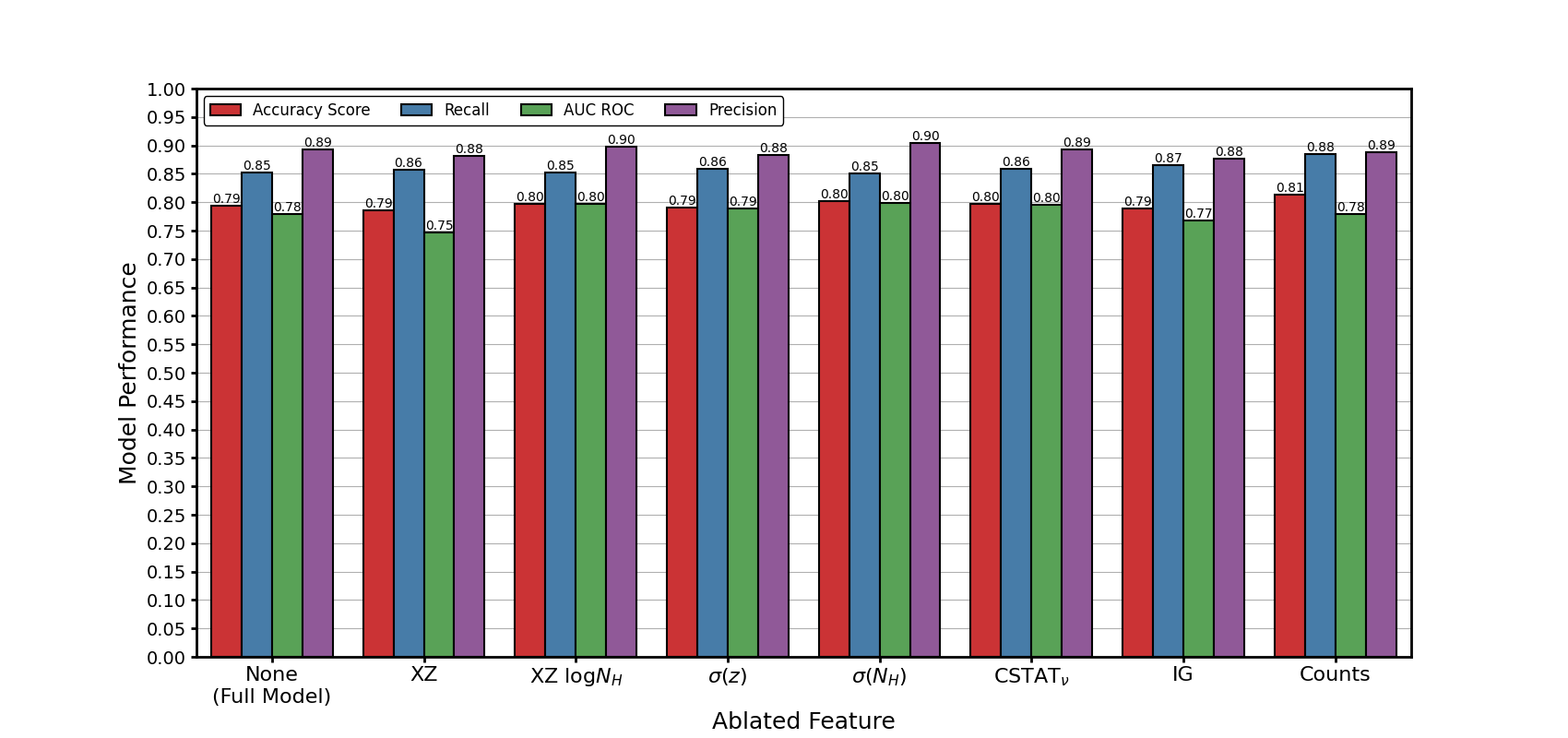}
\caption{Results of the ablation analysis showing model performance when each feature is removed, in terms of the four accuracy metrics used in Table \ref{tab:machine_learning}. Here, all metrics are given as decimals.}
\label{fig:ablation}
\end{figure*}

For the ablation analysis, we removed each of the 7 features individually, fitted the resulting 6-feature model, and compared its performance to that of the full model. As shown in Figure \ref{fig:ablation}, no such iterations yielded considerably poorer results, suggesting that all features are of comparable importance. This is consistent with the findings of our information content analysis, and therefore we conclude that our 7 features each demonstrate sufficient importance for inclusion in the model.

\begin{table}
\centering
 \begin{tabular}{ r | l  }
 
 \hline
 Feature & Information \\
  Name &  Content (bits) \\
  \hline
  XZ & 0.250 \\

  XZ log$N_H$ & 0.208 \\

  $\sigma(z)$ & 0.191 \\

  $\sigma(N_H)$ & 0.181 \\

  CSTAT$_\nu$ & 0.279 \\

  IG & 0.194 \\

  Counts & 0.199 \\
  
  \hline
  \textbf{Mean} & \textbf{0.214} \\
  \hline

 \end{tabular}
 
 \caption{The amount of information, in bits, contained in each feature.}
 \label{tab:attribute_info}
\end{table}

\subsubsection{Cross-Validation}

Our $\sim$500-source training set is small in the context of neural networks, particularly deep learning, in which data sets ideally contain many thousands of examples. This makes our relatively small training set sensitive to the two randomized processes involved in fitting the model, namely the random split of the initial data into the training and testing sets and the initialization of the neural network's fit parameters (``weights"). The Known-z subset is especially sensitive, as it composes just $\sim$10\% of the data. To ensure that our favorable model performance did not arise from unusually convenient splitting or initialization, we employed two cross-validation procedures to explore the impact of varying these randomization processes. Cross-validation is the practice of training and testing the model multiple times (e.g., using a different split of the data into training and testing sets each time) and comparing the results to investigate any associated effects on the performance.

\begin{figure*}[t!]
\includegraphics[width=10cm]{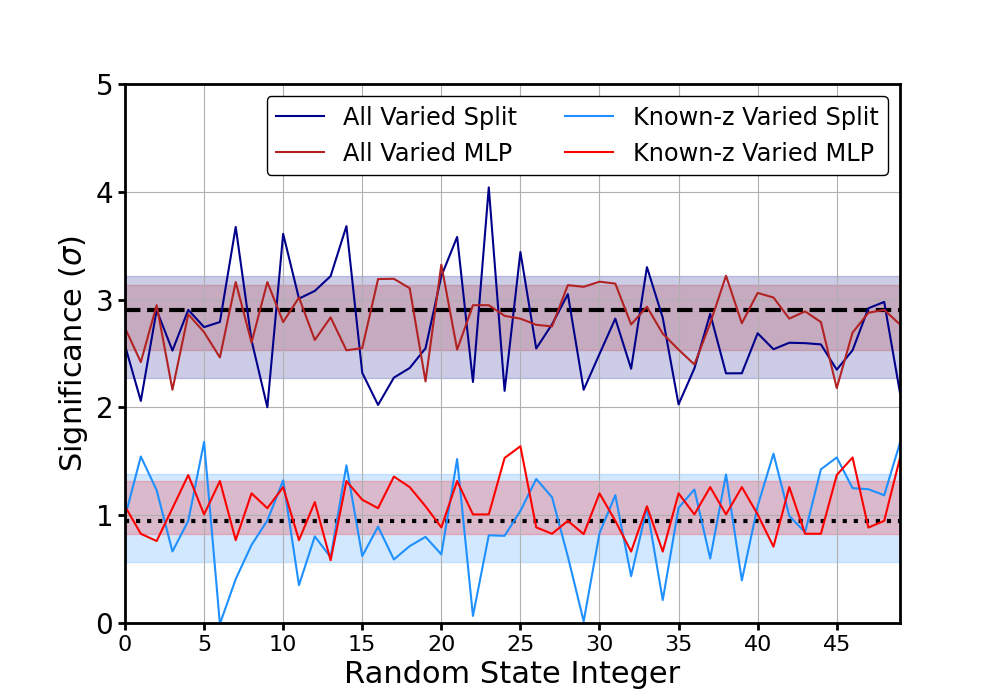}
\includegraphics[width=10cm]{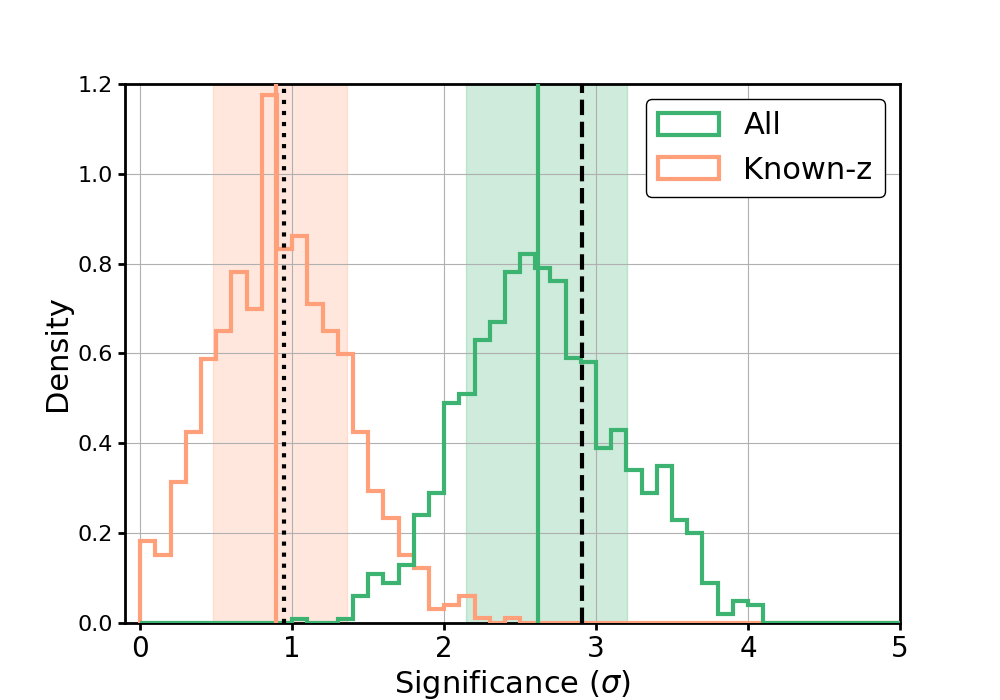}
\caption{Cross-validation results. Shaded regions show 1$\sigma$ intervals. Black dotted and dashed lines show the significance of our final model on the total testing set and Known-z sources in the testing set, respectively. \textbf{Left:} Varying one random state integer while holding the other constant. ``Varied Split" denotes varying the random state for splitting the data set into training and testing, while ``Varied MLP" denotes varying the MLP initialization random state. \textbf{Right:} Choosing randomized pairs of random state integers. Solid vertical lines represent the median value of the distribution with the same color.}
\label{fig:cross-validation}
\end{figure*}

Each random process is ``seeded" by a random state integer. Our first method of cross-validation involved holding one of the random state integers constant while varying the other on the interval [0,50). The second method involved 1000 iterations of selecting unique pairs of random state integers, each independently randomized on the interval [0,50). In both methods, performance was evaluated using the statistical significance of the resulting \edit1{precision} on both the total testing set and the testing set's Known-z subset. As shown in Figure \ref{fig:cross-validation}, there is substantial variance in model performance, as anticipated due to our relatively small data set.

However, the median significance is \edit1{nearly} $\sim$3$\sigma$ for the total testing set and $\sim$1$\sigma$ for Known-z, which is highly consistent with the results of our final MLP model. Indeed, its performance falls well within the 1$\sigma$ ranges of the corresponding cross-validation distributions, further instilling confidence in our results. Furthermore, the distinct distributions representing the total data set and the Known-z sources confirms our previous suspicion that the classifier's lower significance on Known-z compared to the total testing set is systematic, which we hypothesized as arising from Known-z's low redshifts. In any case, due to our reported model's consistency with the mean performance of models found in these cross-validation analyses, we can thus move forward with that model.

\subsubsection{Comparison to Other Models}

The final method we used to evaluate our model was to compare it with other models. We considered two additional formal frameworks (both implemented using \texttt{Scikit-learn}), namely logistic regression and random forest models. For each framework, we considered three sets of features (Table \ref{tab:feature_lists}): the set of 7 features present in our final model (``final"), a set with three additional features (``all features"), and that same set of 10 features except with XZ and Counts replaced by log$(1+\mathrm{XZ})$ and log(Counts), respectively (``all features and substitutions"). We also considered 2 variations of the MLP neural network by using the ``all features" and ``all features and substitutions" feature sets. 

A full breakdown of all models' performances can be found in a table in the Appendix, along with direct comparisons to our final model's performance metrics. The ROC curves for all models are shown in an accompanying figure. Also found in that figure is a manually-constructed ROC curve, achieved by varying the IG threshold from 2--6.7 bits (reflecting the discussion in Section \ref{fit error rate}), which offers a preliminary baseline for comparison with our model before considering the other machine learning frameworks. This curve shows that filtering with higher IG delivers a less favorable AUC ROC than our final model which, in addition to our previous finding that a higher IG filter would sacrifice a major percentage of the data, further motivates our choice to abandon additional IG filters in favor of machine learning. 

Logistic regression in machine learning generally involves fitting a logistic function to two-class training data to yield a classifier model (see, e.g., \citealt{logistic}). Its simplicity makes it a good baseline for evaluating more complex algorithms such as our deep neural network. For example, if a neural network's performance is only marginally better than logistic regression, it would suggest that the more complex model was not necessary for modeling the classifier. As shown in the Appendix, none of the logistic regression models rivaled the significance of our model's \edit1{precision} on the total testing set or on the Known-z sources. Other accuracy metrics were comparable, and in some cases better, but the substantially worse \edit1{precision} performance justifies the use of a complex model.

A random forest classifier repeatedly samples portions of the feature space and subsets of the training set to build an ensemble of decision trees, which represents the ``forest," and averaging these trees yields the classifier (see, e.g., \citealt{randomforest}). Like logistic regression, random forest is generally simpler and less computationally expensive than a deep neural network, allowing it to serve as another baseline for evaluating our neural network. The random forest models performed better than logistic regression in terms of \edit1{precision}, but fell short of our MLP's success on both the total testing set and the Known-z sources, further justifying our choice of a more complex model.

The two alternate MLP models yielded results more similar to that of the final MLP, but still did not match our model's performance. While the ``all features and substitutions" MLP very narrowly outperformed our model's \edit1{precision} on the total testing set (achieving a value of $\sim$89.5\%, an improvement of $\sim$0.2\% over our MLP's $\sim$89.3\%), the ``all features" MLP fell slightly short and neither of the alternate MLPs was able to match our model's \edit1{precision} on Known-z \edit1{(though it should be noted that the ``all features" MLP's Known-z precision, despite being lower than our MLP's by $\sim$0.2\%, had a negligibly higher significance stemming from its higher recall)}. Therefore, we consider our final MLP model's overall performance to be the most favorable among all models considered.

It is important to note that the slightly poorer performances by these MLPs on may simply be a result of the data set's previously discussed sensitivity to model initialization and train/test splitting, so it does not imply that they will always perform definitively worse than our final MLP. However, due to the slight perceived advantage of our model, and to minimize the dimension of the feature space, we conclude that the 7-feature final MLP is the best among all models considered, and the trained model has been made publicly available for download.\footnote{\url{https://github.com/DominicSicilian/XZ_MLP}}

\subsection{Applying the Model}\label{applying model}

Upon applying this finalized MLP classifier as an additional filter on the data set by keeping all positively-classified sources and removing all negatively-classified sources, we saw retention of nearly two-thirds of the IG-filtered data set, keeping $\sim$64\% of those sources. The combined IG and MLP filters ultimately left us with $\sim$53\% of the initial data set, preserving the majority of all AGN originally considered.\edit1{ While cutting down the data set to just over half its original size may, at first, appear detrimental, it is important to note that the two layers of filtering (using IG and the MLP classifier) are specifically designed to, ideally, remove only the poor redshift estimates.}

\edit1{So, to put these retention values more rigorously into context, and in particular to evaluate how well the MLP filter retains good redshift estimates, we can use the recall of the machine learning algorithm on the testing set, as it directly gives the percentage of good redshift estimates that were correctly classified as such in that data. While this can only be computed unambiguously for the testing set, it serves as a good estimate for the No-z data set. As seen in Table \ref{tab:machine_learning}, the model's recall on the testing set was $\sim$85\%, indicating good overall retention of $+$ cases. Specifically, it suggests we may have lost $\sim$15\% of good redshift estimates in the original data set. Taking both the recall and precision on the testing set into account, we see that, while $\sim$15\% of good estimates may have been lost, the trade-off is that the success rate of the resulting data set increased by a statistically significant amount ($\sim$3$\sigma$ over the initial pass rate, from $\sim$83\% to $\sim$89\%, as described in Section \ref{ML performance}). Therefore, while we may have sacrificed a small amount of good XZ redshift estimates, we eliminated enough poor estimates to obtain a substantially cleaner overall set of redshifts. A breakdown of success rates and source retention percentages among Known-z sources across the different data sets and subsets is discussed below and can be found in Table \ref{tab:counter_info}.}


\begin{table*}
\centering
 \begin{tabular}{ c || c | c | c | c | c | c | c | c | c | c }
 
 \hline
 Set Name & \multicolumn{10}{c}{Features} \\
 
  \hline\hline
  
 Final & XZ & $\sigma(z)$& XZ log$N_H$ & $\sigma(N_H)$ & CSTAT$_\nu$ & IG & Counts & - & - & - \\
 
  \hline
  
  All Features & XZ & $\sigma(z)$& XZ log$N_H$ & $\sigma(N_H)$ & CSTAT$_\nu$ & IG & Counts & $S/B$ & log$T_e$ & HR \\
  
  \hline
  
  All Feat. \& Sub. & log$(1+\mathrm{XZ})$ & $\sigma(z)$& XZ log$N_H$ & $\sigma(N_H)$ & CSTAT$_\nu$ & IG & log(Counts) & $S/B$ & log$T_e$ & HR \\
  
  \hline

 \end{tabular}
 
 \caption{The three feature sets used for model comparison. \textbf{``Final"} refers to the 7 features used in our final model. \textbf{``All Features"} refers to the final model's 7 features plus three additional features: source-to-background counts ratio ($S/B$), the log of exposure time (log$T_e$), and HR. \textbf{``All Feat. \& Sub."} refers to the same 10 features, except XZ is substituted with log$(1+\mathrm{XZ})$ and Counts is substituted with log(Counts).}
 \label{tab:feature_lists}
\end{table*}


\subsection{Fully-Filtered Known-z}\label{ffkz}

\begin{figure*}[t!]
\includegraphics[width=9.5cm]{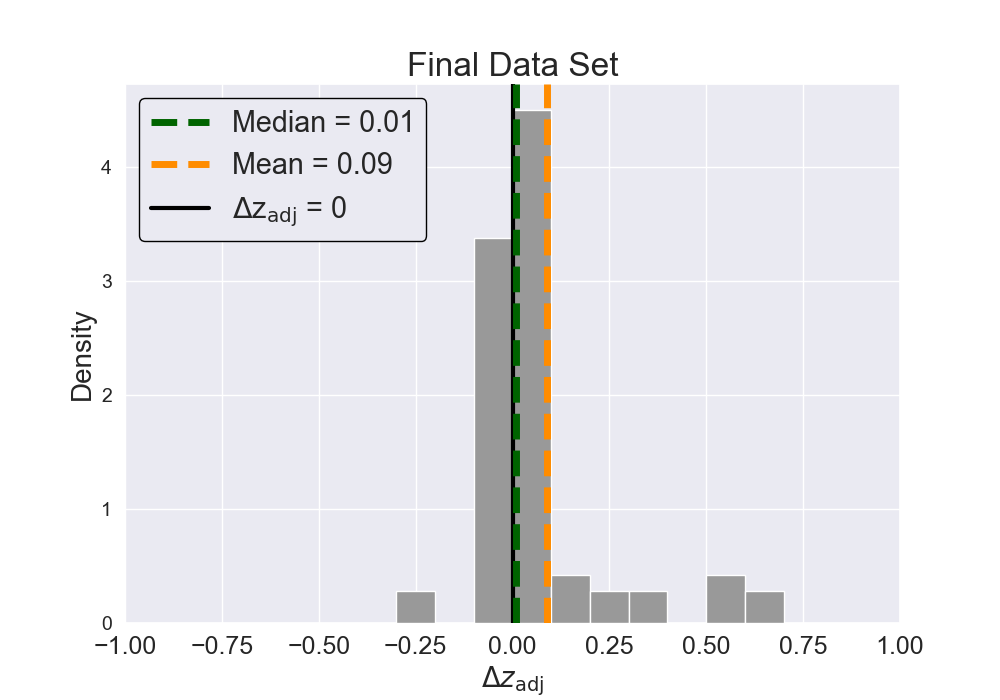}
\includegraphics[width=10.cm]{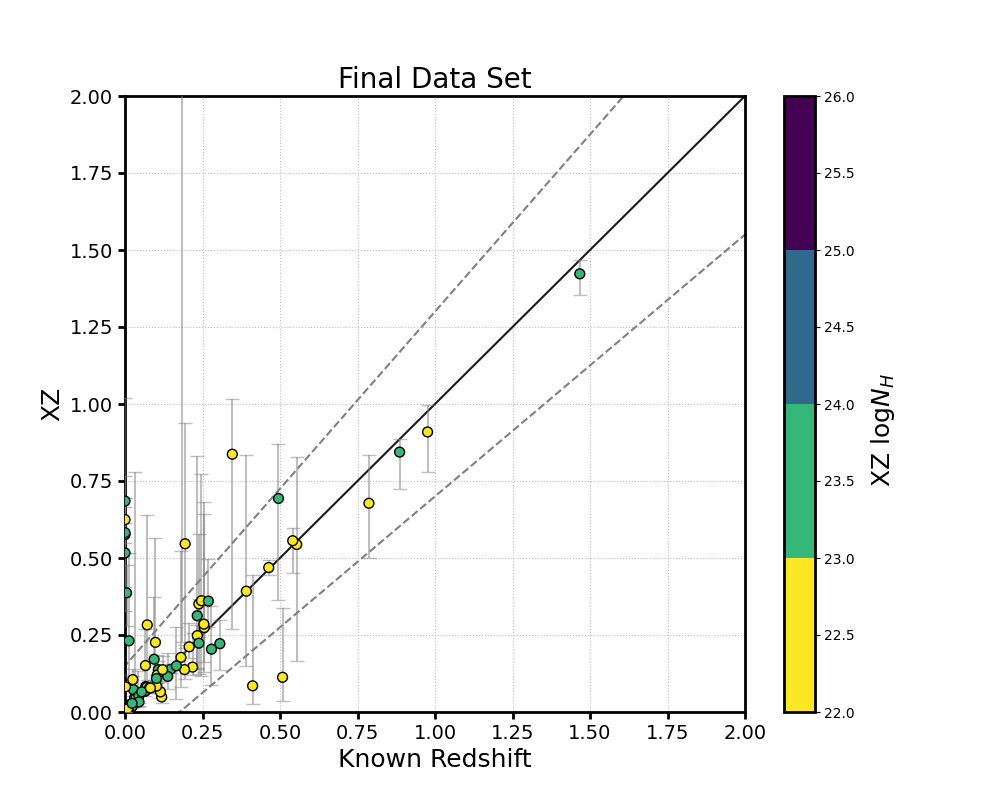}
\caption{\textbf{Left:} The distribution of $\Delta z_{\mathrm{adj}}$ in the final Known-z data subset after machine learning filtering, using the same format as the upper panel of Figure \ref{fig:Delta_z_histogram}. \textbf{Right:} Best-fit XZ redshifts plotted against the documented multiwavelength values for Known-z sources in the final data set after machine learning filtering, using the same format as Figure \ref{fig:IG_only_delz_scatter}.}
\label{fig:newDelz_histogram}
\end{figure*}

 Here, we briefly revisit the Known-z results upon applying the new MLP classifier. Since a large portion of Known-z was included in the training set, examining the filtered Known-z results is not particularly profound, but can at least serve as another realization of the algorithm's success. The primary illustrations of this success are the improved mean of $\Delta z_{\mathrm{adj}}$ to $\sim$0.09 (from $\sim$0.29; still with a median of $\sim$0.01; see the left panel of Figure \ref{fig:newDelz_histogram} compared to the top panel of Figure \ref{fig:Delta_z_histogram}) and the improvements of all accuracy tiers (visualized by right panel of Figure \ref{fig:newDelz_histogram}, compared to Figure \ref{fig:IG_only_delz_scatter}).

In particular, the basic success rate has risen to $\sim$92\% (from $\sim$79\%), mild success increased to $\sim$82\% (from $\sim$68\%), and top-tier success reached $\sim$56\% (from $\sim$49\%). The catastrophic failure rate was reduced to less than half its initial value, as it decreased to $\sim$4\% (from $\sim$11\%). Most notably, the $\sim$82\% mild success rate now exceeds \cite{simmonds 2018}'s $\sim$75\% by a fairly considerable amount, and the top-tier success rate approximately matches that of \cite{simmonds 2018}. As discussed, this data set was far less favorable for XZ than that of \cite{simmonds 2018}, so seeing comparable success rates between the two data sets after applying our new filter is an excellent demonstration of the machine learning classifier's capabilities. \edit1{Furthermore, when considering Section \ref{xz failure analysis}'s pass/fail criteria (as used in the XZ failure analysis and machine learning procedure) for Known-z, we find that after applying the MLP filter, the pass rate increased considerably, from $\sim$70\% to $\sim$85\% (also note that all performance metrics discussed here can be found organized in Table \ref{tab:counter_info}).}

Therefore, \edit1{taking all of this into consideration,} we conclude that at the cost of the MLP classifier removing \edit1{a small minority (approximately $\sim$15\%, as estimated in Section \ref{applying model}) of good redshift estimates}, it provided the great benefit of \edit1{elevating XZ's performance to match \cite{simmonds 2018}} in the AGN that were retained, \edit1{despite the abundance of} sources with redshifts and obscurations that fall outside XZ's ideal ranges.

\subsection{Caveats}

\edit1{While we draw positive conclusions from the simulations, XZ failure analysis, and the machine learning procedure, we acknowledge several caveats that have not yet been fully addressed. Below, we present and discuss these caveats, which are grouped into two categories, namely Model Systematics and Training Set Size.}

\subsubsection{Model Systematics}

\edit1{As mentioned in Section \ref{simulation section}, since the simulations are 1.) based on the physically descriptive \cite{brightman 2011} model, 2.) more numerous than the relatively small Known-z data set, and 3.) contain a more uniform range of parameter values than Known-z, we consider the simulated data to be a far more robust resource for all of the resulting analysis. However, it could be argued that the implementation of the model in the simulations is too simple, namely that there are few free parameters (e.g., with a frozen $\Gamma = 1.9$), which potentially diminishes the physical accuracy of the simulated spectra. Additionally, those concerns could be extended to the use of the simulations in the analysis since, as we have acknowledged, it may introduce some level of systematic bias, particularly when training the neural network on a simulation-dominated data set. It is possible, then, that the neural network would learn more about how to identify XZ successes in the simulated data than in actual observed AGN.}

\edit1{We believe all such concerns have merit, but do not undermine the results presented here. Although there are, undoubtedly, more problem classes that could be investigated and resolved---either with more realistic simulations or real test data---this work already resolves an interesting and important class of failures that occur both in our simulations and, as expected, in real data, such as the Known-z spectra we have analyzed here. This is evidenced by the improvement of accuracy on the Known-z sources after applying the MLP classifier, where (as shown in Table \ref{tab:counter_info}) the pass rate (in terms of Section \ref{xz failure analysis}'s pass/fail criteria) greatly increased from $\sim$69.5\% to $\sim$84.5\% and all tiers of success rates (as defined in Table \ref{tab:XZ_metrics}) have been improved to levels consistent with the findings of \cite{simmonds 2018}. Even when isolating only the Known-z sources in the testing set (to avoid bias from sources that helped train the MLP model), we showed in Table \ref{tab:machine_learning} that there was a statistically marginal ($\sim$1$\sigma$) yet noticeable increase in the pass rate (from $\sim$75.0\% to $\sim$82.9\%).} 

\edit1{Furthermore, specifically regarding the simulated models, we closely followed precedent set by, e.g., \cite{simmonds 2018} and \cite{peca 2021} in our choices of frozen parameters. In the case of $\Gamma$, we froze the parameter at 1.9 just as both \cite{simmonds 2018} and \cite{peca 2021} have done, largely due to its degeneracy with $z$ and $N_H$, both of which are already degenerate with each other. Such degeneracy is minimized in simulated data by eliminating $\Gamma$ as a free parameter to focus primarily on $z$ and $N_H$. In addition, as suggested by \cite{peca 2021} (and others ref. therein), given the low-counts nature of the majority of our data set, it is possible that using a more complex model would not have a positive impact on the work. Specifically, \cite{peca 2021} discusses the negative impact that more complicated models can have on estimating X-ray redshifts. Therefore, taking all concerns and counterpoints into consideration, we conclude that the lingering systematic effects of the simulations and models do not cause a substantial detriment to the analysis within the scope of this work and the problem classes it studies.}

\subsubsection{Training Set Size}

\edit1{An issue we have discussed is the size of our training data set, which we accounted for through our cross-validation analysis, but we did not yet fully address it at its roots. As mentioned, it is a relatively small training set for a deep neural network, which is a consequence of only producing 1000 simulations. Hence, it could be argued that this issue could be rectified by simply producing a larger set of simulations.}

\edit1{However, we do not believe a larger set of simulations would substantially improve the results. This is due to the limited number of possible meaningfully different simulations. As described in Section \ref{simulation section}, there are three free simulation inputs, namely counts (effectively spanning 150 to well over 5000, in increments of 1), $z$ (spanning 0.01--3 in increments of 0.01), and log$N_H$ (spanning 22--26, in increments of 0.01). While there are, therefore, thousands of unique combinations of those inputs, our analysis in Section \ref{xz failure analysis} suggests that there are relatively few meaningfully distinct simulations (i.e., distinguishable by the XZ procedure in its successes and failures).}

\edit1{It is clear by visual inspection of the input counts, $z$, and log$N_H$ XZ failure rote plots in Figure \ref{fig:err_hist_plots} that there is a limited number of distinguishable failure cases. Specifically, there appear to be roughly 5--11 unique intervals of counts showing changing failure rates compared to surrounding intervals. Likewise, there appear to be approximately 3--12 such intervals for $z$ and around 5--15 for log$N_H$. For example, with $z$, there are above-average failure rates between 0--0.75, below-average rates between 0.75 and $\sim$2.25, and above-average rates above 2.25, with some additional possible sub-classes of intervals viewable within each of those intervals.}

\edit1{From these estimates, we can approximate that there are between $\sim$75 and $\sim$1980 combinations of input parameters that are distinguishable by XZ in terms of its failure rate. Our choice of 1000 simulations, then, falls almost exactly in the center of this range, meaning that it approximately represents the maximum number of meaningfully different spectra. Therefore, if we were to produce substantially more simulations, particularly if we increased the training set size by orders of magnitude, the vast majority of them would effectively be duplicate entries in the training and testing set. This would be highly detrimental to the machine learning algorithm, as it would cause the model to become overfit to the same set of $\sim$1000 combinations we have already evaluated, thus depleting its predictive power on general data sets. Incidentally, this issue, too, could be further investigated using more realistic, more complex models, as the larger number of free parameters may offer a more robust set of meaningfully different spectra.}

\subsubsection{Takeaways}

\edit1{While these counter-arguments do not wholly eliminate our aforementioned concerns about systematic bias towards the simulations, the observed effectiveness of the machine learning model on the Known-z data set shows that the procedure in its current form provides useful results within the scope of the work. Due to the clear success, despite lingering concerns, we consider this analysis to be a strong foundational work for future studies of the XZ method and X-ray redshifts that can build upon the procedures employed here.}

\section{Results}\label{results}

Below, we describe our results. In Section \ref{no-z}, we provide the X-ray redshift catalog for our No-z sources, followed in Section \ref{all results} by a breakdown of the total combined Known-z and No-z data set to further validate and characterize the methodology.

\subsection{Redshift Catalog for No-z}\label{no-z}
After applying the MLP classifier to the IG-filtered No-z, we were left with a final No-z data set of 121 sources, all of which were X-ray selected. Given the performance of our MLP classifier on the total machine learning testing set, Known-z sources in the testing set, and the total Known-z data set, we can estimate that between $\sim$82\% and $\sim$89\% the redshifts documented in this catalog are consistent with the mild- to top-tier success standards. The distribution of the catalog's redshift values in the final data set is given in Figure \ref{fig:par_hist}, which shows that we find redshifts mostly between zero and $\sim$1.2, with various higher redshift results, up to $z \sim 5$. We find 74 new redshifts between 0.5 and 3.0, a notable result since obscured phases in this range are where most supermassive black hole growth occurs (see, e.g., \citealt{ueda 2003}; \citealt{treister 2012}; \citealt{buchner 2015}), making our catalog abundant with ideal sources for studying AGN accretion and evolution. We report this catalog of redshifts in the Appendix, where we offer a basic breakdown of the sources, including key features such as the CSC source names, coordinates, counts, and various model characteristics.

\subsection{Characterizing the Overall Data Set}\label{all results}

\begin{figure}[h!]

\includegraphics[width=9.cm]{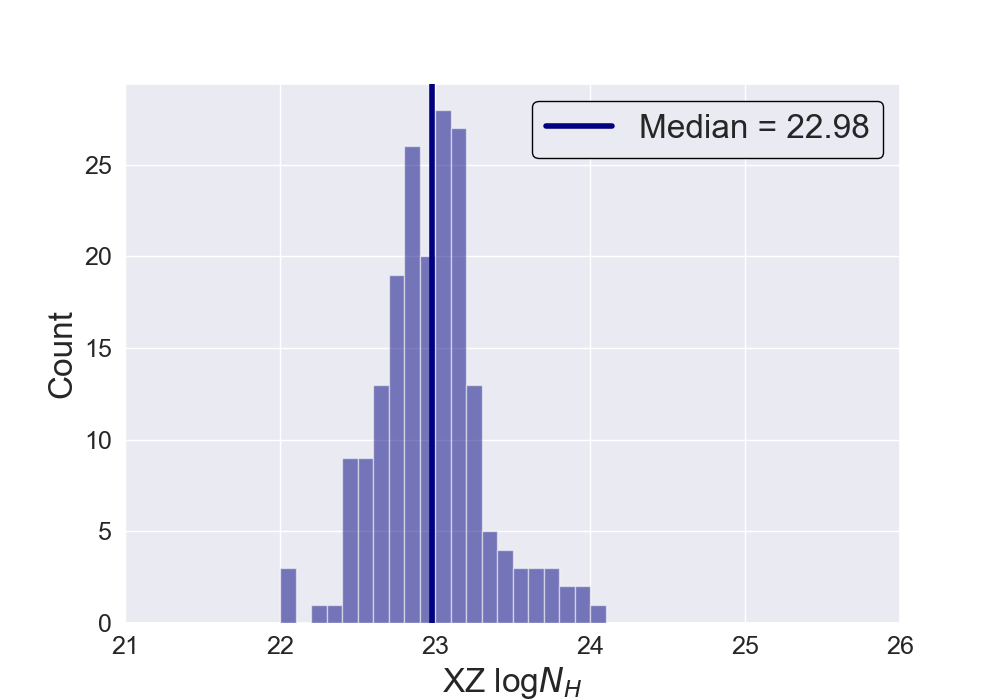}
\includegraphics[width=9.cm]{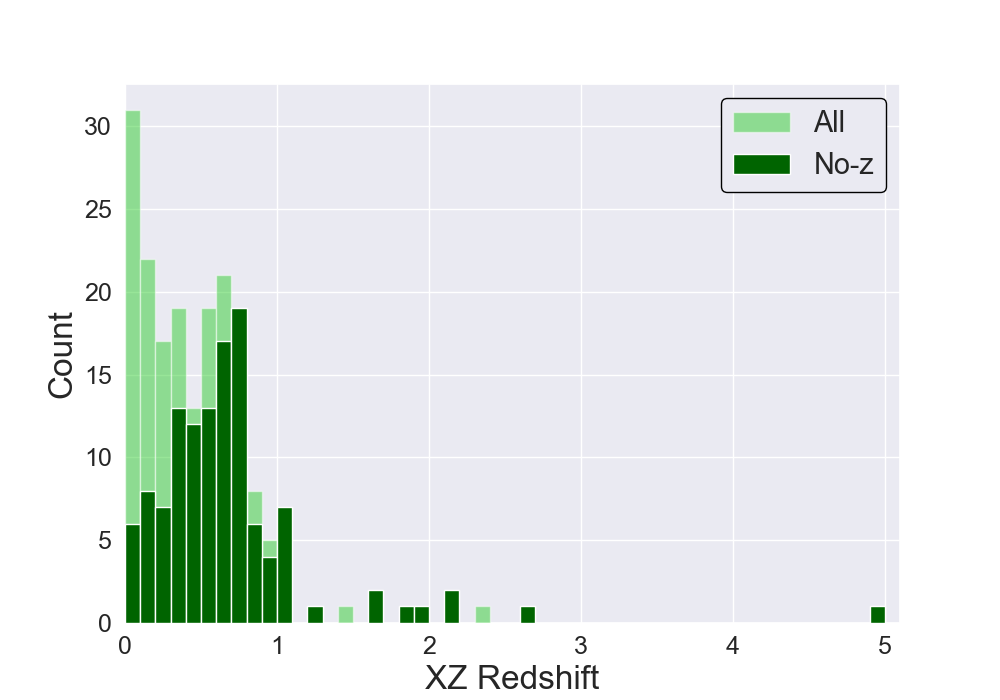}
\includegraphics[width=9.cm]{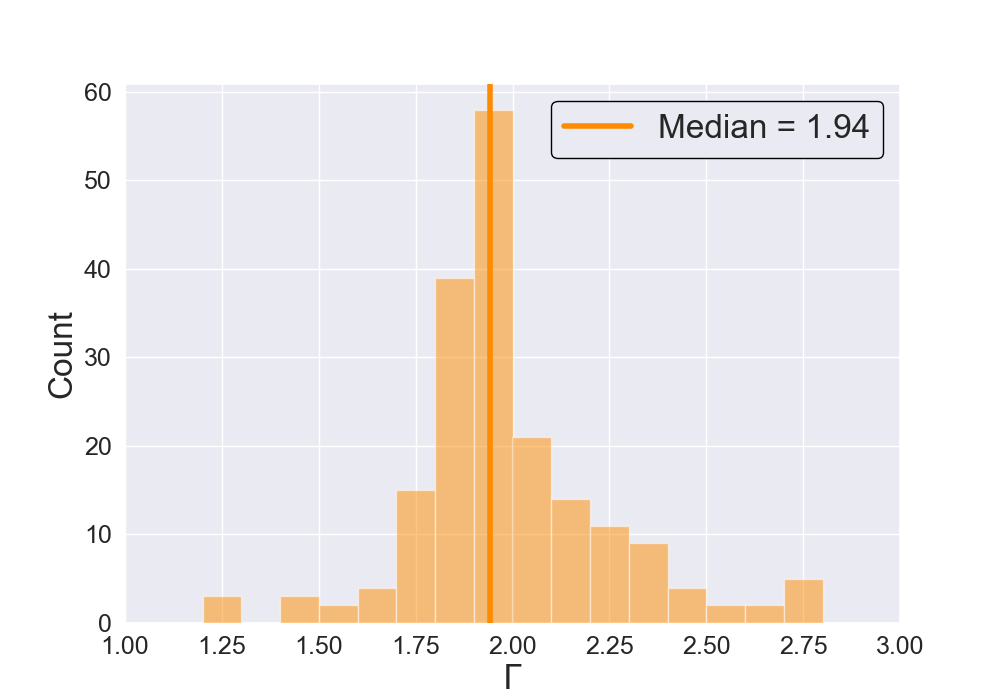}

\caption{The distribution of XZ-modeled best-fit parameter values for the fully-filtered data set. For a given spectrum, the best-fit value of each parameter is the median of that parameter's posterior probability distribution. Parameters shown are \textbf{Upper:} log$N_H$, \textbf{Middle:} redshift, and \textbf{Lower:} power law photon index ($\Gamma$). The posterior distributions from which these best-fit values were obtained can be found in the Appendix.}
\label{fig:par_hist}
\end{figure}

In Table \ref{tab:counter_info}, the number of sources in the data set at each stage of filtering is given. The Known-z success rates and catastrophic failure rates described above are also given in Table \ref{tab:counter_info}. In addition, the aforementioned Figure \ref{fig:par_hist} contains the total MLP-filtered data set's $N_H$ distribution, showing that most sources have a log$N_H$ between 22 and $\sim$24. Some sources exhibit highly-obscured log$N_H \sim 24$, and the distribution has a median value of log$N_H \sim 23$. We found one possible Compton-thick candidate, which we analyze and discuss further in Section \ref{CT}.

\begin{table}
\centering
 \begin{tabular}{ c | c c c c}
 
 \hline

  & Raw & IG & Final \\ 

 \hline 

\textbf{Total} & \textbf{363} & \textbf{302} & \textbf{192} \\ 

Known-z & 121 & 105 & 71 \\ 

No-z & 242 & 197 & 121 \\ 

\hline 

Total MW & 59 & 44 & 28 \\ 

Total X & 304 & 258 & 164 \\ 

Known-z MW & 55 & 44 & 28 \\ 

Known-z X & 66 & 61 & 43 \\ 

No-z MW & 4 & 0 & 0 \\ 

No-z X & 238 & 197 & 121 \\ 

\hline 

Basic Rate &  -  & 79.0\% & 91.5\% \\ 

Mild Rate &  -  & 67.6\% & 81.7\% \\ 

Top Rate &  -  & 48.6\% & 56.3\% \\ 

Cat. Fail &  -  & 10.5\% & 4.2\% \\ 

\hline

Pass Rate & - & 69.5\% & 84.5\% \\

 \hline

 \end{tabular}
 
 \caption{A summary of the initial (\textbf{Raw}), IG-filtered (\textbf{IG}), and MLP-filtered (\textbf{Final}) data sets. \edit1{Tiered success rates (\textbf{Basic Rate}, \textbf{Mild Rate}, and \textbf{Top Rate}) and the catastrophic failure rate (\textbf{Cat. Fail}) here are evaluated on the Known-z portion of each data set. \textbf{Pass Rate} refers to the pass/fail criteria set in Section \ref{xz failure analysis} for the machine learning procedure. Note that, as in the text, all such rates were assessed only on the IG-filtered and Final data sets.}}
 \label{tab:counter_info}
\end{table}

\begin{figure}[h!]

\includegraphics[width=9.cm]{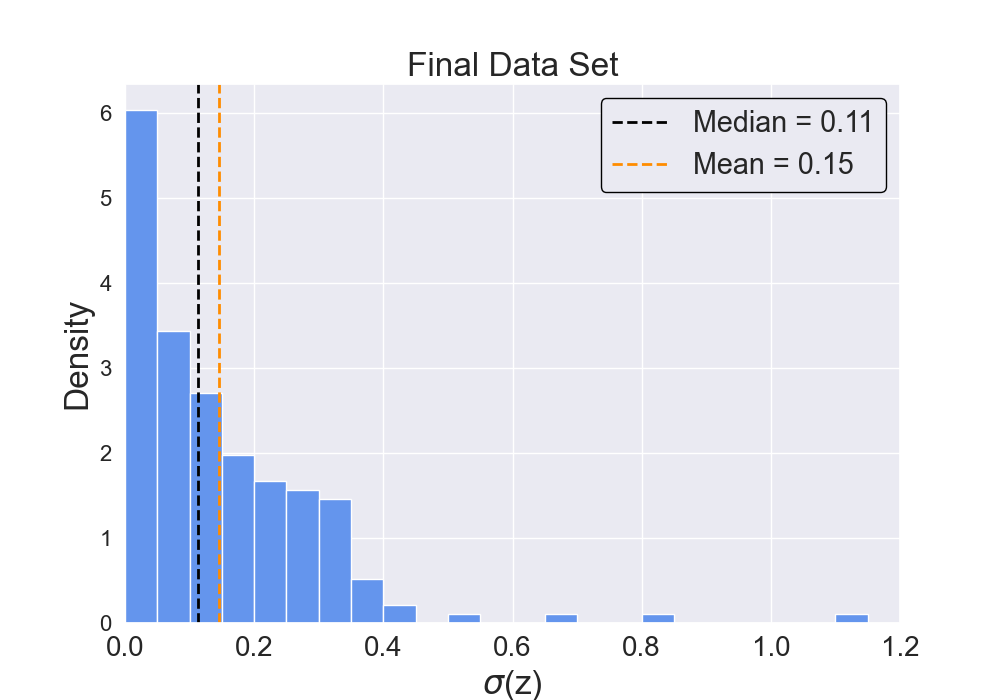}
\includegraphics[width=9.cm]{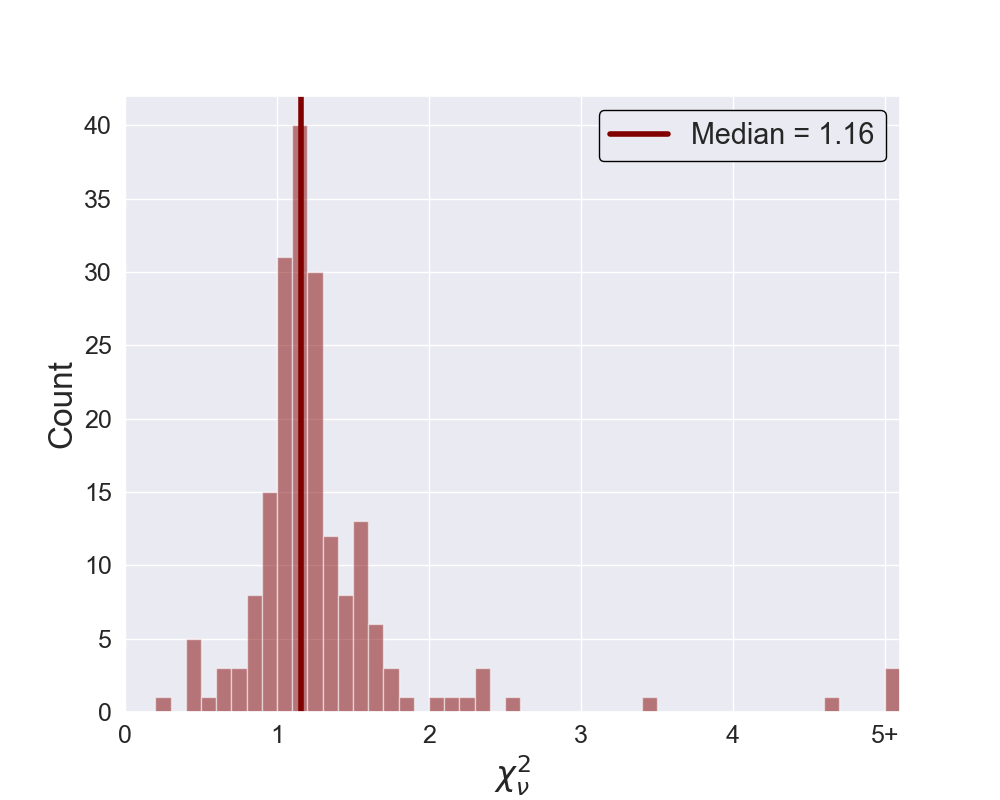}

\caption{\textbf{Upper:} The XZ redshift uncertainty distribution for the total fully-filtered data set, using the same format as seen in the lower panel of Figure \ref{fig:Delta_z_histogram}. \textbf{Lower:} The distribution of the reduced test statistic $\chi_{\nu}^2$. The median and mean values are represented by solid and dotted lines, respectively.}
\label{fig:par_hist2}
\end{figure}

The total data set's $\sigma(z)$ after the machine learning filter is shown in Figure \ref{fig:par_hist2}, which shows slight improvement over the initial IG-filtered distribution (Figure \ref{fig:Delta_z_histogram}), with the mean improving from $\sim$0.19 to $\sim$0.15 and the median improving from $\sim$0.13 $\sim$0.11. As previously discussed, the strong tendency towards zero, median of $\sim$0.1, and mean of $\sim$0.2 are consistent with the expectations set by \cite{simmonds 2018} and are largely controlled by filtering for high IG.

\begin{figure}[h!]
\includegraphics[width=10.cm]{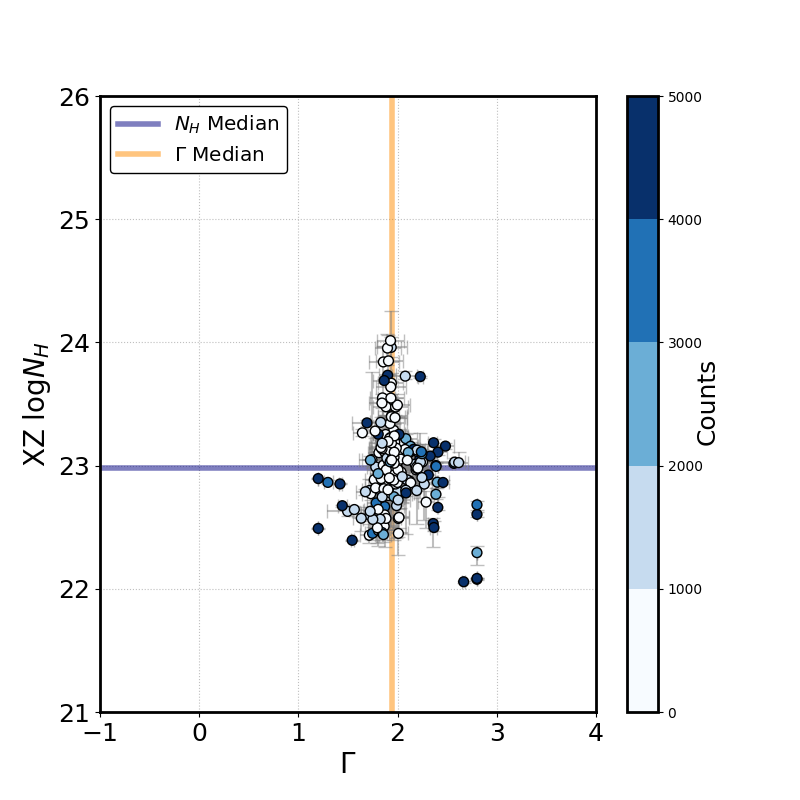}
\caption{The distribution of best-fit XZ log$N_H$ vs. $\Gamma$ for the total fully-filtered data set.}
\label{fig:NH_vs_Gamma}
\end{figure}

Another parameter of interest shown in Figure \ref{fig:par_hist} is $\Gamma$, where we find a strong tendency towards the median value of $\sim$1.9, which is consistent with expectations for obscured AGN. This is a somewhat trivial result, as XZ uses a Gaussian distribution centered at \edit2{1.95} as the prior for $\Gamma$ and low-count spectra may not substantially inform $\Gamma$. The $N_H$ and $\Gamma$ distributions have been revisualized in Figure \ref{fig:NH_vs_Gamma}, where they are plotted together.

For the best-fit parameters, reflected in Figure \ref{fig:par_hist}'s histograms, it is important to note that these values are, themselves, the medians of distributions. In particular, BXA's Nested Sampling fitting algorithm yields a posterior probability distribution for each parameter, and the best-fit value is simply the median of that parameter's posterior distribution. To fully and properly visualize the parameters, the Appendix contains a figure showing the posterior probability distributions for XZ, log$N_H$, and $\Gamma$ for all sources in the final data set.

The goodness of fit is also highly favorable in general, shown in Figure \ref{fig:par_hist2}. The reduced fit statistic ($\chi_{\nu}^2$) converges well around a median of $\sim$1.16, with few poor fits. Most fits exhibit $\chi_{\nu}^2 < 1.4$, with the vast majority of all sources being fitted within the $\sim$90\% confidence interval, which corresponds to $\chi_{\nu}^2 < 1.7$ and offers further confidence in our results.

\subsection{AGN Flux Variability}\label{variability}

A possible objection to our handling of data, alluded to in Section \ref{data selection}, is the stacking of spectra for sources with multiple observations (``multi-spectra sources" hereafter) due to the potential for some such sources to be variable AGN. The goal of the spectral stacking process is to optimize the XZ modeling results by maximizing the counts for a given multi-spectra source, but it is possible that the results may instead be negatively impacted if the source is a flux-variable AGN, which could potentially lead to issues such as a poorly-constrained redshift or an inaccurate redshift estimate.

\begin{figure}
\centering
\includegraphics[width=9.cm]{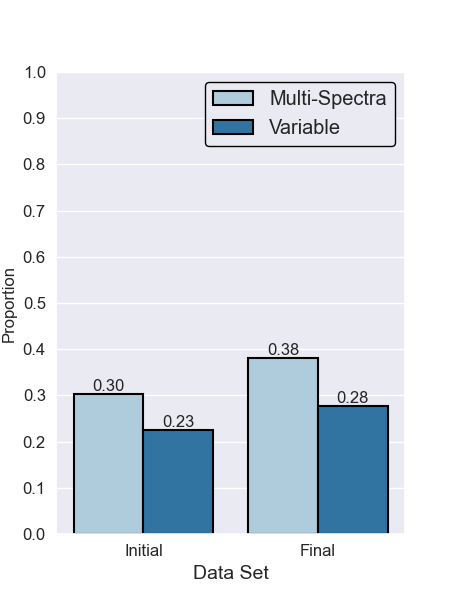}
\caption{Proportion of multi-spectra sources and multi-spectra variable sources in both the initial and final data sets.}
\label{fig:varplot}
\end{figure}

In our data set, however, the overall trend suggests that any XZ modeling issues related to AGN variability are non-existent or minor, and that the increase of statistics from the stacking process outweighs any such issues. This is demonstrated in Figure \ref{fig:varplot}, which shows the proportion of multi-spectra sources and variable multi-spectra sources (as identified by the CSC master source variability flag) in the data set, both for the initial data set (after AGN selection, but before any XZ modeling or filtering) and the final data set (the fully-filtered data).

As seen in the figure, the data set was initially $\sim$30\% multi-spectra sources, a proportion that increased to $\sim$38\% in the final data set. Since the final data set has been filtered using IG, which generally removes poorly constrained redshifts, and using the MLP classifier, which removes inaccurate redshift estimates, the increased proportion of multi-spectra is consistent with expectations that the increased counts of multi-spectra sources would lead to more favorable XZ results. This is evident in Table $\ref{tab:multispec_counts}$, which shows that multi-spectra sources have substantially higher mean and median counts than single-spectrum sources in both the initial and final data sets.

The proportion of variable AGN in the total data set also increased, from $\sim$23\% to $\sim$28\%. Moreover, the proportion of variable AGN among the multi-spectra sources saw only a slight decrease (from $\sim$77\% to $\sim$74\%) consistent with a statistical fluctuation ($\sim$0.3$\sigma$), showing that the variable multi-spectra AGN demonstrated approximately the same level of XZ success as their non-variable multi-spectra counterparts. These results suggest that if stacking variable AGN spectra has a negative impact on the XZ modeling process, any such impact is minor and overcome by the increased counts in multi-spectra sources. Therefore, we conclude that AGN variability introduces no serious issues to our data reduction or analysis procedures. This finding is strongly reinforced by the fact that \cite{simmonds 2018} made extensive use of stacked spectra in the original formulation of XZ.

\begin{table}
\centering
\begin{tabular}{c || c | c || c | c}

 \hline
 \textbf{Data} & \multicolumn{2}{c||}{\textbf{Single Spectrum}} & \multicolumn{2}{c}{\textbf{Multi-Spectra}}\\
 \cline{2-5}
  \textbf{Set} & Mean & Median & Mean & Median \\
\hline 
\hline

Initial & 992 & 413 & 9777 & 2392 \\ 

\hline

Final & 1436 & 542 & 8115 & 2519 \\

\hline

\end{tabular}
\caption{Mean and median counts in the 0.5--7 keV Chandra broad band for sources with one observation (Single Spectrum) and with multiple stacked spectra (Multi-Spectra), given for both the initial and fully-filtered final data sets.} 

\label{tab:multispec_counts}
\end{table}

\section{Beyond Redshifts}\label{discussion}

Thus far, we have been almost exclusively concerned with measuring and estimating redshifts. Here, we offer a discussion of results focused on other parameters. We first examine our general obscuration findings in Section \ref{cscXZ}, then focus on the possible Compton-thick candidate in Section \ref{CT}.

\subsection{CSC versus XZ: Obscuration}\label{cscXZ}

\begin{figure}[h!]
\includegraphics[width=10.cm]{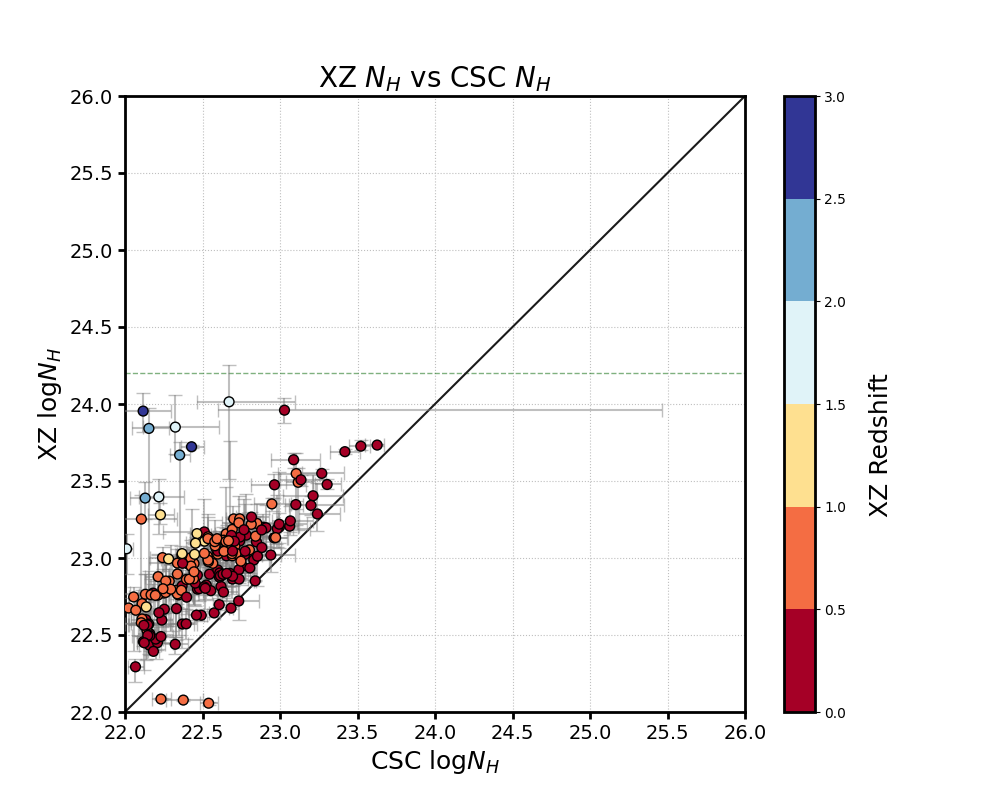}
\caption{XZ-computed Obscuration vs. the CSC-documented value for all sources in the fully-filtered data set. The black solid line represents XZ $N_H$ $=$ CSC $N_H$ and the color map indicates the best-fit XZ redshift. The green horizontal dashed line represents the Compton-thick log$N_H = 24.2$ threshold.}
\label{fig:logNH}
\end{figure}

In Figure \ref{fig:logNH}, we have revisualized the basic $N_H$ distribution from Figure \ref{fig:par_hist} by comparing the XZ-fitted values to those documented in the CSC. In particular, we used the CSC per-observation $N_H$ for each spectrum, taking the mean value for sources with multiple spectra, though typically the per-observation $N_H$ value is virtually identical for all such spectra, so the averaging process is generally trivial. As reflected by Figure \ref{fig:logNH}, it is clear that there is generally good agreement between the two $N_H$ measurements, with a moderate Pearson correlation coefficient of $\sim$0.6 and a mean difference of just $\sim$0.4 between them.

This is favorable for the CSC's reported obscuration values, as the XZ AGN model is substantially different from the CSC's simpler model. The CSC employs a simple absorbed power law for all X-ray sources, which assumes $z = 0$ and does not account for the AGN structure that is considered by XZ. That the CSC recovers, on average, approximately the same obscuration values as XZ strongly suggests that the CSC's values are fairly reliable estimates of the true source $N_H$.

In some individual cases, however, we have found considerable differences between the XZ best-fit $N_H$ and the CSC $N_H$, suggesting XZ may be more effective at detecting high obscuration. In a particularly notable case, we have identified a possible Compton-thick candidate (2CXO J215141.1-055049). A Compton-thick AGN is defined as $N_H \geq 1.5 \times 10^{24}$ cm$^{-2} = \sigma_T^{-1}$, where $\sigma_T$ is the Thomson cross-section. This definition corresponds to log$N_H \geq 24.2$, and for the source in question, the best-fit log$N_H \sim 24.0$ nearly reaches that threshold, with the 1$\sigma$ error range extending beyond log$N_H = 24.2$, thus demanding consideration as a Compton-thick candidate.

\begin{figure*}[t!]

\includegraphics[width=12.cm]{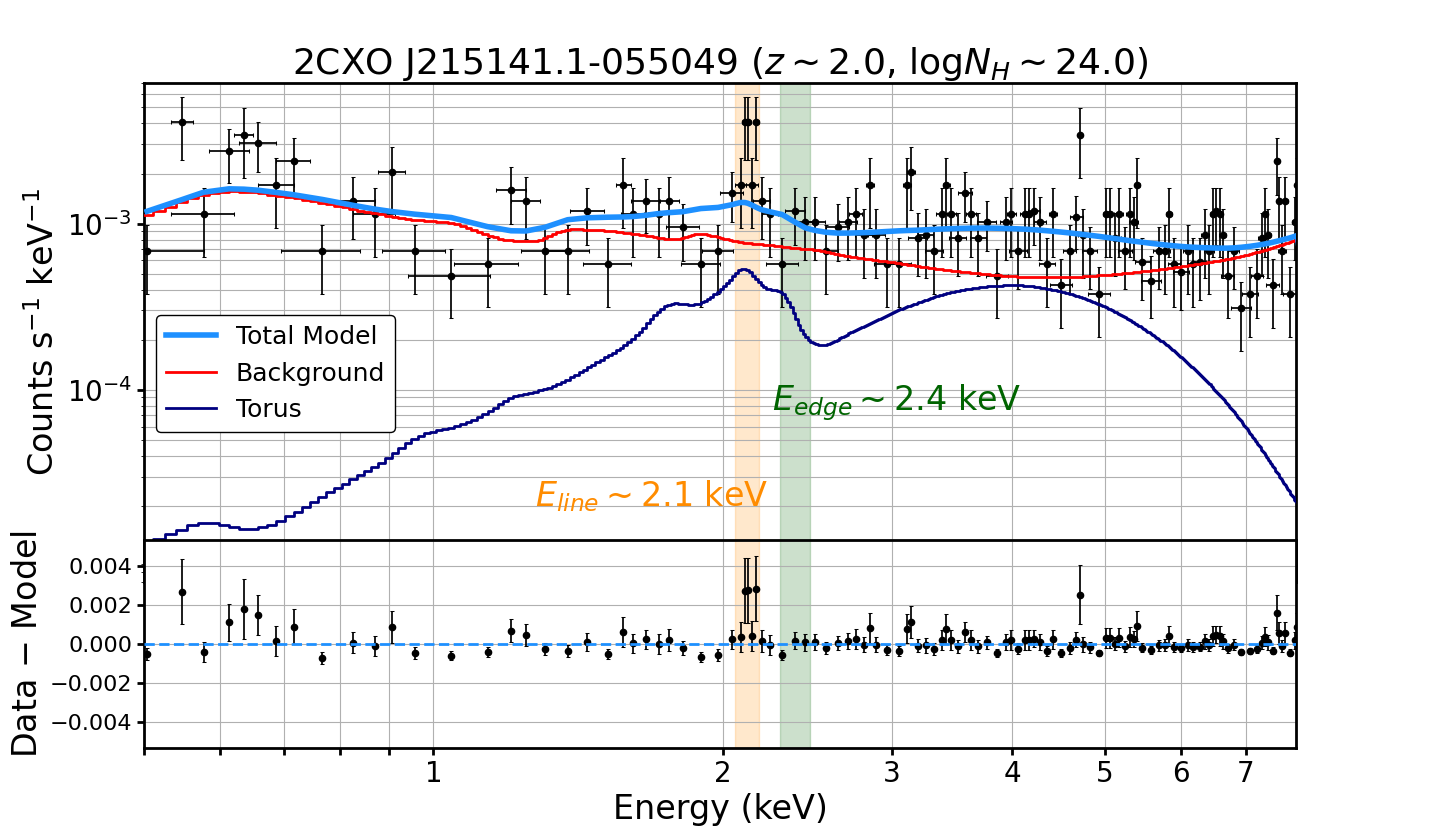}
\includegraphics[width=6.6cm]{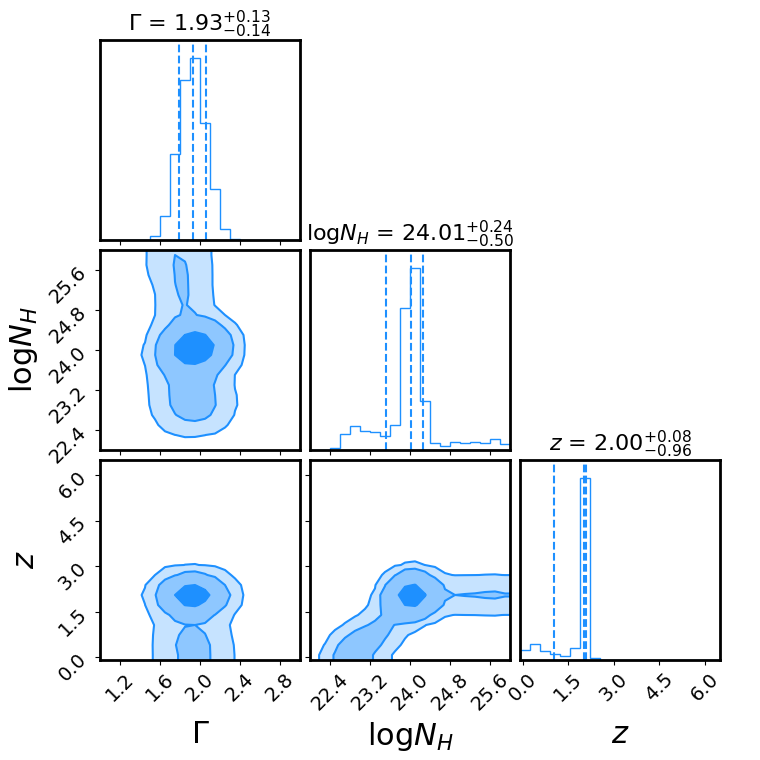}

\caption{\textbf{Left:} Spectrum and best-fit model for the Compton-thick candidate. The main model is plotted in sky blue with residuals. The two dominant model components, namely the background and torus, are also shown. The redshifted 6.4 keV Fe K$\alpha$ line and 7.1 keV Fe K$\alpha$ edge are highlighted. \textbf{Right:} Corner plot showing posterior parameter distributions for $\Gamma$, log$N_H$, and $z$.}
\label{fig:all_CT}
\end{figure*}

\subsection{Compton-Thick Candidate}\label{CT}

The spectrum of the Compton-thick candidate fitted with the XZ AGN model is shown Figure \ref{fig:all_CT}, and a breakdown of the source's characteristics is given in Table \ref{tab:CT_catalog} showing, among other quantities, its high 2--10 keV X-ray luminosity (log$L_X \sim 46$) and considerable CSC-documented hardness ratio (HR $\sim 0.999$). As mentioned, the 1$\sigma$ error range on its best-fit log$N_H \sim 24.0$ includes values beyond the Compton-thick threshold, and XZ found a best-fit $z \sim 2$. This redshift estimate is assigned a high probability of accuracy by the machine learning algorithm ($P_+ \sim 0.99$), which suggests it has a $\sim$99\% chance of meeting our mild- to top-tier success thresholds. Our cross-match attempts did not yield a counterpart for this source in any of the databases considered (Gaia, SDSS, WISE, and Simbad), so it does not have a redshift or any other multiwavelength data associated with it, and it has only one available spectrum in the CSC. The XZ model produced a fairly good fit, with a test statistic $\chi_{\nu}^2 \sim 1.39$. Also, as shown in Figure \ref{fig:all_CT}'s accompanying corner plots, key parameters $\Gamma$, log$N_H$, and $z$ are fairly constrained, with the posterior distributions showing strong peaks around the best-fit values.

The spectrum, containing 605 source counts in the \textit{Chandra} broad band, has a source-to-background counts ratio of just $\sim$0.1, as it is dominated by background and hence the spectrum has a peculiar flat shape. As a result, the total XZ model for this source is mostly background, although the torus model is also prominent, while the scattering and \texttt{APEC} components offer virtually no contribution. This is illustrated in Figure \ref{fig:all_CT}, where we have plotted the total model, as well as the background and torus components, showing the negligible nature of the other components. The torus model appears promising due to its high hardness ratio, which is consistent with the CSC-reported value and suggests the source is a highly-absorbed, possibly Compton-thick AGN. However, to arrive at a more definitive conclusion, additional source counts are required to overcome the current background-dominated spectrum's low counts.
 
 There also appears to be an excess at $\sim$6.5 keV, just beyond the highest possible observed energy for an AGN's Fe K$\alpha$ line (assuming $z \geq 0$), with its rest-frame $E = 6.4$ keV. There is no known 6.5 keV line in either the Chandra instrumental background \citep{bartalucci 2014} or the CXB (e.g., \citealt{hickox 2006}) and we do not expect any lines at that energy for a $z \sim 2$ AGN. Accordingly, fitting the excess with an emission component does not significantly improve the model, and is found to be a consistent only with a statistical fluctuation, as its significance is between $\sim$0 and $\sim$1$\sigma$ across four different methods of significance testing. Namely, we employed $\Delta\chi^2$, the Bayesian Information Criterion (BIC; \citealt{schwarz 1978}), the Akaike Information Criterion (AIC; \citealt{aic}), and the Bayes Factor (\citealt{kass 1995}; \citealt{friel 2012}). Note that the BIC approach merely approximates the Bayes Factor and was done using CSTAT, while the Bayes Factor was computed directly using the statistical evidence ($Z$) supplied by the BXA modeling procedure.

Taking all of this into account, it is clear that, due to the potentially Compton-thick obscuration level, background-dominated spectrum, and lack of counterparts, this source is of interest and warrants future study. It is important to further investigate this source and potentially verify its status as a Compton-thick AGN, due to the important role such highly-obscured cases may have in our overall understanding of AGN. Despite the expected abundance of Compton-thick AGN, thought to compose $\sim$10-50\% of the total AGN population (see \citealt{akylas 2009}, \citealt{vignali 2010}, \citealt{alexander 2011}, \citealt{buchner 2015}, \citealt{lanzuisi 2015}, \citealt{azadi 2017}), relatively few have been observed due to their high obscuration (e.g., \citealt{comastri 2010}, \citealt{vignali 2014}). X-ray surveys are known to reveal highly-obscured AGN via the hard band (e.g., \citealt{rees 1984}, \citealt{ueda 2003}), so confirming the Compton-thick nature of this AGN would be a useful continuation of this impactful X-ray trend.

\begin{table}
\centering
\begin{tabular}{l || l }
\hline

\multirow{3}{7em}{Source Name} & \multirow{3}{7em}{2CXO J215141.1 -055049} \\ 

  &   \\ 

  &   \\ 

\hline\hline

R.A. (deg) & 327.92138 \\ 

Dec (deg) & -5.84711 \\ 

$N_\mathrm{s}$ & 1 \\ 

Counts & 605 \\ 

CSC log$N_H$ & 22.67$_{-0.21}^{+0.42}$ \\ 

HR & 0.999 \\ 

$S/B$ & 0.114 \\

$\chi_{\nu}^2$ & 1.39 \\ 

log$L_X$ & 45.92$_{-1.36}^{+0.94}$ \\ 

$\Gamma$ & 1.93$_{-0.14}^{+0.13}$ \\ 

XZ log$N_H$ & 24.01$_{-0.50}^{+0.24}$ \\ 

XZ & 1.998$_{-0.96}^{+0.08}$ \\ 

$P_+$ & 0.9854 \\ 

\hline

 \end{tabular}

\caption{A breakdown of the Compton-thick candidate's characteristics. $N_\mathrm{s}$ is the number of available spectra; $S/B$ is the source-to-background counts ratio; log$L_X$ is the rest-frame 2--10 keV luminosity; and $P_+$ is MLP-computed probability that the redshift estimate is successful.} 

\label{tab:CT_catalog}

\end{table}

\section{Summary \& Closing Remarks}\label{conclusion}

In this work we have pushed and extended the limits of XZ with the help of the CSC, simulations, and machine learning techniques. Using a combined data set of simulated spectra and CSC-searched, largely X-ray selected obscured AGN with documented multiwavelength redshifts, we analyzed XZ's failure rate, specifically its dependence on various fit and spectral characteristics. This culminated in the formulation of a machine learning algorithm in which we used 7 features to classify a given XZ fit as either a success or failure using a multi-layer perceptron neural network.

We showed that our classifier exhibits a statistically significant $\sim$3$\sigma$ improvement over applying only the existing redshift IG $\geq$ 2 bits filter. After screening our entire data set with the machine learning model, we were left with $\sim$53\% of the original data set ($\sim$64\% of the IG-filtered data). Despite our aggressive treatment and added filtering, this still leaves us with a higher percentage of the original sources than what was seen in \cite{simmonds 2018}'s CDF-S analysis, where the IG filter alone left only $\sim$23\% of the original sources. We are left with a lower percentage than \cite{peca 2021}'s $\sim$70\%, though in that case the 54-source data set was from a deep \textit{Chandra} field with much more favorable redshifts, with $z \geq 0.5$ for $\sim$85\% of sources versus $\sim$47\% for our data set. With the fully-filtered data set, we produced redshift estimates for 121 CSC-searched obscured AGN with no cross-matched documented redshift values, all of which were selected using only X-ray data. The majority of these redshifts were found between 0.5--3, the ideal range for studying supermassive black hole growth due to its prevalence in that era of cosmological history. \edit1{Examining the XZ best-fit $N_H$ values, we have identified a possible Compton-thick source that will be the subject of further investigation.}

The findings of this work further demonstrate the combined capabilities of X-ray observatories and data science techniques, such as machine learning, in astrophysics. Through our use of X-ray spectra and a neural network, we reinforced the work of \cite{buchner 2014} and \cite{simmonds 2018}, among others, in showing that redshifts can be reliably estimated for obscured AGN using only X-rays, particularly for X-ray AGN selected without successfully-matched multiwavelength counterparts. As indicated by our final machine learning model's \edit1{precision}, we expect nearly 90\% of our 121 new redshift values to be consistent with hypothetical spectroscopic or photometric measurements, illustrating the reliability of our XZ and neural network combination.

We have also further demonstrated the wide applicability of the \textit{Chandra} Source Catalog, which supplied us with all sources and their corresponding fully-reduced X-ray data products. This work, then, is an example of the large-scale science and data analysis that can be performed with the CSC. This is all while the CSC2 only contains observations up to 2014. Hence, these capabilities will continue to grow with future CSC releases, as it catalogs more recent data. Moreover, we expect the impact of this work to go beyond the machine learning model we have produced, serving more generally as a foundational building block for future computations of redshifts using only X-ray data. In particular, given the amount of progress our methodology was able to make using a relatively small training set by deep learning standards, as well as the moderate success of other machine learning frameworks we tested, we can reasonably hypothesize that larger future data sets will produce more advanced and more effective models.

In the midst of recently-launched and future missions such as Athena, this fortification of XZ with machine learning algorithms will allow for wide-reaching analysis on the large data sets these observatories can offer in their early days. For example, it will be difficult to match counterparts with obscured AGN detected by Athena due to its large PSF. With a machine learning-infused XZ, AGN can be selected using the observed hardness ratios and redshifts can be reliably estimated for the obscured sources among them, bypassing the wait required when relying on multiwavelength counterparts. Furthermore, the machine learning aspect of this approach was formulated for \textit{Chandra} data, hence it can and should be thoroughly examined in the context of these new instruments, as well as in the context of other observatories such as XMM-Newton, to maximize its reach of applicability.

At minimum, this work serves as a proof-of-concept, showing the success of using machine learning to enhance the XZ method, opening the door for further similar improvements. Thus, with the enormous amount of X-ray data and machine learning techniques on the proverbial horizon, the future of X-ray redshifts and large-scale X-ray data analysis is remarkably luminous.

\acknowledgements
DS acknowledges the University of Miami for supplying funding throughout the completion of this work. DS also acknowledges Martin Elvis, Giuseppina Fabbiano, and J. Rafael Mart\'{i}nez-Galarza for useful, enlightening discussions. FC acknowledges the support of NASA Contract NAS8-03060. The authors thank the anonymous referee for helpful commentary.
\newline
\software{Astropy (\citealt{astropy 2013}; \citealt{astropy 2018}), Bayesian X-ray Analysis \citep{buchner 2014}, CIAO \citep{antonella ciao}, corner \citep{corner}, Matplotlib \citep{matplotlib}, NumPy \citep{numpy}, pandas (\citealt{pandas}; \citealt{pandas2}), Scikit-learn \citep{pedregosa 2012}, Sherpa \citep{sherpa}, XSPEC 12.10.1f \citep{arnaud 1996}}

\newpage 

\appendix

\begin{figure*}[htp]
\includegraphics[width=6cm]{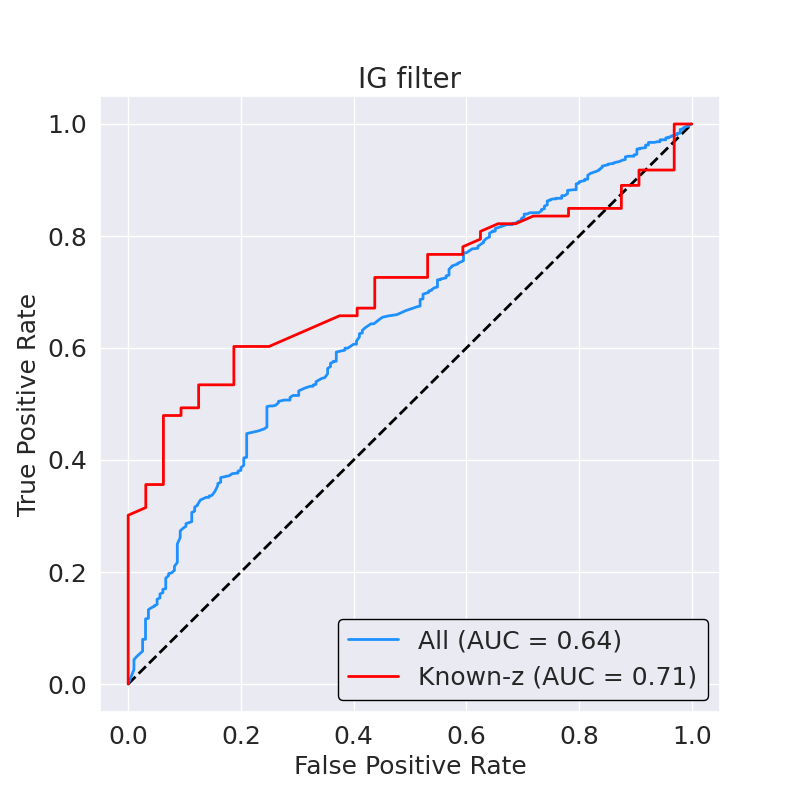}
\includegraphics[width=6cm]{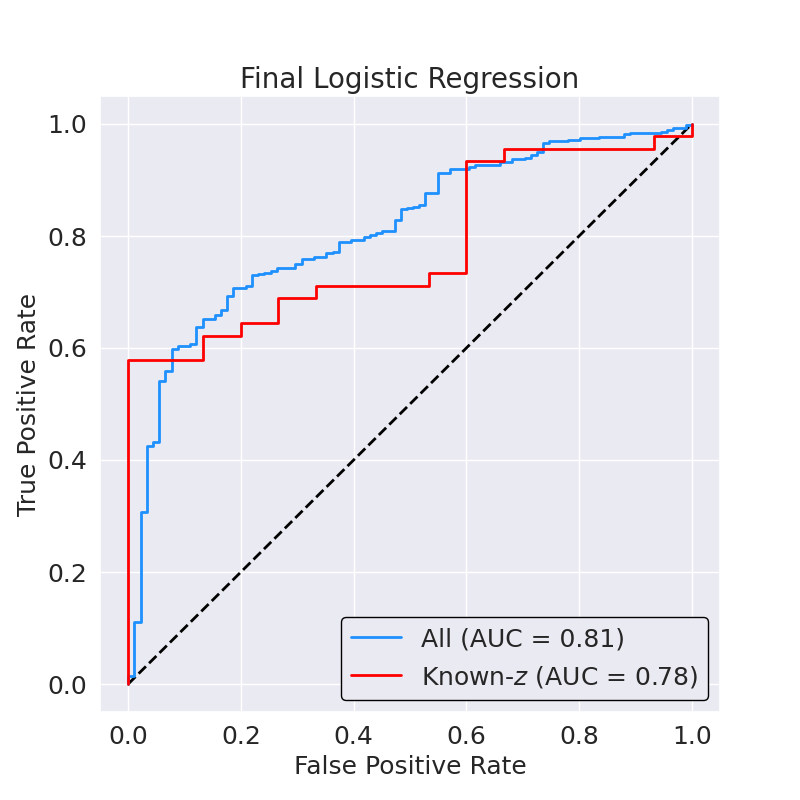}
\includegraphics[width=6cm]{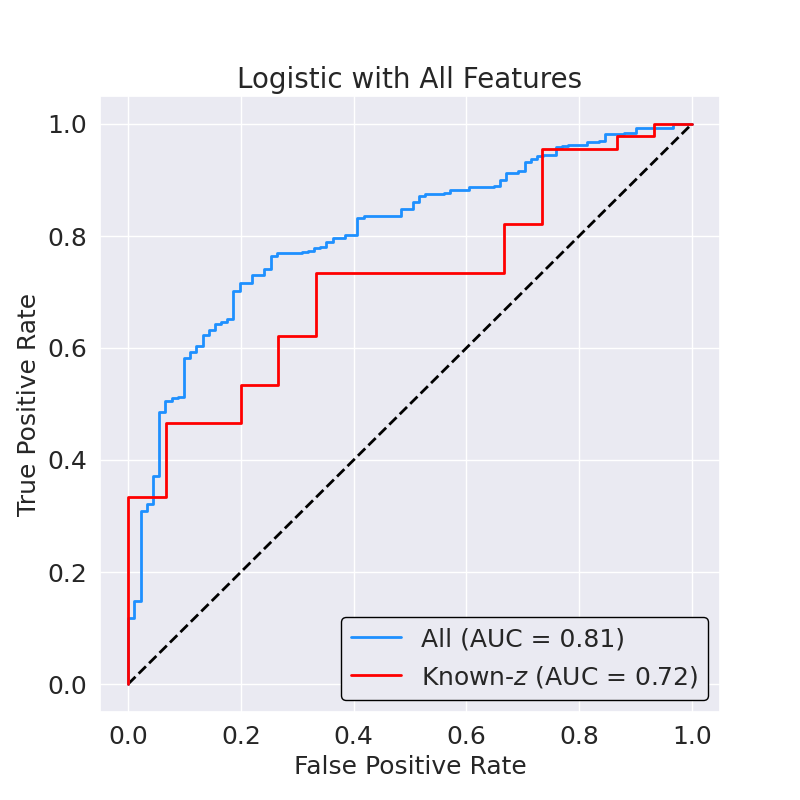}
\includegraphics[width=6cm]{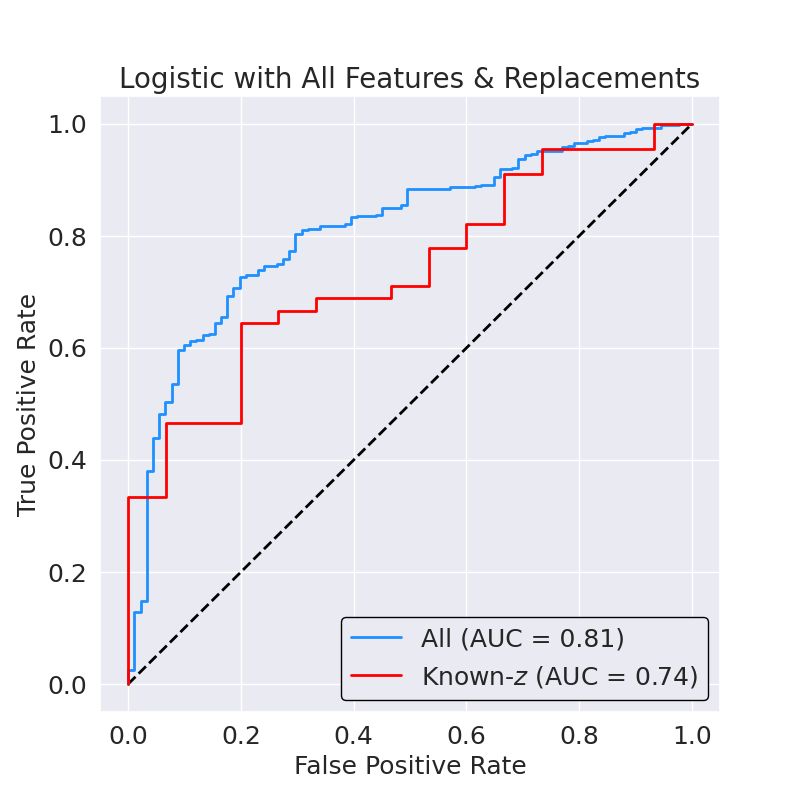}
\includegraphics[width=6cm]{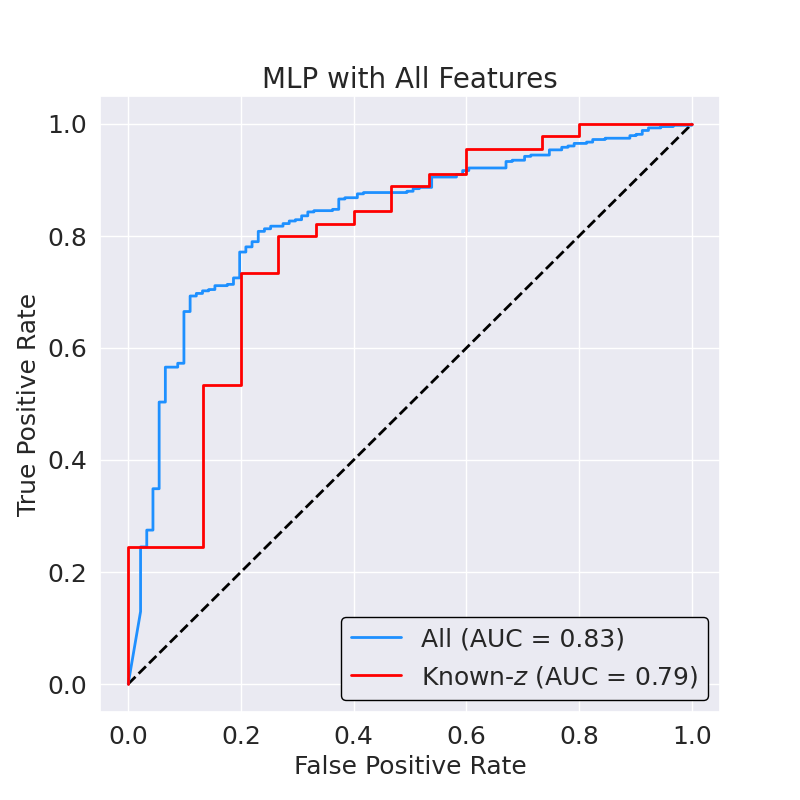}
\includegraphics[width=6cm]{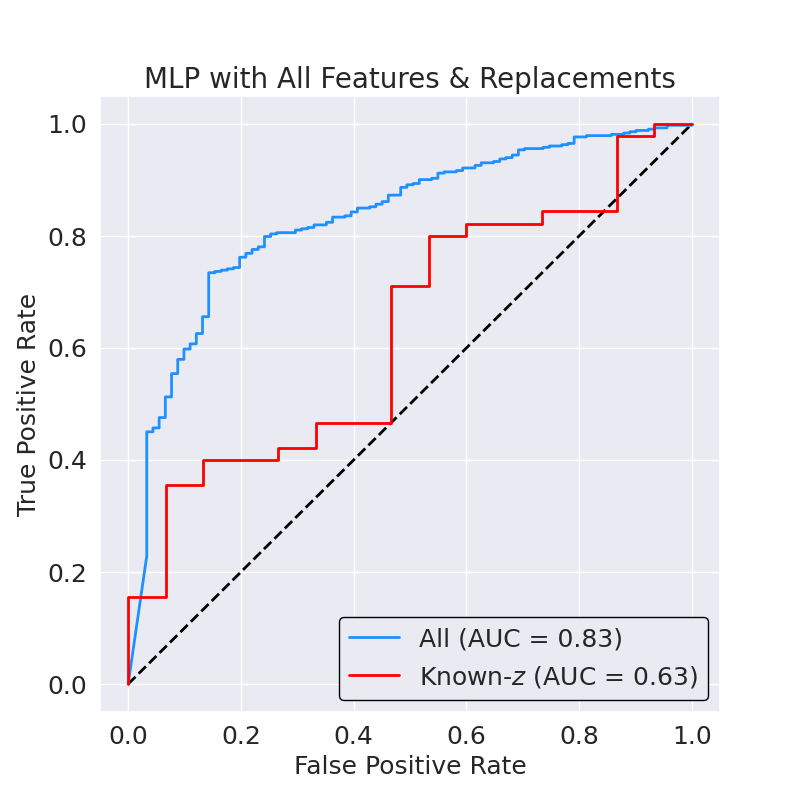}
\includegraphics[width=6cm]{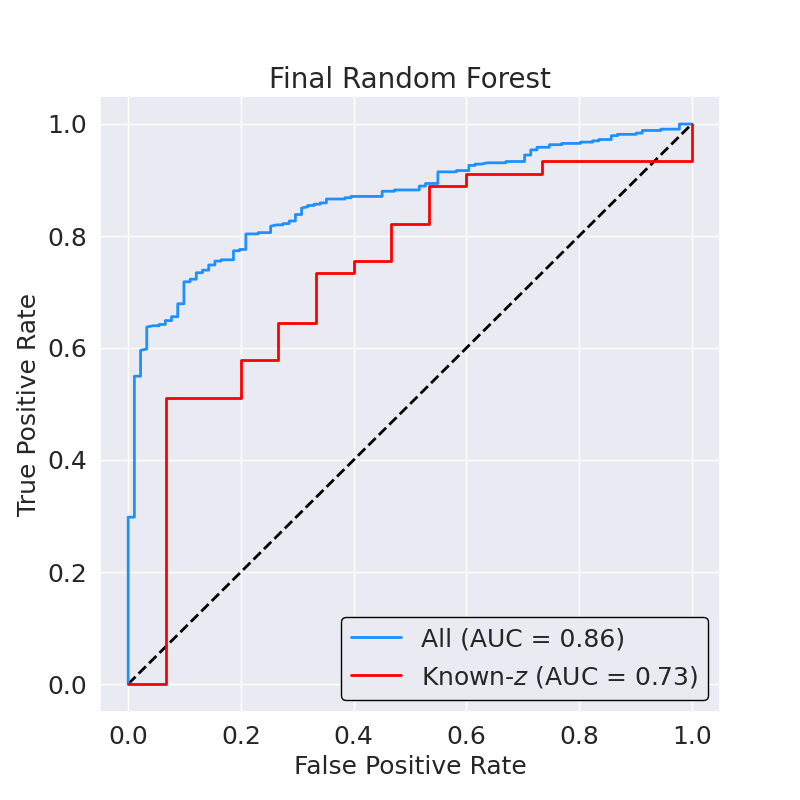}
\includegraphics[width=6cm]{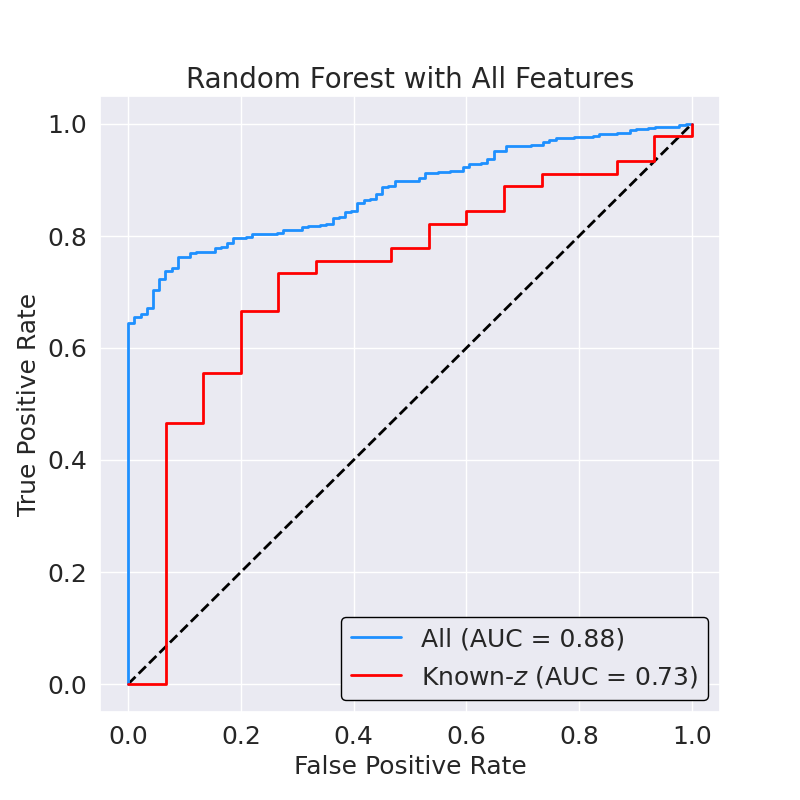}
\includegraphics[width=6cm]{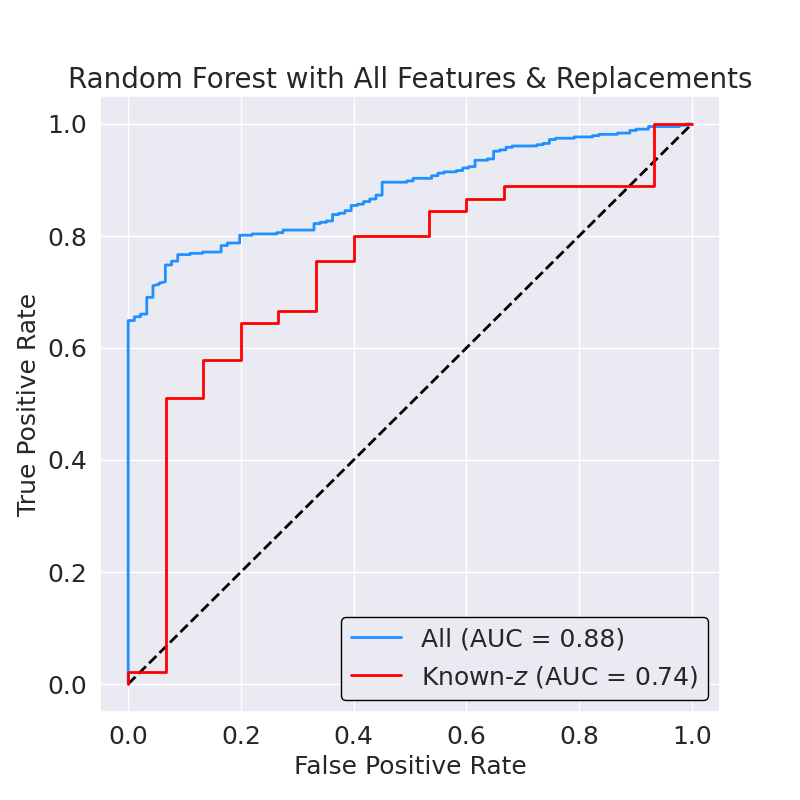}

\caption{ROC curves for all alternate classifiers, shown on both the total testing set and its Known-z sources. A black dotted line represents the baseline case of AUC = 0.5 for comparison.}
\label{fig:other_ROCs}
\end{figure*}

\begin{figure*}[htp]
\includegraphics[width=6.25cm]{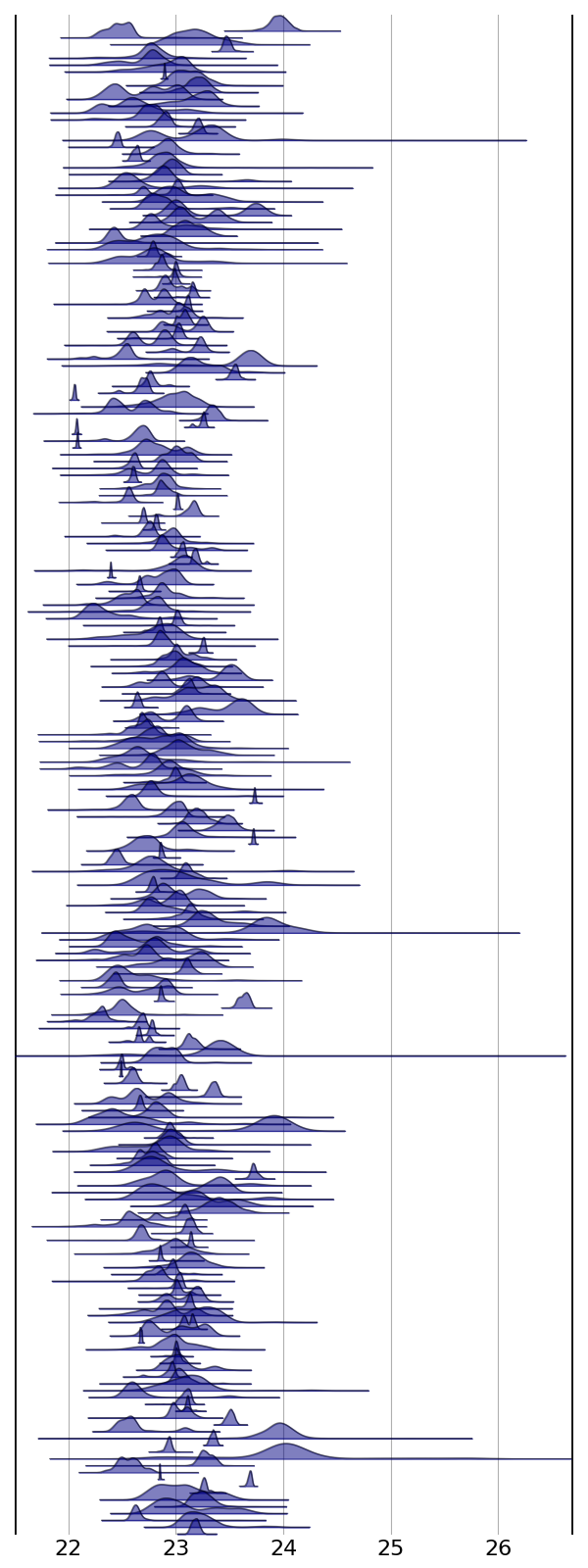}
\includegraphics[width=6.25cm]{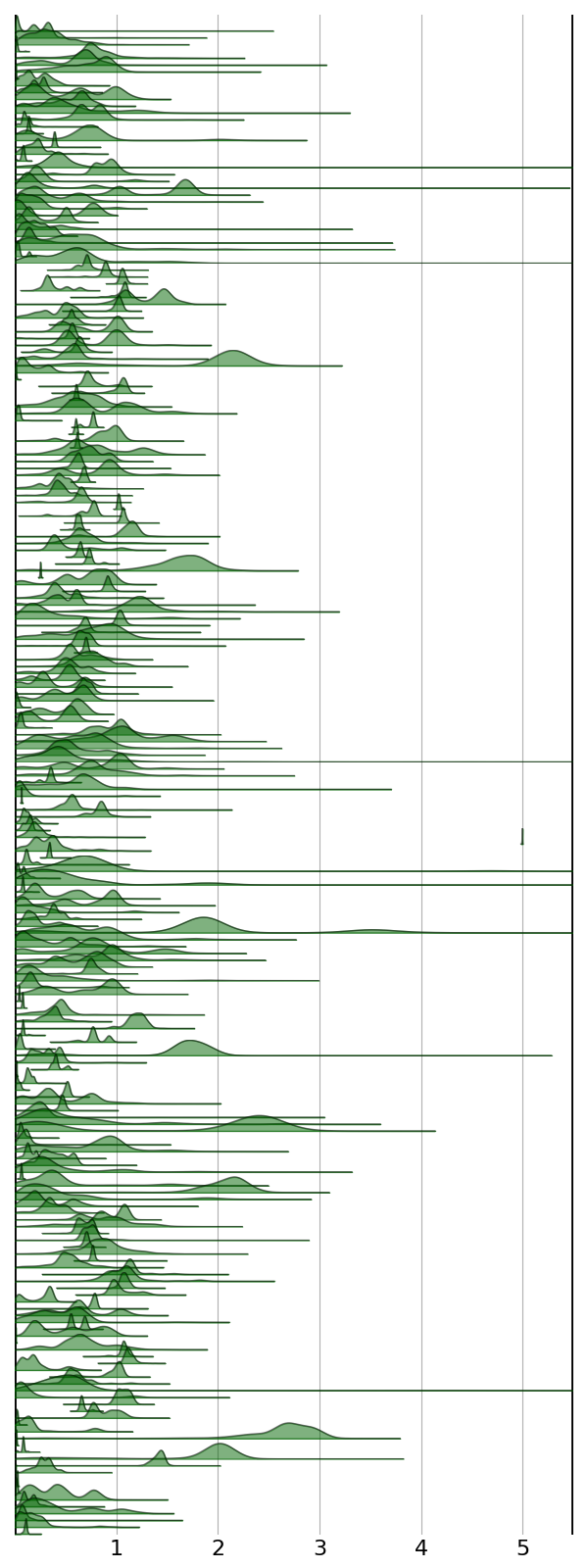}
\includegraphics[width=6.25cm]{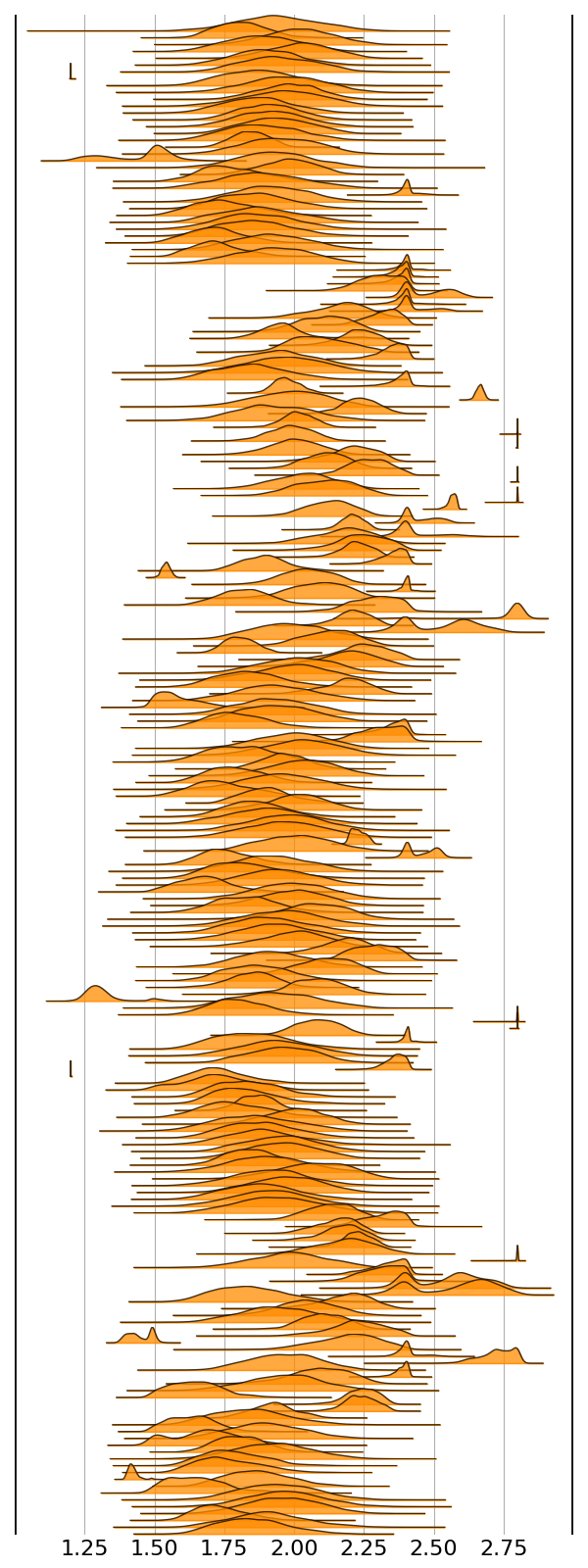}

\caption{The XZ model's posterior distribution of parameters for each spectrum in the fully-filtered data set. Parameters shown are \textbf{Left:} log$N_H$, \textbf{Middle:} redshift, and \textbf{Right:} power law photon index ($\Gamma$).}
\label{fig:par_ridgelines}
\end{figure*}

The machine learning model comparison mentioned in Section \ref{machine learning} can be found below in Table \ref{tab:attribute_test}, while the corresponding ROC curves can be found in Figure \ref{fig:other_ROCs}. The redshift catalog mentioned in Section \ref{no-z} can be found in Table \ref{tab:noz_catalog}. Posterior distributions for parameters can be found in Figure \ref{fig:par_ridgelines}.


\begin{longtable*}{c | c || c c c | c c c}

 \hline
 Model & Testing & Accuracy & Recall & AUC & Model & Null & Model \\
 
  & Subset & Score & & ROC & Precision  & Precision & Significance \\
  \hline\hline


 \multirow{4}{6em}{Final LR } & Total & 83.2\% & 97.7\% & 0.81 & 84.4\% & 82.6\% & \textbf{0.78$\sigma$} \\
 &   & \textit{+3.8\%} & \textit{+12.5\%} & \textit{+0.03} & \textit{-4.9\%} &   & \textit{-2.13}$\sigma$ \\
  \cline{2-8} 

 & Known-z & 76.7\% & 95.6\% & 0.78 & 78.2\% & 75.0\% & \textbf{0.40$\sigma$} \\
 &   & \textit{+6.7\%} & \textit{+20.0\%} & \textit{+0.05} & \textit{-4.7\%} &   & \textit{-0.55}$\sigma$ \\
 \hline\hline 


 \multirow{4}{6em}{LR, All Feat.} & Total & 83.6\% & 97.9\% & 0.81 & 84.6\% & 82.6\% & \textbf{0.86$\sigma$} \\
 &   & \textit{+4.2\%} & \textit{+12.7\%} & \textit{+0.03} & \textit{-4.7\%} &   & \textit{-2.04}$\sigma$ \\
  \cline{2-8} 

 & Known-z & 78.3\% & 95.6\% & 0.72 & 79.6\% & 75.0\% & \textbf{0.59$\sigma$} \\
 &   & \textit{+8.3\%} & \textit{+20.0\%} & \textit{-0.01} & \textit{-3.3\%} &   & \textit{-0.36}$\sigma$ \\
 \hline\hline 


 \multirow{4}{6em}{LR, All Feat. \& Sub.} & Total & 83.4\% & 97.2\% & 0.81 & 84.9\% & 82.6\% & \textbf{0.97$\sigma$} \\
 &   & \textit{+4.0\%} & \textit{+12.0\%} & \textit{+0.03} & \textit{-4.5\%} &   & \textit{-1.93}$\sigma$ \\
  \cline{2-8} 

 & Known-z & 78.3\% & 95.6\% & 0.74 & 79.6\% & 75.0\% & \textbf{0.59$\sigma$} \\
 &   & \textit{+8.3\%} & \textit{+20.0\%} & \textit{+0.01} & \textit{-3.3\%} &   & \textit{-0.36}$\sigma$ \\
 \hline\hline 


 \multirow{4}{6em}{Final RF } & Total & 82.8\% & 93.1\% & 0.86 & 87.0\% & 82.6\% & \textbf{1.92$\sigma$} \\
 &   & \textit{+3.4\%} & \textit{+7.9\%} & \textit{+0.08} & \textit{-2.3\%} &   & \textit{-0.99}$\sigma$ \\
  \cline{2-8} 

 & Known-z & 70.0\% & 93.3\% & 0.73 & 73.7\% & 75.0\% & \textbf{-0.16$\sigma$} \\
 &   & \textit{0.0\%} & \textit{+17.8\%} & \textit{+0.01} & \textit{-9.2\%} &   & \textit{-1.11}$\sigma$ \\
 \hline\hline 


 \multirow{4}{6em}{RF, All Feat.} & Total & 83.8\% & 93.8\% & 0.88 & 87.5\% & 82.6\% & \textbf{2.13$\sigma$} \\
 &   & \textit{+4.4\%} & \textit{+8.5\%} & \textit{+0.10} & \textit{-1.8\%} &   & \textit{-0.77}$\sigma$ \\
  \cline{2-8} 

 & Known-z & 73.3\% & 84.4\% & 0.73 & 80.9\% & 75.0\% & \textbf{0.72$\sigma$} \\
 &   & \textit{+3.3\%} & \textit{+8.9\%} & \textit{+0.01} & \textit{-2.1\%} &   & \textit{-0.23}$\sigma$ \\
 \hline\hline 


 \multirow{4}{6em}{RF, All Feat. \& Sub.} & Total & 83.8\% & 93.8\% & 0.88 & 87.5\% & 82.6\% & \textbf{2.13$\sigma$} \\
 &   & \textit{+4.4\%} & \textit{+8.5\%} & \textit{+0.10} & \textit{-1.8\%} &   & \textit{-0.77}$\sigma$ \\
  \cline{2-8} 

 & Known-z & 71.7\% & 88.9\% & 0.74 & 76.9\% & 75.0\% & \textbf{0.24$\sigma$} \\
 &   & \textit{+1.7\%} & \textit{+13.3\%} & \textit{+0.01} & \textit{-6.0\%} &   & \textit{-0.71}$\sigma$ \\
 \hline\hline 


 \multirow{4}{6em}{MLP, All Feat.} & Total & 81.9\% & 89.4\% & 0.83 & 88.8\% & 82.6\% & \textbf{2.68$\sigma$} \\
 &   & \textit{+2.5\%} & \textit{+4.2\%} & \textit{+0.05} & \textit{-0.6\%} &   & \textit{-0.23}$\sigma$ \\
  \cline{2-8} 

 & Known-z & 81.7\% & 95.6\% & 0.79 & 82.7\% & 75.0\% & \textbf{0.99$\sigma$} \\
 &   & \textit{+11.7\%} & \textit{+20.0\%} & \textit{+0.07} & \textit{-0.2\%} &   & \textit{+0.04}$\sigma$ \\
 \hline\hline 


 \multirow{4}{6em}{MLP, All Feat. \& Sub.} & Total & 82.1\% & 88.7\% & 0.83 & 89.5\% & 82.6\% & \textbf{3.02$\sigma$} \\
 &   & \textit{+2.7\%} & \textit{+3.5\%} & \textit{+0.05} & \textit{+0.2\%} &   & \textit{+0.11}$\sigma$ \\
  \cline{2-8} 

 & Known-z & 70.0\% & 82.2\% & 0.63 & 78.7\% & 75.0\% & \textbf{0.45$\sigma$} \\
 &   & \textit{0.0\%} & \textit{+6.7\%} & \textit{-0.09} & \textit{-4.2\%} &   & \textit{-0.50}$\sigma$ \\
 \hline\hline

 \caption{Performance metrics for all alternative models. The comparison to the final MLP model's performance (Table \ref{tab:machine_learning}) is reflected in italics. For metrics expressed as percents, we show the \textbf{change in percentage}, which is not to be confused with the percent change.}
 \label{tab:attribute_test}
\end{longtable*}

\newpage


\begin{longtable*}{c | c c c c c c c c c c }
 \hline

Name & R.A. & Dec & $N_\mathrm{s}$ & Cts. & HR & log$L_X$ & $\Gamma$ & XZ & XZ & $P_+$  \\ 

  & (deg) & (deg) &   &   &   &   &   & log$N_H$ &   &    \\ 

\hline

2CXO J000735.6+193206 & 1.89858 & 19.53504 & 1 & 1507 & 0.79 & 46.01$_{-3.26}^{+2.28}$ & 1.94$_{-0.14}^{+0.16}$ & 23.96$_{-0.08}^{+0.08}$ & 0.023$_{-0.009}^{+0.019}$ & 1.00  \\ 

2CXO J000917.2-321938 & 2.32171 & -32.32732 & 1 & 1190 & 0.93 & 44.42$_{-1.39}^{+1.80}$ & 1.81$_{-0.09}^{+0.09}$ & 22.47$_{-0.13}^{+0.11}$ & 0.216$_{-0.133}^{+0.127}$ & 1.00  \\ 

2CXO J002404.9-720455 & 6.02058 & -72.08210 & 1 & 236 & 0.77 & 44.28$_{-1.90}^{+1.92}$ & 2.02$_{-0.12}^{+0.12}$ & 23.17$_{-0.19}^{+0.22}$ & 0.316$_{-0.176}^{+0.221}$ & 1.00  \\ 

2CXO J004429.0-202538 & 11.12100 & -20.42745 & 1 & 671 & 0.84 & 44.68$_{-0.91}^{+1.11}$ & 2.03$_{-0.12}^{+0.11}$ & 22.77$_{-0.12}^{+0.11}$ & 0.755$_{-0.156}^{+0.183}$ & 1.00  \\ 

2CXO J004732.0-251721 & 11.88350 & -25.28933 & 1 & 375 & 0.92 & 44.34$_{-1.41}^{+1.62}$ & 1.89$_{-0.11}^{+0.12}$ & 22.76$_{-0.29}^{+0.10}$ & 0.659$_{-0.395}^{+0.105}$ & 1.00  \\ 

2CXO J004736.3-251638 & 11.90133 & -25.27741 & 1 & 266 & 0.94 & 44.48$_{-1.27}^{+1.36}$ & 1.94$_{-0.13}^{+0.14}$ & 22.96$_{-0.26}^{+0.13}$ & 0.753$_{-0.343}^{+0.178}$ & 1.00  \\ 

2CXO J004853.3-732458 & 12.22238 & -73.41613 & 1 & 340 & 0.94 & 44.26$_{-1.52}^{+1.65}$ & 1.87$_{-0.13}^{+0.13}$ & 23.11$_{-0.14}^{+0.15}$ & 0.197$_{-0.117}^{+0.154}$ & 1.00  \\ 

2CXO J013432.0+303454 & 23.63346 & 30.58174 & 3 & 825 & 0.88 & 44.03$_{-1.82}^{+1.33}$ & 1.98$_{-0.11}^{+0.10}$ & 22.76$_{-0.35}^{+0.26}$ & 0.621$_{-0.441}^{+0.375}$ & 0.97  \\ 

2CXO J013436.4+304713 & 23.65179 & 30.78721 & 1 & 167 & 1.00 & 44.22$_{-1.14}^{+1.40}$ & 1.97$_{-0.15}^{+0.14}$ & 23.22$_{-0.24}^{+0.12}$ & 0.615$_{-0.300}^{+0.090}$ & 0.90  \\ 

2CXO J014829.4+384754 & 27.12275 & 38.79856 & 1 & 320 & 0.88 & 44.41$_{-1.83}^{+1.63}$ & 1.87$_{-0.12}^{+0.12}$ & 22.60$_{-0.26}^{+0.30}$ & 0.396$_{-0.296}^{+0.436}$ & 1.00  \\ 

2CXO J015752.6+374231 & 29.46933 & 37.70874 & 1 & 710 & 0.87 & 44.61$_{-0.94}^{+1.19}$ & 1.91$_{-0.11}^{+0.11}$ & 22.76$_{-0.10}^{+0.09}$ & 0.702$_{-0.104}^{+0.159}$ & 1.00  \\ 

2CXO J020931.2-101135 & 32.38008 & -10.19318 & 3 & 766 & 0.97 & 43.44$_{-1.09}^{+1.61}$ & 1.92$_{-0.12}^{+0.12}$ & 22.90$_{-0.06}^{+0.05}$ & 0.104$_{-0.026}^{+0.044}$ & 1.00  \\ 

2CXO J022751.9+281438 & 36.96633 & 28.24408 & 1 & 205 & 0.87 & 44.74$_{-1.61}^{+1.55}$ & 1.88$_{-0.12}^{+0.14}$ & 23.25$_{-0.49}^{+0.16}$ & 0.680$_{-0.563}^{+0.139}$ & 1.00  \\ 

2CXO J025551.3+192620 & 43.96375 & 19.43898 & 4 & 3608 & 0.88 & 44.27$_{-0.50}^{+1.19}$ & 1.84$_{-0.07}^{+0.07}$ & 22.46$_{-0.03}^{+0.02}$ & 0.392$_{-0.012}^{+0.012}$ & 0.97  \\ 

2CXO J031103.6-270156 & 47.76538 & -27.03228 & 1 & 359 & 0.93 & 44.23$_{-1.50}^{+1.87}$ & 1.95$_{-0.13}^{+0.14}$ & 22.92$_{-0.10}^{+0.10}$ & 0.209$_{-0.109}^{+0.071}$ & 1.00  \\ 

2CXO J032220.9-512618 & 50.58729 & -51.43835 & 1 & 279 & 0.90 & 44.76$_{-1.48}^{+1.95}$ & 1.93$_{-0.13}^{+0.13}$ & 22.89$_{-0.13}^{+0.12}$ & 0.432$_{-0.117}^{+0.108}$ & 1.00  \\ 

2CXO J034257.7-352258 & 55.74058 & -35.38300 & 1 & 300 & 0.77 & 44.25$_{-2.57}^{+1.93}$ & 1.88$_{-0.14}^{+0.15}$ & 22.57$_{-0.10}^{+0.34}$ & 0.141$_{-0.041}^{+0.276}$ & 0.98  \\ 

2CXO J034518.4+315425 & 56.32675 & 31.90718 & 2 & 837 & 1.00 & 44.59$_{-1.99}^{+1.72}$ & 1.91$_{-0.13}^{+0.14}$ & 23.02$_{-0.17}^{+0.33}$ & 0.251$_{-0.149}^{+0.397}$ & 1.00  \\ 

2CXO J053503.9-052941 & 83.76642 & -5.49476 & 1 & 233 & 0.92 & 44.14$_{-1.56}^{+1.45}$ & 1.93$_{-0.14}^{+0.13}$ & 22.78$_{-0.27}^{+0.19}$ & 0.579$_{-0.367}^{+0.236}$ & 1.00  \\ 

2CXO J053504.5-052013 & 83.76892 & -5.33713 & 5 & 2677 & 0.91 & 44.37$_{-0.36}^{+0.47}$ & 2.40$_{-0.03}^{+0.01}$ & 22.87$_{-0.05}^{+0.04}$ & 0.704$_{-0.082}^{+0.036}$ & 1.00  \\ 

2CXO J053509.3-052141 & 83.78888 & -5.36150 & 3 & 3218 & 0.91 & 45.06$_{-0.37}^{+0.30}$ & 2.39$_{-0.04}^{+0.02}$ & 23.00$_{-0.02}^{+0.03}$ & 0.896$_{-0.033}^{+0.035}$ & 1.00  \\ 

2CXO J053509.6-052355 & 83.79029 & -5.39885 & 5 & 3515 & 0.84 & 44.85$_{-0.32}^{+0.22}$ & 2.39$_{-0.04}^{+0.02}$ & 22.99$_{-0.02}^{+0.02}$ & 1.063$_{-0.027}^{+0.031}$ & 1.00  \\ 

2CXO J053511.1-052237 & 83.79663 & -5.37710 & 8 & 4766 & 0.99 & 44.05$_{-0.49}^{+0.55}$ & 2.31$_{-0.08}^{+0.07}$ & 22.92$_{-0.06}^{+0.14}$ & 0.338$_{-0.039}^{+0.190}$ & 1.00  \\ 

2CXO J053511.5-052447 & 83.79817 & -5.41332 & 6 & 4315 & 0.94 & 45.01$_{-0.29}^{+0.16}$ & 2.48$_{-0.08}^{+0.09}$ & 23.16$_{-0.03}^{+0.03}$ & 1.075$_{-0.071}^{+0.028}$ & 0.80  \\ 

2CXO J053511.8-052100 & 83.79933 & -5.35002 & 7 & 4687 & 0.96 & 44.95$_{-0.29}^{+0.16}$ & 2.41$_{-0.02}^{+0.09}$ & 23.11$_{-0.03}^{+0.02}$ & 1.024$_{-0.038}^{+0.032}$ & 1.00  \\ 

2CXO J053512.1-052447 & 83.80050 & -5.41326 & 2 & 1601 & 0.97 & 44.30$_{-0.95}^{+0.85}$ & 2.18$_{-0.11}^{+0.08}$ & 23.01$_{-0.19}^{+0.10}$ & 0.485$_{-0.257}^{+0.118}$ & 0.99  \\ 

2CXO J053512.8-052133 & 83.80358 & -5.35935 & 7 & 4127 & 0.99 & 44.48$_{-0.27}^{+0.34}$ & 2.33$_{-0.07}^{+0.05}$ & 23.08$_{-0.06}^{+0.04}$ & 0.562$_{-0.055}^{+0.060}$ & 1.00  \\ 

2CXO J053513.5-051745 & 83.80650 & -5.29594 & 3 & 2124 & 0.94 & 44.52$_{-0.77}^{+0.47}$ & 2.11$_{-0.10}^{+0.09}$ & 23.14$_{-0.27}^{+0.13}$ & 0.867$_{-0.414}^{+0.163}$ & 0.90  \\ 

2CXO J053513.7-052135 & 83.80717 & -5.35991 & 5 & 5893 & 0.97 & 44.66$_{-0.97}^{+0.39}$ & 1.96$_{-0.07}^{+0.07}$ & 23.03$_{-0.03}^{+0.03}$ & 0.566$_{-0.022}^{+0.015}$ & 1.00  \\ 

2CXO J053514.2-052304 & 83.80950 & -5.38446 & 6 & 2185 & 1.00 & 44.14$_{-0.82}^{+0.42}$ & 2.08$_{-0.09}^{+0.12}$ & 23.22$_{-0.23}^{+0.04}$ & 0.633$_{-0.312}^{+0.036}$ & 1.00  \\ 

2CXO J053514.3-052308 & 83.80962 & -5.38558 & 5 & 4056 & 0.66 & 44.16$_{-0.79}^{+1.02}$ & 2.36$_{-0.05}^{+0.04}$ & 22.53$_{-0.19}^{+0.05}$ & 0.564$_{-0.242}^{+0.063}$ & 1.00  \\ 

2CXO J053514.4-051824 & 83.81017 & -5.30680 & 1 & 440 & 0.90 & 46.46$_{-1.30}^{+0.73}$ & 1.94$_{-0.13}^{+0.12}$ & 23.67$_{-0.66}^{+0.09}$ & 2.117$_{-1.241}^{+0.121}$ & 1.00  \\ 

2CXO J053514.5-052315 & 83.81058 & -5.38772 & 1 & 184 & 1.00 & 43.44$_{-2.14}^{+2.01}$ & 1.97$_{-0.14}^{+0.14}$ & 23.19$_{-0.12}^{+0.22}$ & 0.134$_{-0.082}^{+0.205}$ & 1.00  \\ 

2CXO J053514.6-052211 & 83.81108 & -5.36972 & 1 & 417 & 1.00 & 42.45$_{-0.79}^{+2.08}$ & 1.85$_{-0.13}^{+0.12}$ & 23.55$_{-0.04}^{+0.03}$ & 0.014$_{-0.003}^{+0.005}$ & 1.00  \\ 

2CXO J053514.7-052322 & 83.81133 & -5.38964 & 5 & 2868 & 1.00 & 44.32$_{-0.33}^{+0.47}$ & 2.39$_{-0.05}^{+0.02}$ & 22.77$_{-0.04}^{+0.05}$ & 0.728$_{-0.041}^{+0.082}$ & 1.00  \\ 

2CXO J053515.6-052338 & 83.81529 & -5.39414 & 6 & 92424 & 1.00 & 45.11$_{-0.11}^{+0.11}$ & 2.67$_{-0.01}^{+0.01}$ & 22.06$_{-0.01}^{+0.01}$ & 0.607$_{-0.014}^{+0.012}$ & 1.00  \\ 

2CXO J053515.7-052338 & 83.81567 & -5.39394 & 1 & 240 & 1.00 & 44.03$_{-1.33}^{+1.55}$ & 1.99$_{-0.14}^{+0.12}$ & 23.01$_{-0.19}^{+0.15}$ & 0.605$_{-0.214}^{+0.215}$ & 1.00  \\ 

2CXO J053515.9-052200 & 83.81654 & -5.36676 & 1 & 176 & 1.00 & 43.14$_{-2.25}^{+2.23}$ & 1.92$_{-0.13}^{+0.14}$ & 23.34$_{-0.06}^{+0.06}$ & 0.037$_{-0.016}^{+0.016}$ & 1.00  \\ 

2CXO J053516.2-052210 & 83.81779 & -5.36952 & 6 & 6318 & 0.98 & 44.97$_{-0.25}^{+0.14}$ & 2.01$_{-0.05}^{+0.06}$ & 23.25$_{-0.07}^{+0.02}$ & 0.769$_{-0.098}^{+0.015}$ & 0.97  \\ 

2CXO J053516.2-052316 & 83.81783 & -5.38789 & 6 & 91928 & -0.00 & 45.13$_{-0.19}^{+0.09}$ & 2.80$_{-0.00}^{+0.00}$ & 22.08$_{-0.01}^{+0.01}$ & 0.603$_{-0.013}^{+0.012}$ & 1.00  \\ 

2CXO J053516.7-052019 & 83.81967 & -5.33883 & 1 & 1902 & 0.09 & 44.97$_{-1.05}^{+0.95}$ & 1.99$_{-0.08}^{+0.08}$ & 22.68$_{-0.10}^{+0.06}$ & 0.936$_{-0.159}^{+0.085}$ & 1.00  \\ 

2CXO J053516.7-052327 & 83.81983 & -5.39106 & 4 & 90771 & 0.19 & 45.10$_{-0.20}^{+0.10}$ & 2.80$_{-0.00}^{+0.00}$ & 22.08$_{-0.01}^{+0.01}$ & 0.617$_{-0.012}^{+0.011}$ & 1.00  \\ 

2CXO J053516.9-051840 & 83.82050 & -5.31131 & 2 & 1583 & 0.79 & 44.49$_{-0.89}^{+0.86}$ & 2.00$_{-0.09}^{+0.09}$ & 22.79$_{-0.13}^{+0.29}$ & 0.713$_{-0.191}^{+0.505}$ & 1.00  \\ 

2CXO J053517.4-052105 & 83.82292 & -5.35166 & 1 & 900 & 0.97 & 44.68$_{-0.60}^{+0.65}$ & 2.22$_{-0.09}^{+0.08}$ & 23.01$_{-0.10}^{+0.12}$ & 0.727$_{-0.147}^{+0.173}$ & 0.78  \\ 

2CXO J053517.6-052153 & 83.82338 & -5.36489 & 3 & 1079 & 0.76 & 44.69$_{-1.02}^{+0.80}$ & 2.27$_{-0.08}^{+0.07}$ & 22.85$_{-0.29}^{+0.08}$ & 0.898$_{-0.452}^{+0.097}$ & 1.00  \\ 

2CXO J053517.9-051613 & 83.82492 & -5.27032 & 9 & 12953 & 0.53 & 44.71$_{-1.57}^{+1.04}$ & 2.80$_{-0.00}^{+0.00}$ & 22.60$_{-0.02}^{+0.02}$ & 0.682$_{-0.026}^{+0.024}$ & 1.00  \\ 

2CXO J053518.5-052149 & 83.82721 & -5.36385 & 2 & 736 & 1.00 & 43.81$_{-0.75}^{+1.23}$ & 2.12$_{-0.12}^{+0.10}$ & 22.88$_{-0.05}^{+0.07}$ & 0.431$_{-0.059}^{+0.074}$ & 1.00  \\ 

2CXO J053519.5-051702 & 83.83150 & -5.28411 & 8 & 8890 & 0.89 & 45.23$_{-0.20}^{+0.38}$ & 2.57$_{-0.02}^{+0.01}$ & 23.02$_{-0.01}^{+0.01}$ & 1.024$_{-0.012}^{+0.013}$ & 1.00  \\ 

2CXO J053519.8-051534 & 83.83254 & -5.25964 & 6 & 2136 & 1.00 & 44.24$_{-0.73}^{+0.37}$ & 2.13$_{-0.09}^{+0.08}$ & 23.16$_{-0.08}^{+0.04}$ & 0.769$_{-0.124}^{+0.022}$ & 1.00  \\ 

2CXO J053521.0-052055 & 83.83788 & -5.34869 & 2 & 438 & 0.97 & 44.08$_{-0.76}^{+1.06}$ & 2.19$_{-0.12}^{+0.09}$ & 22.97$_{-0.10}^{+0.09}$ & 0.654$_{-0.103}^{+0.136}$ & 0.53  \\ 

2CXO J053521.4-051710 & 83.83938 & -5.28624 & 2 & 1184 & 1.00 & 44.39$_{-0.74}^{+0.76}$ & 2.25$_{-0.09}^{+0.09}$ & 22.90$_{-0.05}^{+0.27}$ & 0.414$_{-0.041}^{+0.385}$ & 1.00  \\ 

2CXO J053521.7-051946 & 83.84058 & -5.32948 & 7 & 4739 & 0.98 & 44.62$_{-0.21}^{+0.24}$ & 2.22$_{-0.05}^{+0.06}$ & 23.06$_{-0.03}^{+0.02}$ & 0.642$_{-0.026}^{+0.020}$ & 1.00  \\ 

2CXO J053522.9-051744 & 83.84583 & -5.29574 & 7 & 5642 & 0.98 & 44.95$_{-0.21}^{+0.23}$ & 2.36$_{-0.05}^{+0.04}$ & 23.19$_{-0.03}^{+0.03}$ & 0.736$_{-0.023}^{+0.027}$ & 1.00  \\ 

2CXO J053523.6-052212 & 83.84871 & -5.37015 & 1 & 1197 & 0.66 & 45.99$_{-0.93}^{+0.67}$ & 1.89$_{-0.08}^{+0.08}$ & 23.06$_{-0.16}^{+0.10}$ & 1.656$_{-0.318}^{+0.197}$ & 0.59  \\ 

2CXO J053524.7-052759 & 83.85292 & -5.46651 & 11 & 43360 & 0.91 & 44.36$_{-0.67}^{+0.13}$ & 1.54$_{-0.02}^{+0.01}$ & 22.39$_{-0.01}^{+0.01}$ & 0.254$_{-0.004}^{+0.004}$ & 1.00  \\ 

2CXO J053527.9-051859 & 83.86642 & -5.31665 & 2 & 1226 & 0.93 & 44.39$_{-0.98}^{+0.96}$ & 2.05$_{-0.09}^{+0.10}$ & 22.90$_{-0.19}^{+0.12}$ & 0.770$_{-0.309}^{+0.161}$ & 1.00  \\ 

2CXO J053531.7-052120 & 83.88229 & -5.35570 & 4 & 1046 & 0.97 & 44.06$_{-0.78}^{+0.97}$ & 2.09$_{-0.11}^{+0.10}$ & 22.88$_{-0.07}^{+0.14}$ & 0.412$_{-0.074}^{+0.192}$ & 1.00  \\ 

2CXO J053715.2-691316 & 84.31358 & -69.22131 & 5 & 1664 & 1.00 & 44.29$_{-1.20}^{+1.53}$ & 1.82$_{-0.09}^{+0.09}$ & 22.57$_{-0.13}^{+0.10}$ & 0.469$_{-0.132}^{+0.152}$ & 1.00  \\ 

2CXO J054139.4-015327 & 85.41454 & -1.89086 & 1 & 259 & 0.94 & 44.62$_{-1.29}^{+1.37}$ & 1.98$_{-0.13}^{+0.13}$ & 22.85$_{-0.29}^{+0.15}$ & 0.813$_{-0.444}^{+0.206}$ & 1.00  \\ 

2CXO J054140.0-015335 & 85.41692 & -1.89328 & 1 & 812 & 0.93 & 44.80$_{-0.62}^{+0.97}$ & 2.13$_{-0.10}^{+0.09}$ & 22.86$_{-0.05}^{+0.06}$ & 0.669$_{-0.058}^{+0.078}$ & 1.00  \\ 

2CXO J054141.3-015332 & 85.42229 & -1.89249 & 1 & 6643 & 0.99 & 45.94$_{-0.30}^{+0.17}$ & 1.80$_{-0.05}^{+0.06}$ & 23.25$_{-0.02}^{+0.02}$ & 0.701$_{-0.020}^{+0.017}$ & 0.99  \\ 

2CXO J054141.3-015444 & 85.42246 & -1.91246 & 1 & 620 & 0.99 & 44.73$_{-0.55}^{+0.73}$ & 2.25$_{-0.09}^{+0.09}$ & 23.03$_{-0.06}^{+0.15}$ & 0.575$_{-0.073}^{+0.210}$ & 0.97  \\ 

2CXO J054142.6-015446 & 85.42758 & -1.91279 & 1 & 423 & 0.95 & 44.76$_{-0.80}^{+1.03}$ & 2.20$_{-0.11}^{+0.09}$ & 22.98$_{-0.11}^{+0.11}$ & 0.748$_{-0.148}^{+0.160}$ & 0.54  \\ 

2CXO J054143.8-015622 & 85.43258 & -1.93954 & 1 & 255 & 1.00 & 44.40$_{-0.93}^{+1.26}$ & 2.01$_{-0.14}^{+0.13}$ & 23.08$_{-0.12}^{+0.13}$ & 0.524$_{-0.112}^{+0.183}$ & 1.00  \\ 

2CXO J054144.1-015347 & 85.43388 & -1.89652 & 1 & 299 & 0.99 & 44.73$_{-1.21}^{+0.92}$ & 2.00$_{-0.14}^{+0.12}$ & 23.49$_{-0.31}^{+0.09}$ & 0.522$_{-0.331}^{+0.059}$ & 1.00  \\ 

2CXO J054145.0-015406 & 85.43775 & -1.90183 & 1 & 630 & 0.99 & 44.56$_{-1.32}^{+1.06}$ & 1.83$_{-0.11}^{+0.12}$ & 22.89$_{-0.13}^{+0.31}$ & 0.293$_{-0.138}^{+0.405}$ & 1.00  \\ 

2CXO J054145.8-015411 & 85.44125 & -1.90314 & 1 & 1008 & 0.97 & 45.12$_{-0.38}^{+0.51}$ & 2.19$_{-0.09}^{+0.07}$ & 23.12$_{-0.06}^{+0.04}$ & 0.692$_{-0.057}^{+0.070}$ & 1.00  \\ 

2CXO J054145.9-015626 & 85.44146 & -1.94077 & 1 & 177 & 0.98 & 44.44$_{-1.23}^{+1.40}$ & 1.93$_{-0.13}^{+0.14}$ & 23.23$_{-0.22}^{+0.17}$ & 0.530$_{-0.221}^{+0.168}$ & 1.00  \\ 

2CXO J054146.1-015415 & 85.44217 & -1.90418 & 1 & 1710 & 0.99 & 42.88$_{-1.49}^{+2.90}$ & 1.56$_{-0.05}^{+0.09}$ & 22.64$_{-0.02}^{+0.03}$ & 0.028$_{-0.013}^{+0.019}$ & 1.00  \\ 

2CXO J054146.2-015533 & 85.44262 & -1.92596 & 1 & 162 & 1.00 & 44.54$_{-1.30}^{+1.31}$ & 1.93$_{-0.13}^{+0.14}$ & 23.55$_{-0.33}^{+0.13}$ & 0.580$_{-0.343}^{+0.096}$ & 0.91  \\ 

2CXO J054147.4-015526 & 85.44771 & -1.92406 & 1 & 894 & 0.98 & 44.75$_{-1.51}^{+0.97}$ & 1.94$_{-0.12}^{+0.13}$ & 23.04$_{-0.32}^{+0.08}$ & 0.492$_{-0.390}^{+0.080}$ & 1.00  \\ 

2CXO J054553.6+002526 & 86.47333 & 0.42412 & 1 & 3979 & 0.99 & 43.93$_{-1.08}^{+1.66}$ & 1.78$_{-0.09}^{+0.11}$ & 22.70$_{-0.03}^{+0.05}$ & 0.061$_{-0.020}^{+0.022}$ & 1.00  \\ 

2CXO J060745.9-062457 & 91.94125 & -6.41607 & 1 & 317 & 0.98 & 44.18$_{-1.60}^{+1.41}$ & 2.00$_{-0.13}^{+0.12}$ & 22.86$_{-0.28}^{+0.21}$ & 0.577$_{-0.363}^{+0.276}$ & 1.00  \\ 

2CXO J060748.0-062230 & 91.95012 & -6.37521 & 1 & 175 & 1.00 & 44.10$_{-1.09}^{+1.22}$ & 2.03$_{-0.13}^{+0.12}$ & 23.05$_{-0.13}^{+0.23}$ & 0.525$_{-0.137}^{+0.332}$ & 1.00  \\ 

2CXO J060758.3-061405 & 91.99292 & -6.23473 & 1 & 317 & 0.89 & 44.16$_{-1.55}^{+1.71}$ & 1.80$_{-0.11}^{+0.13}$ & 22.64$_{-0.18}^{+0.23}$ & 0.427$_{-0.218}^{+0.312}$ & 1.00  \\ 

2CXO J060814.0-062559 & 92.05842 & -6.43320 & 1 & 2198 & 0.75 & 45.38$_{-1.23}^{+0.87}$ & 1.96$_{-0.08}^{+0.08}$ & 22.75$_{-0.35}^{+0.06}$ & 0.992$_{-0.578}^{+0.078}$ & 1.00  \\ 

2CXO J061049.4-061141 & 92.70600 & -6.19498 & 1 & 323 & 0.92 & 44.74$_{-0.99}^{+1.16}$ & 2.04$_{-0.12}^{+0.11}$ & 22.95$_{-0.14}^{+0.17}$ & 0.763$_{-0.166}^{+0.270}$ & 1.00  \\ 

2CXO J074012.5+742850 & 115.05242 & 74.48080 & 1 & 378 & 0.86 & 44.81$_{-1.45}^{+1.59}$ & 1.95$_{-0.13}^{+0.14}$ & 23.08$_{-0.31}^{+0.16}$ & 0.641$_{-0.397}^{+0.160}$ & 1.00  \\ 

2CXO J094533.6-141526 & 146.39038 & -14.25735 & 1 & 329 & 0.93 & 44.07$_{-1.13}^{+1.53}$ & 1.96$_{-0.14}^{+0.14}$ & 23.07$_{-0.08}^{+0.11}$ & 0.202$_{-0.052}^{+0.084}$ & 1.00  \\ 

2CXO J095550.3+694036 & 148.95975 & 69.67680 & 8 & 25931 & 0.67 & 47.06$_{-2.02}^{+1.08}$ & 2.23$_{-0.02}^{+0.03}$ & 23.72$_{-0.01}^{+0.01}$ & 4.998$_{-0.003}^{+0.001}$ & 1.00  \\ 

2CXO J095553.3+694101 & 148.97250 & 69.68387 & 2 & 1012 & 1.00 & 43.85$_{-1.47}^{+1.61}$ & 2.01$_{-0.13}^{+0.12}$ & 22.72$_{-0.13}^{+0.12}$ & 0.318$_{-0.137}^{+0.118}$ & 1.00  \\ 

2CXO J095554.6+694101 & 148.97771 & 69.68362 & 9 & 7385 & 1.00 & 44.21$_{-0.29}^{+0.77}$ & 2.46$_{-0.06}^{+0.06}$ & 22.86$_{-0.01}^{+0.02}$ & 0.342$_{-0.013}^{+0.011}$ & 1.00  \\ 

2CXO J112729.9+565249 & 171.87496 & 56.88042 & 1 & 485 & 0.80 & 45.14$_{-1.18}^{+1.09}$ & 2.00$_{-0.12}^{+0.11}$ & 22.97$_{-0.28}^{+0.09}$ & 0.905$_{-0.448}^{+0.083}$ & 1.00  \\ 

2CXO J122549.6+131052 & 186.45700 & 13.18126 & 1 & 287 & 1.00 & 45.47$_{-1.03}^{+0.98}$ & 1.91$_{-0.13}^{+0.14}$ & 23.85$_{-0.17}^{+0.21}$ & 1.871$_{-0.165}^{+1.461}$ & 1.00  \\ 

2CXO J123029.4+413927 & 187.62279 & 41.65769 & 3 & 1114 & 0.91 & 44.31$_{-0.99}^{+0.93}$ & 2.19$_{-0.09}^{+0.08}$ & 22.80$_{-0.27}^{+0.29}$ & 0.770$_{-0.376}^{+0.520}$ & 0.55  \\ 

2CXO J123038.4+413831 & 187.66017 & 41.64219 & 1 & 847 & 0.72 & 45.17$_{-0.90}^{+0.88}$ & 2.29$_{-0.09}^{+0.08}$ & 22.70$_{-0.21}^{+0.09}$ & 0.934$_{-0.354}^{+0.134}$ & 1.00  \\ 

2CXO J124211.1+323236 & 190.54629 & 32.54336 & 1 & 1139 & 0.98 & 45.25$_{-0.60}^{+0.49}$ & 2.12$_{-0.10}^{+0.09}$ & 23.11$_{-0.05}^{+0.05}$ & 0.756$_{-0.050}^{+0.063}$ & 0.98  \\ 

2CXO J130525.4-492831 & 196.35621 & -49.47554 & 4 & 2325 & 0.93 & 43.62$_{-1.10}^{+1.44}$ & 1.86$_{-0.09}^{+0.08}$ & 22.44$_{-0.06}^{+0.05}$ & 0.160$_{-0.041}^{+0.025}$ & 1.00  \\ 

2CXO J130540.7-492603 & 196.42000 & -49.43420 & 2 & 1226 & 0.88 & 44.34$_{-1.40}^{+0.97}$ & 2.08$_{-0.09}^{+0.09}$ & 22.80$_{-0.35}^{+0.13}$ & 0.799$_{-0.515}^{+0.184}$ & 1.00  \\ 

2CXO J132524.1-425959 & 201.35083 & -42.99981 & 3 & 786 & 0.82 & 43.74$_{-1.19}^{+1.82}$ & 1.80$_{-0.09}^{+0.11}$ & 22.50$_{-0.12}^{+0.09}$ & 0.442$_{-0.174}^{+0.063}$ & 1.00  \\ 

2CXO J132525.7-430055 & 201.35729 & -43.01546 & 3 & 2055 & 0.20 & 44.51$_{-1.40}^{+0.86}$ & 2.80$_{-0.00}^{+0.00}$ & 22.29$_{-0.10}^{+0.05}$ & 0.379$_{-0.121}^{+0.058}$ & 1.00  \\ 

2CXO J132526.1-430132 & 201.35912 & -43.02568 & 3 & 3442 & 0.08 & 45.13$_{-1.44}^{+0.90}$ & 2.80$_{-0.00}^{+0.00}$ & 22.68$_{-0.05}^{+0.03}$ & 1.202$_{-0.085}^{+0.077}$ & 1.00  \\ 

2CXO J132959.0+471318 & 202.49612 & 47.22181 & 6 & 12364 & 0.67 & 44.99$_{-0.39}^{+0.40}$ & 2.41$_{-0.01}^{+0.01}$ & 22.66$_{-0.02}^{+0.09}$ & 0.776$_{-0.032}^{+0.148}$ & 1.00  \\ 

2CXO J133613.3+374830 & 204.05562 & 37.80859 & 1 & 516 & 0.75 & 45.89$_{-1.09}^{+0.94}$ & 1.94$_{-0.13}^{+0.14}$ & 23.40$_{-0.24}^{+0.12}$ & 1.699$_{-0.314}^{+0.181}$ & 0.98  \\ 

2CXO J133629.1-295122 & 204.12133 & -29.85625 & 3 & 570 & 0.97 & 44.09$_{-1.37}^{+1.62}$ & 1.96$_{-0.12}^{+0.12}$ & 22.88$_{-0.12}^{+0.12}$ & 0.322$_{-0.159}^{+0.122}$ & 1.00  \\ 

2CXO J133707.1-295101 & 204.27962 & -29.85046 & 9 & 11200 & 0.81 & 44.24$_{-0.42}^{+0.78}$ & 2.37$_{-0.04}^{+0.03}$ & 22.50$_{-0.02}^{+0.05}$ & 0.405$_{-0.034}^{+0.092}$ & 1.00  \\ 

2CXO J140248.1+541350 & 210.70075 & 54.23071 & 4 & 1186 & 0.98 & 44.59$_{-0.49}^{+0.32}$ & 1.83$_{-0.11}^{+0.12}$ & 23.35$_{-0.04}^{+0.04}$ & 0.516$_{-0.027}^{+0.023}$ & 0.98  \\ 

2CXO J142333.7+240247 & 215.89071 & 24.04657 & 1 & 703 & 0.89 & 44.19$_{-1.75}^{+1.57}$ & 2.01$_{-0.11}^{+0.10}$ & 22.45$_{-0.18}^{+0.31}$ & 0.361$_{-0.194}^{+0.447}$ & 1.00  \\ 

2CXO J145358.8+033216 & 223.49538 & 3.53783 & 1 & 1338 & 0.98 & 44.75$_{-1.01}^{+1.16}$ & 1.84$_{-0.09}^{+0.10}$ & 22.75$_{-0.11}^{+0.11}$ & 0.382$_{-0.123}^{+0.182}$ & 1.00  \\ 

2CXO J150344.4-415722 & 225.93508 & -41.95616 & 1 & 203 & 0.96 & 44.16$_{-1.66}^{+1.50}$ & 1.92$_{-0.13}^{+0.13}$ & 22.90$_{-0.18}^{+0.17}$ & 0.360$_{-0.196}^{+0.163}$ & 1.00  \\ 

2CXO J151427.2-151039 & 228.61333 & -15.17765 & 1 & 599 & 0.86 & 45.95$_{-0.79}^{+0.61}$ & 1.97$_{-0.13}^{+0.12}$ & 23.39$_{-0.13}^{+0.10}$ & 2.110$_{-0.247}^{+0.151}$ & 1.00  \\ 

2CXO J162629.6-241905 & 246.62362 & -24.31823 & 1 & 335 & 0.96 & 44.61$_{-0.81}^{+0.95}$ & 2.19$_{-0.12}^{+0.10}$ & 22.99$_{-0.15}^{+0.13}$ & 0.858$_{-0.210}^{+0.201}$ & 0.83  \\ 

2CXO J162649.2-242003 & 246.70508 & -24.33417 & 1 & 1350 & 0.86 & 45.44$_{-0.29}^{+0.23}$ & 2.58$_{-0.18}^{+0.10}$ & 23.03$_{-0.03}^{+0.03}$ & 1.079$_{-0.041}^{+0.043}$ & 1.00  \\ 

2CXO J162654.4-242620 & 246.72688 & -24.43910 & 2 & 1118 & 0.91 & 44.96$_{-0.25}^{+0.30}$ & 2.62$_{-0.22}^{+0.10}$ & 23.02$_{-0.03}^{+0.08}$ & 1.003$_{-0.055}^{+0.195}$ & 0.96  \\ 

2CXO J162700.0-242640 & 246.75004 & -24.44462 & 2 & 468 & 1.00 & 43.84$_{-1.66}^{+1.61}$ & 1.83$_{-0.12}^{+0.13}$ & 23.15$_{-0.23}^{+0.08}$ & 0.319$_{-0.273}^{+0.036}$ & 1.00  \\ 

2CXO J162715.4-242640 & 246.81433 & -24.44445 & 1 & 1290 & 0.98 & 45.21$_{-0.38}^{+0.42}$ & 2.20$_{-0.10}^{+0.08}$ & 23.13$_{-0.05}^{+0.02}$ & 0.780$_{-0.049}^{+0.022}$ & 1.00  \\ 

2CXO J162717.5-242856 & 246.82325 & -24.48235 & 1 & 176 & 1.00 & 44.18$_{-1.48}^{+1.52}$ & 1.94$_{-0.13}^{+0.14}$ & 23.18$_{-0.24}^{+0.20}$ & 0.446$_{-0.251}^{+0.210}$ & 1.00  \\ 

2CXO J162727.0-243217 & 246.86279 & -24.53827 & 1 & 1072 & 0.99 & 44.57$_{-1.32}^{+0.80}$ & 2.23$_{-0.10}^{+0.10}$ & 23.03$_{-0.29}^{+0.24}$ & 0.547$_{-0.359}^{+0.327}$ & 1.00  \\ 

2CXO J162732.6-243324 & 246.88621 & -24.55675 & 1 & 336 & 0.98 & 44.40$_{-0.87}^{+1.05}$ & 2.20$_{-0.13}^{+0.10}$ & 22.98$_{-0.15}^{+0.17}$ & 0.654$_{-0.176}^{+0.283}$ & 1.00  \\ 

2CXO J162738.9-244020 & 246.91238 & -24.67238 & 1 & 760 & 1.00 & 44.02$_{-1.41}^{+1.25}$ & 1.97$_{-0.13}^{+0.12}$ & 23.02$_{-0.11}^{+0.14}$ & 0.174$_{-0.105}^{+0.124}$ & 1.00  \\ 

2CXO J162750.3-243149 & 246.95996 & -24.53043 & 1 & 192 & 0.92 & 44.39$_{-1.66}^{+1.92}$ & 1.97$_{-0.14}^{+0.13}$ & 23.11$_{-0.21}^{+0.15}$ & 0.484$_{-0.211}^{+0.140}$ & 1.00  \\ 

2CXO J190141.5-365831 & 285.42317 & -36.97537 & 5 & 2641 & 0.85 & 45.71$_{-0.39}^{+0.27}$ & 2.25$_{-0.06}^{+0.06}$ & 23.10$_{-0.05}^{+0.04}$ & 1.070$_{-0.066}^{+0.067}$ & 1.00  \\ 

2CXO J190148.0-365722 & 285.45008 & -36.95620 & 8 & 3919 & 0.99 & 45.34$_{-0.17}^{+0.22}$ & 2.24$_{-0.06}^{+0.06}$ & 23.11$_{-0.02}^{+0.02}$ & 0.661$_{-0.017}^{+0.025}$ & 1.00  \\ 

2CXO J190150.6-365809 & 285.46112 & -36.96935 & 4 & 2398 & 0.94 & 45.43$_{-0.45}^{+0.47}$ & 1.93$_{-0.08}^{+0.07}$ & 23.03$_{-0.07}^{+0.10}$ & 0.867$_{-0.105}^{+0.164}$ & 0.76  \\ 

2CXO J203348.6-593641 & 308.45279 & -59.61160 & 1 & 1328 & 0.97 & 44.00$_{-1.70}^{+1.95}$ & 1.63$_{-0.09}^{+0.08}$ & 22.57$_{-0.10}^{+0.09}$ & 0.137$_{-0.105}^{+0.076}$ & 1.00  \\ 

2CXO J204242.0-503830 & 310.67508 & -50.64181 & 1 & 251 & 0.86 & 45.90$_{-0.68}^{+0.48}$ & 1.90$_{-0.13}^{+0.13}$ & 23.95$_{-0.14}^{+0.12}$ & 2.693$_{-0.330}^{+0.232}$ & 0.96  \\ 

2CXO J215141.1-055049 & 327.92138 & -5.84711 & 1 & 605 & 1.00 & 45.92$_{-1.36}^{+0.94}$ & 1.93$_{-0.14}^{+0.13}$ & 24.01$_{-0.50}^{+0.24}$ & 1.998$_{-0.961}^{+0.081}$ & 0.99  \\ 

2CXO J215748.2-694153 & 329.45104 & -69.69824 & 2 & 1196 & 0.85 & 44.03$_{-1.33}^{+2.18}$ & 1.75$_{-0.10}^{+0.13}$ & 22.56$_{-0.10}^{+0.10}$ & 0.287$_{-0.085}^{+0.076}$ & 1.00  \\ 

2CXO J223846.8+751132 & 339.69538 & 75.19250 & 1 & 985 & 0.99 & 42.92$_{-0.66}^{+1.79}$ & 1.64$_{-0.10}^{+0.11}$ & 23.27$_{-0.02}^{+0.03}$ & 0.020$_{-0.007}^{+0.009}$ & 1.00  \\ 

2CXO J230139.0-400719 & 345.41262 & -40.12216 & 1 & 405 & 0.90 & 44.70$_{-2.06}^{+1.83}$ & 1.94$_{-0.14}^{+0.14}$ & 23.04$_{-0.23}^{+0.30}$ & 0.388$_{-0.258}^{+0.359}$ & 1.00  \\ 

2CXO J230748.9-223953 & 346.95412 & -22.66498 & 1 & 171 & 0.96 & 43.57$_{-1.46}^{+1.89}$ & 1.97$_{-0.13}^{+0.13}$ & 23.24$_{-0.09}^{+0.10}$ & 0.142$_{-0.059}^{+0.066}$ & 1.00  \\

 \hline

\caption{The X-ray redshift catalog for the 121 sources in our fully-filtered No-z data set. $N_\mathrm{s}$ is the number of available spectra; Cts. is the number of \textit{Chandra} broad band spectral counts; log$L_X$ is the rest-frame 2--10 keV luminosity; and $P_+$ is MLP-computed probability that the redshift estimate is successful.}

\label{tab:noz_catalog}
\end{longtable*}

\newpage

\listofchanges
\end{document}